\documentclass[iop, revtex4]{emulateapj}

\usepackage{natbib,graphicx}
\usepackage{amsmath}

\usepackage{hyperref}

\usepackage{graphicx}

\begin{document}

%\linenumbers

\title{Classical Novae at Radio Wavelengths}
\author{ Laura Chomiuk\altaffilmark{1}, 
Justin D.\ Linford\altaffilmark{2}, 
 Elias Aydi\altaffilmark{1}, Keith W.\ Bannister\altaffilmark{3}, Miriam I.\ Krauss\altaffilmark{4}, Amy J.\ Mioduszewski\altaffilmark{2}, Koji Mukai\altaffilmark{5,6},  Thomas J.\ Nelson\altaffilmark{7}, Michael P.\ Rupen\altaffilmark{8}, Stuart D.\ Ryder\altaffilmark{9,10}, Jennifer L.\ Sokoloski\altaffilmark{11}, Kirill V.\ Sokolovsky\altaffilmark{1,12}, Jay Strader\altaffilmark{1}, 
Miroslav D.\ Filipovi\'c\altaffilmark{13}, Tom Finzell\altaffilmark{14}, Adam Kawash\altaffilmark{1}, Erik C.\ Kool\altaffilmark{15}, Brian D.\ Metzger\altaffilmark{11,16},   Miriam M.\ Nyamai\altaffilmark{17}, Val\'erio A.~R.~M.\ Ribeiro\altaffilmark{18},   Nirupam Roy\altaffilmark{19},  Ryan Urquhart\altaffilmark{1} \& Jennifer Weston\altaffilmark{20}     
}
\altaffiltext{1}{Center for Data Intensive and Time Domain Astronomy, Department of Physics and Astronomy, Michigan State University, East Lansing, MI 48824, USA}
\altaffiltext{2}{National Radio Astronomy Observatory, P.O. Box O, Socorro, NM 87801, USA}
\altaffiltext{3}{CSIRO Astronomy and Space Science, PO Box 76, Epping, New South Wales 1710, Australia }
\altaffiltext{4}{Pasadena City College, 1570 E. Colorado Blvd., Pasadena, CA 91106, USA} 
\altaffiltext{5}{Center for Space Science and Technology, University of Maryland Baltimore County, Baltimore, MD 21250, USA}
\altaffiltext{6}{CRESST and X-ray Astrophysics Laboratory, NASA/GSFC, Greenbelt MD 20771 USA}
\altaffiltext{7}{Department of Physics and Astronomy, University of Pittsburgh, 3941 O?Hara St., Pittsburgh, PA 15260, USA}
\altaffiltext{8}{Herzberg Institute of Astrophysics, National Research Council of Canada, Penticton, BC V2A 6J9, Canada}
\altaffiltext{9}{Department of Physics and Astronomy, Macquarie University, NSW 2109, Australia}
\altaffiltext{10}{Macquarie University Research Centre for Astronomy, Astrophysics \& Astrophotonics, Sydney, NSW 2109, Australia}

\altaffiltext{11}{Columbia Astrophysics Laboratory, Columbia University, New York, NY, USA} 
\altaffiltext{12}{Sternberg Astronomical Institute, Moscow State University, Universitetskii~pr.~13, 119992 Moscow, Russia}
\altaffiltext{13}{Western Sydney University, Locked Bag 1797, Penrith South DC, NSW 2751, Australia}
\altaffiltext{14}{Department of Computational Math, Science, and Engineering, Michigan State University, East Lansing, MI 48824, USA}
\altaffiltext{15}{The Oskar Klein Centre, Department of Astronomy, Stockholm University, AlbaNova, SE-106 91 Stockholm, Sweden}
\altaffiltext{16}{Center for Computational Astrophysics, Flatiron Institute, 162 W. 5th Avenue, New York, NY 10011, USA}
\altaffiltext{17}{Department of Astronomy, University of Cape Town, Private Bag X3, Rondebosch 7701, South Africa}
\altaffiltext{18}{Instituto de Telecomunica\c{c}\~oes, Campus Universit\'ario de Santiago, 3810-193 Aveiro, Portugal}
\altaffiltext{19}{Department of Physics, Indian Institute of Science, Bangalore 560012,India}
\altaffiltext{20}{Federated IT, 1201 Wilson Blvd, 27th Floor, Arlington, VA 22209, USA}
\email{chomiuk@pa.msu.edu}

\begin{abstract}
We present radio observations (1--40 GHz) for 36 classical novae, representing data from over five decades compiled from the literature, telescope archives, and our own programs. Our targets display a striking diversity in their optical parameters (e.g., spanning optical fading timescales, $t_2 =$ 1--263 days), and we find a similar diversity in the radio light curves. Using a brightness temperature analysis, we find that radio emission from novae is a mixture of thermal and synchrotron emission, with non-thermal emission observed at earlier times.
We identify high brightness temperature emission ($T_B > 5 \times 10^4$ K) as an indication of synchrotron emission in at least 9 (25\%) of the novae.  We find a class of synchrotron-dominated novae with mildly evolved companions, exemplified by V5589~Sgr and V392~Per, that appear to be a bridge between classical novae with dwarf companions and symbiotic binaries with giant companions.
Four of the novae in our sample have two distinct radio maxima (the first dominated by synchrotron and the later by thermal emission), and in four cases the early synchrotron peak is temporally coincident with a dramatic dip in the optical light curve, hinting at a common site for particle acceleration and dust formation. 
We publish the light curves as tables and encourage use of these data by the broader community in multi-wavelength studies and modeling efforts.

\end{abstract}
\keywords{Cataclysmic variable stars (203), Novae (1127), White dwarf stars (1799), Galactic radio sources (571), Radio transient sources (2008)}

\section{Introduction} \label{intro}
Classical novae are thermonuclear eruptions on the surfaces of  white dwarf stars, occurring in a layer of material accreted from a binary companion \citep{Gallagher&Starrfield78, Bode&Evans08}. Novae are typically discovered and primarily observed as optical transient events, but
% which brighten as the ejecta expand, and fade as they become optically thin and drop in density. 
novae emit detectable radiation across the entire electromagnetic spectrum, from radio to $\gamma$-ray wavelengths. The long-standing canonical picture is that the bolometric luminosity of a nova is determined by the central white dwarf, which can continue nuclear fusion in residual material on its surface for days to years after eruption, maintaining near-Eddington luminosities ($\sim 10^{38}$ erg s$^{-1}$). This bolometric luminosity originally emerges at optical wavelengths, but as the ejecta expand, the peak of the spectral energy distribution shifts blueward before settling in the extreme UV/super-soft X-ray regime \citep{Gallagher&Code74, Schwarz+11}.

The recent discovery of GeV $\gamma$-rays from classical novae has led to a growing appreciation for the role of shocks and non-thermal radiation in shaping the emission signatures of novae \citep{Ackermann+14, Chomiuk+21}. The relatively high $\gamma$-ray luminosities observed from novae ($\sim 10^{35}-10^{36}$ erg s$^{-1}$) imply shock luminosities that rival the Eddington luminosity of a white dwarf
%, $\sim 10^{38}$ erg s$^{-1}$ representing data from over 5 decades compiled from the literature, telescope archives, and our own programs.
\citep{Metzger+15}. Signatures of shocks are also increasingly appreciated at wavelengths including (i) X-ray, where we observe relatively hard emission from hot shocked gas  \citep{Mukai+08, Gordon+21}; (ii) optical, where light curve features occur simultaneously with $\gamma$-ray features \citep{Li+17, Aydi+20}; and (iii) infrared, where shocks might lead to dust formation \citep{Derdzinski+17}.

Our understanding of radio emission from novae has also been undergoing a recent, dramatic shift. Radio observations of novae date back to the 1970s (Figure \ref{fig:lcpub1}), when the radio signatures were interpreted as thermal free-free emission from the warm, ionized, expanding ejecta \citep{Seaquist&Palimaka77, Hjellming+79}. Much as at optical wavelengths---but more simply, because line absorption/emission should not be a significant issue---the radio luminosity at a given frequency increases until the optical depth declines to $\tau \approx 1$ within the ejecta, at which point the light curve turns over and begins to fade as the ejecta continue to drop in density. Radio light curves evolve slowly, over years, and most of these first radio observations did not commence until months after explosion. The thermal radio light curves of novae hold great promise as tracers of the ejecta mass and kinetic energy, as they should relatively directly and simply trace the ionized ejecta \citep{Bode&Evans08}.

As time went on, observations of novae pushed to earlier times in the eruption, over a larger radio frequency range, and sampled more diverse eruptions. Hints that radio emission from novae may be more complex than purely thermal came with V1370~Aql and QU~Vul (Figure \ref{fig:lcnew1}; \citealt{Snijders+87, Taylor+87}). The high brightness temperatures ($>10^5$ K) of early emission observed in these novae indicated a role for non-thermal emission. Since 2010 and the advent of the upgraded Karl G.~Jansky Very Large Array (VLA), more examples of novae with high-brightness-temperature radio emission and other evidence of synchrotron emission arrived with V1723 Aql (\citealt{Krauss+11, Weston+16a}; Figure \ref{fig:lcpub2}), V5589 Sgr (\citealt{Weston+16b}; Figure \ref{fig:lcpub2}), and V1324 Sco (\citealt{Finzell+18};  Figure \ref{fig:lcpub3}). Hints also emerged that even the thermal emission from novae is complex, sometimes showing multiple, aspherical components (as in V959 Mon; \citealt{Chomiuk+14}; Figure \ref{fig:lcpub3}), or prolonged/delayed ejection of material (as in T~Pyx; \citealt{Nelson+14}; Figure \ref{fig:lcpub2}).

It is now clear that radio emission from novae is diverse and complex, but we still do not fully understand how radio properties map to other properties of the nova eruption and host binary.  Do all novae show some synchrotron emission? What determines the synchrotron luminosity and timing? Can we use thermal radio light curves to derive ejecta masses with unprecedented accuracy? How do aspherical ejecta morphology and inclination effects impact the radio light curve? These are just some of the questions we can work to answer, if we study the radio properties of novae for a large, diverse population.

Here, we compile radio observations of Galactic classical novae, foregrounding observations obtained with the VLA (both before and after its 2010 upgrade), but also including data from other telescopes like the Australia Telescope Compact Array (ATCA) and the Westerbork Synthesis Radio Telescope (WSRT). 
%This work represents the culmination of a decade of effort by the ENova collaboration, which had the goal of observing and understanding the radio emission of novae with the Jansky VLA.
 The dataset represents the work of five decades, with eruptions spanning 1968--2020 and radio observations spanning 1970--2020 (Tables \ref{tab:props} and \ref{tab:radio}). By collecting published data, reducing archival data, and acquiring new data, we present radio light curves for 36 novae (24 with new data presented here), and include poorer-quality data for 7 more objects in the Appendix (mostly non-detections). 
In \S \ref{sec:data}, we describe the nova sample and the data acquisition and reduction. In \S \ref{sec:discussion}, we analyze these light curves by considering not only radio flux density, but also luminosity and brightness temperature. We consider the relative roles of thermal and non-thermal emission in the radio light curves, and conclude in \S \ref{sec:conclusion}. 

%should we note here that this is just the first paper in a series?

\section{Data} \label{sec:data}

\subsection{Sample}\label{sec:sample}
We worked to collect all radio light curves for novae, including both published data sets and archival unpublished VLA data. We searched the VLA archive for all novae that erupted since 1980, using the Galactic nova catalogs curated by the Central Bureau for Astronomical Telegrams\footnote{\url{http://www.cbat.eps.harvard.edu/nova_list.html}} and K.\ Mukai\footnote{\url{https://asd.gsfc.nasa.gov/Koji.Mukai/novae/novae.html}}. Most of the  recent VLA observations (after $\sim$2006) were obtained by our ENova collaboration, and we also include a few novae that our collaboration targeted with ATCA.
The sample of novae with sufficient radio data to enable light curve construction, including those already published, is listed in Tables \ref{tab:props} and  \ref{tab:radio}. Their diverse multi-wavelength properties are cataloged and discussed in Table \ref{tab:props} and \S \ref{sec:mwprop}. Novae are selected for radio campaigns for a multitude of reasons, including high optical brightness, GeV $\gamma$-ray detection, or recurrent nova status.

\begin{figure*}
{\includegraphics[width = 0.48\textwidth]{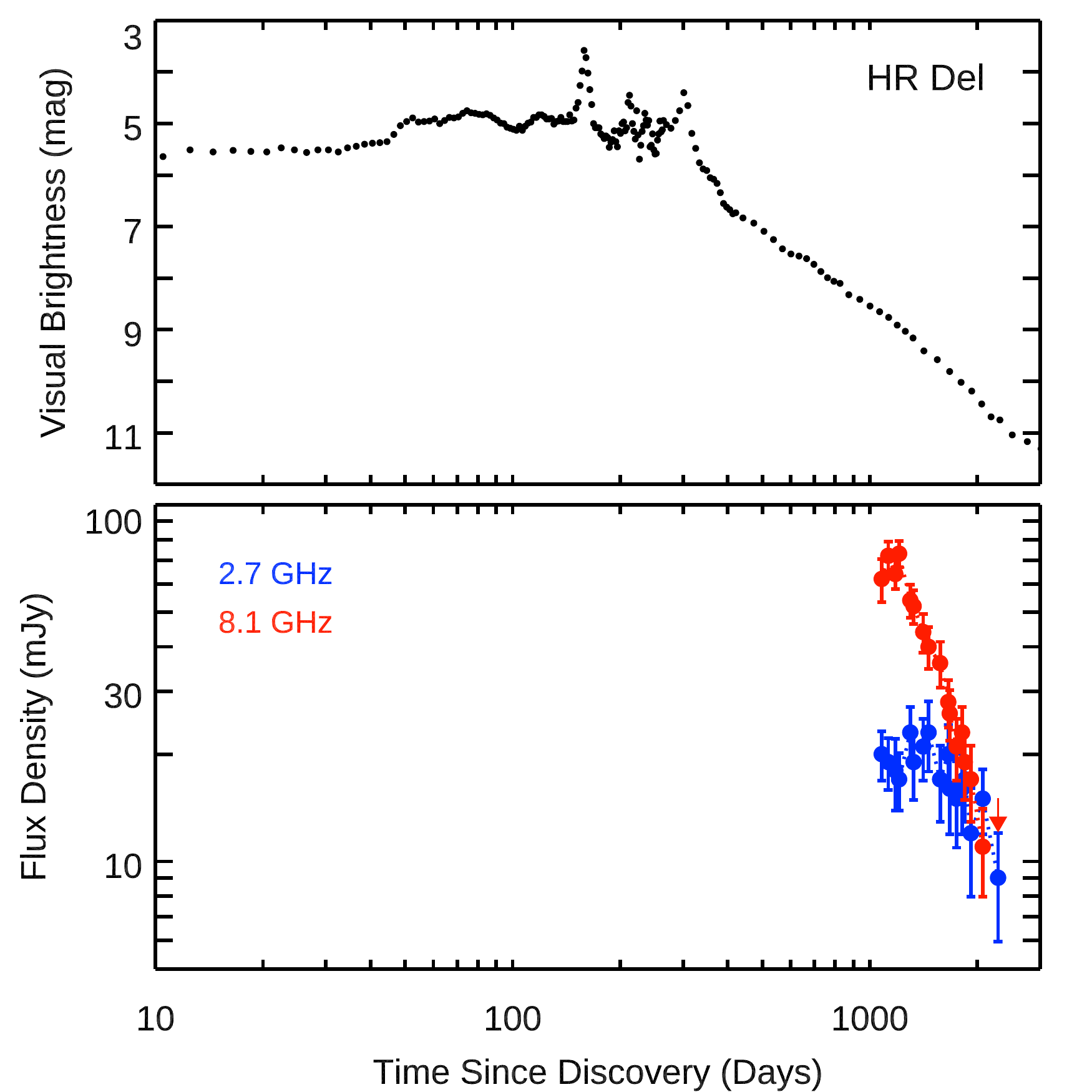}
\includegraphics[width = 0.48\textwidth]{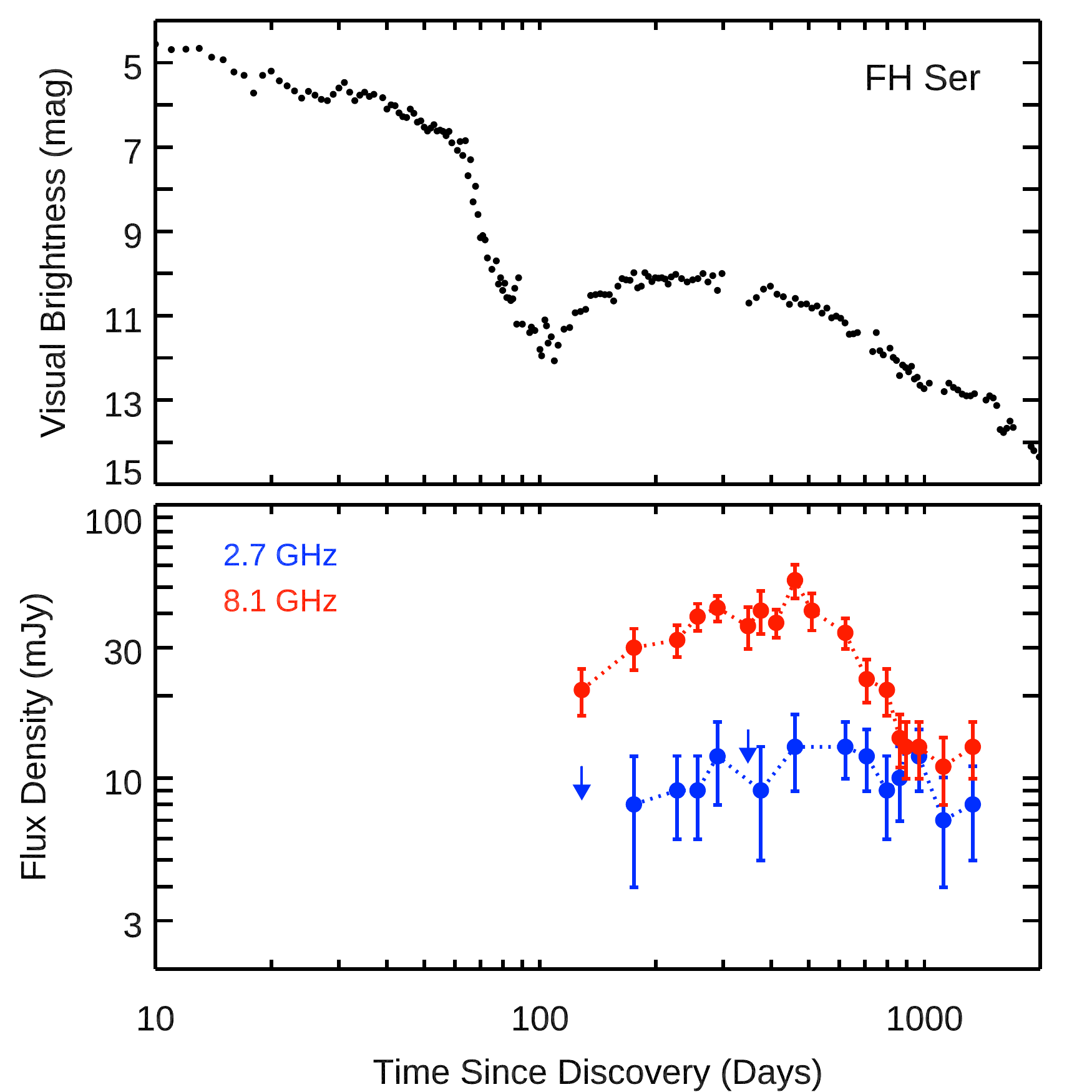}\\
\includegraphics[width = 0.48\textwidth]{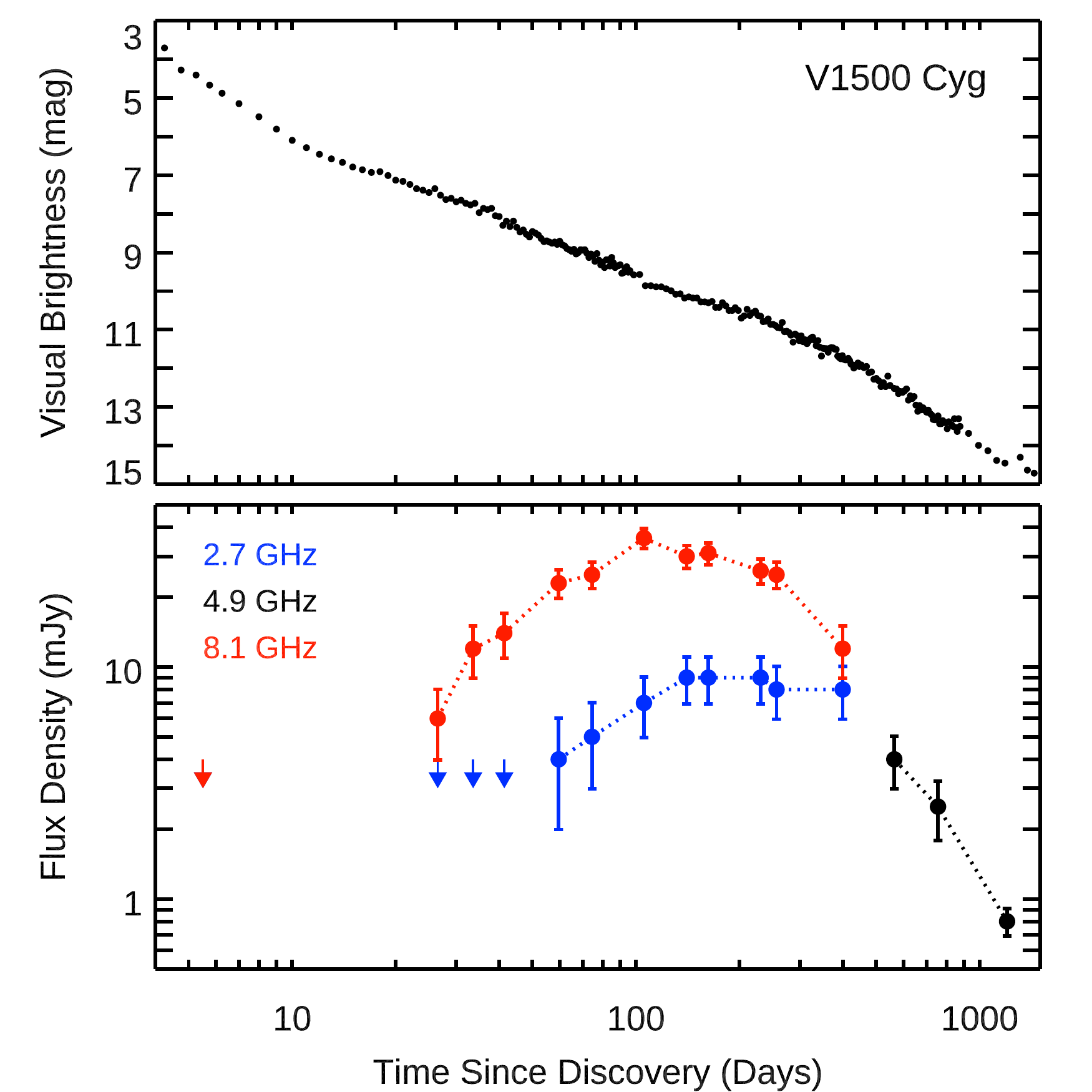}
\includegraphics[width = 0.48\textwidth]{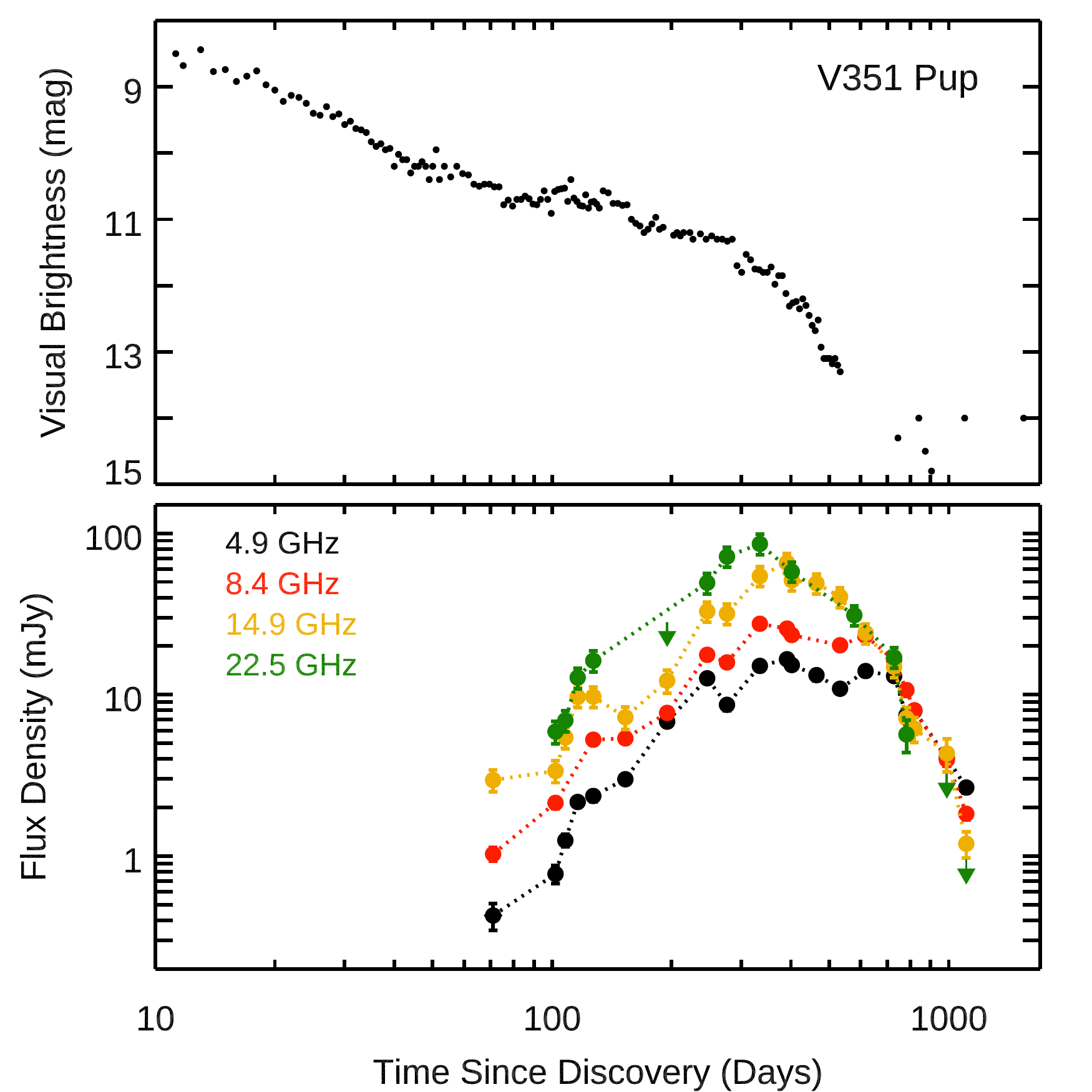}}
\caption{Previously published radio light curves, accompanied by optical light curves, for four novae (clockwise from top left): HR~Del (1967), FH~Ser (1970), V351~Pup (1991), and V1500~Cyg (1975). Radio epochs with non-detections are plotted with arrows. There are sporadic additional high-frequency ($\gtrsim$15 GHz) points not plotted here. The visual-band light curves are binned AAVSO data published by \citet{Strope+10}.}
\label{fig:lcpub1}
\end{figure*}

We do not consider recurrent novae (those observed in human history to have more than one nova explosion) as a unique subclass, but instead consider them as the extreme end of a continuum of recurrence times; all novae should recur if one waits long enough. We therefore include all novae with main sequence or sub-giant companions, whether they are ``recurrent" 
or ``classical" (only one outburst observed thus far). We exclude V445 Pup, as it is the only known helium nova and has a distinct synchrotron-dominated radio light curve \citep{Nyamai+21}.
We also exclude novae suspected to have more evolved red giant companions, but encourage a future study of embedded novae like RS~Oph, V745~Sco, V407~Cyg, V1534~Sco, V1535~Sco, and V3890~Sgr, keeping in mind that in some of these systems the orbital period is not yet known. We \emph{do} include V392~Per here, which has a binary orbital period of 3.4 days indicative of a companion that has just begun its ascent up the red giant branch and has luminosity class III/IV \citep{Munari+20}. As Munari et al.\ point out, V392~Per therefore serves as a useful touchstone bridging novae in cataclysmic variables and novae in symbiotic systems, and we include it here for this reason. 
%Jay notes that v1535 sco is actually more like v392 per in its soar data. but we haven't published that yet.

\begin{figure*}
{\includegraphics[width = 0.48\textwidth]{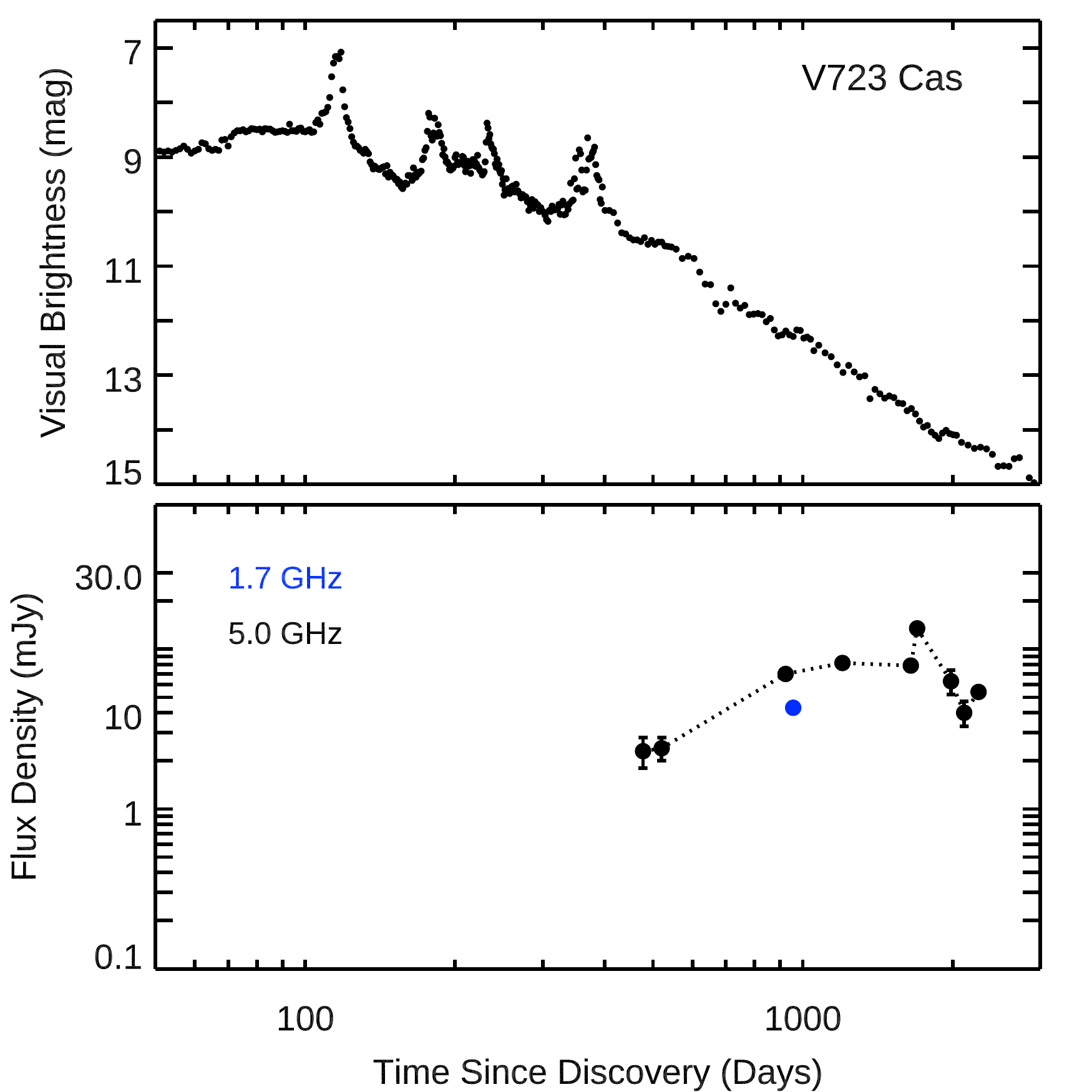}
\includegraphics[width = 0.48\textwidth]{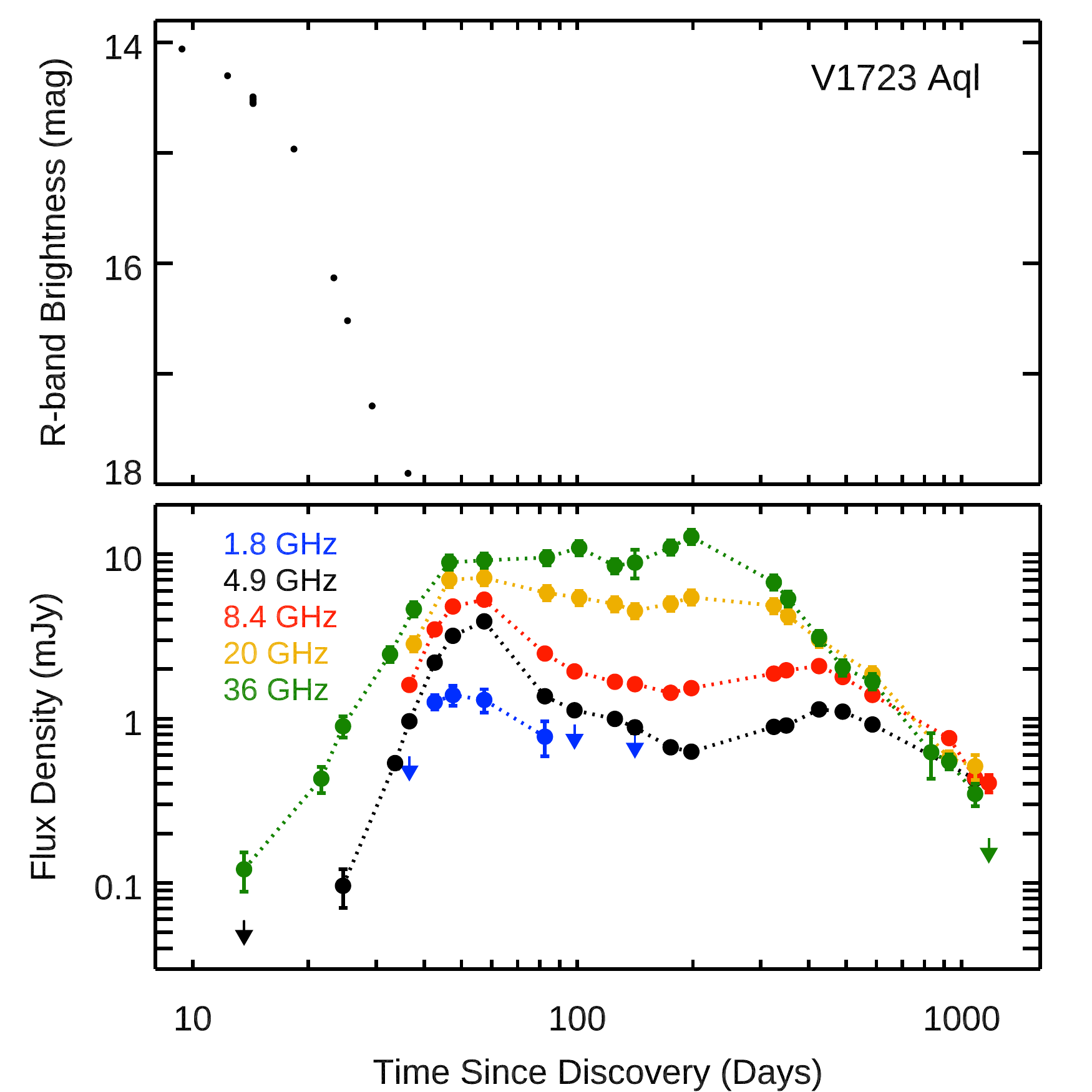}\\
\includegraphics[width = 0.48\textwidth]{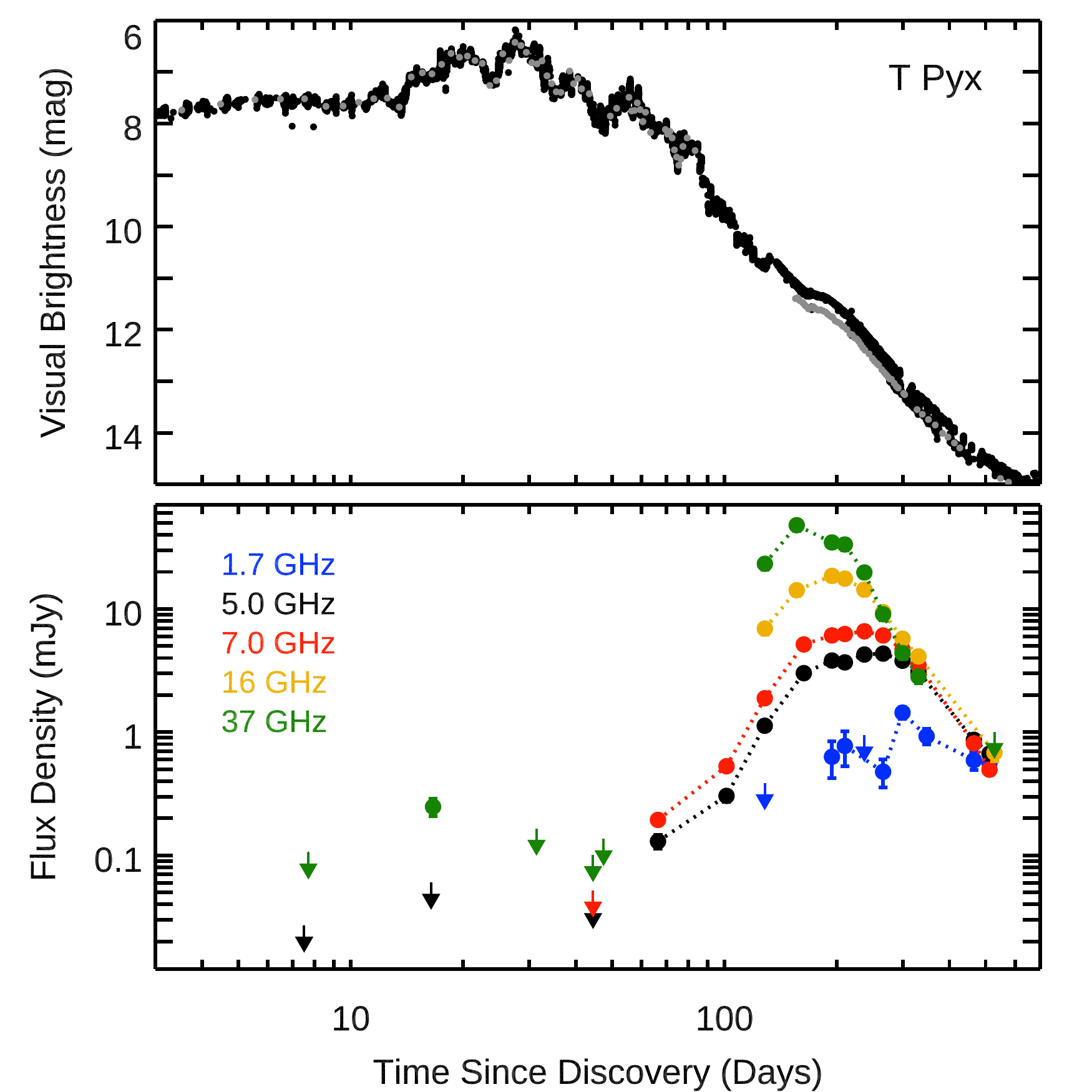}
\includegraphics[width = 0.48\textwidth]{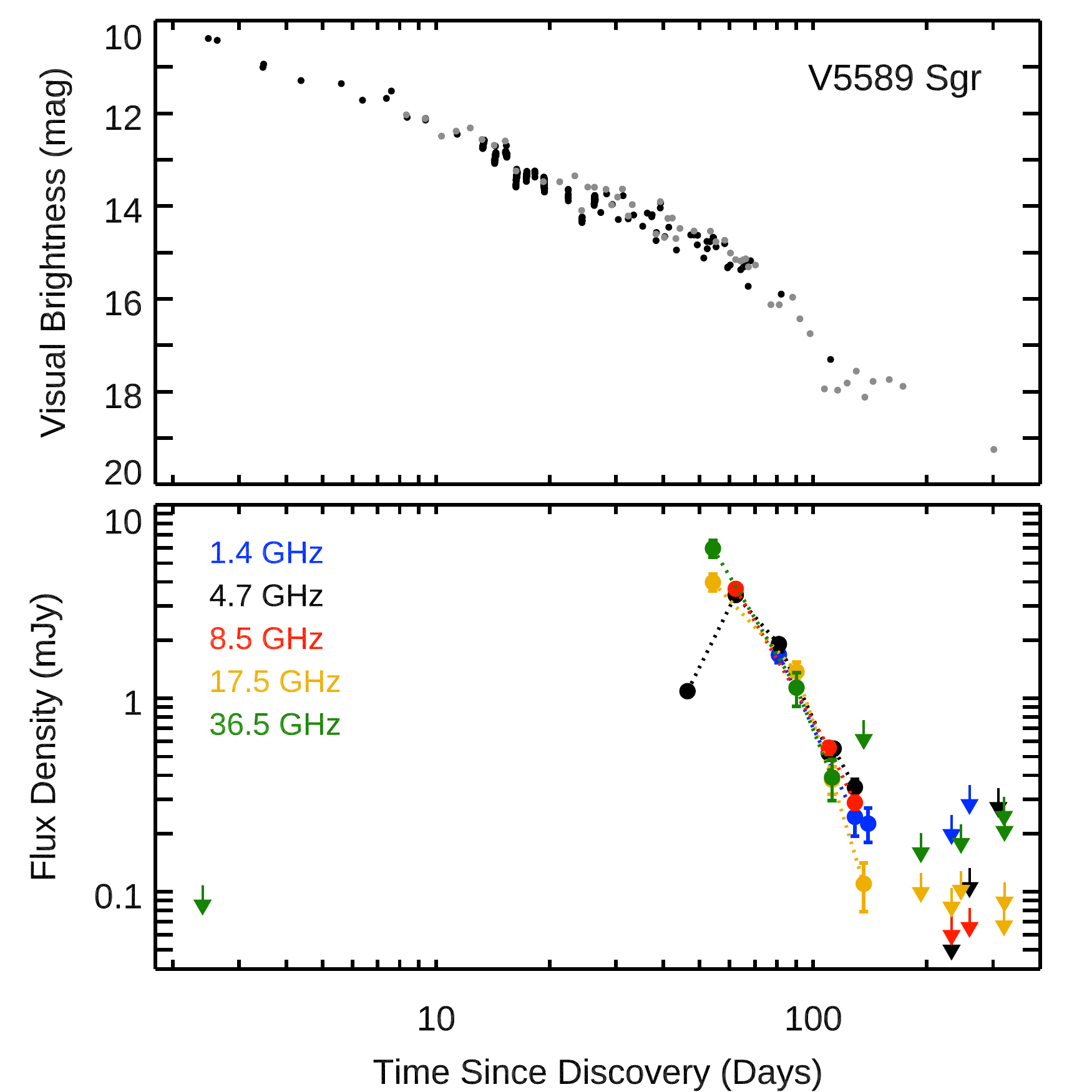}}
\caption{Previously published radio light curves, accompanied by optical light curves, for four novae (clockwise from top left): V723~Cas (1995), V1723~Aql (2010), V5589~Sgr (2012), and T~Pyx (2011). Epochs with non-detections are plotted with downward-facing arrows. Optical light curves are made using AAVSO data (black points; \citealt{Kafka20}) and the Stony Brook/SMARTS Atlas of Southern Novae (grey points; \citealt{Walter+12}). While most are plotted at $V$-band, V1723 Aql is plotted in $R$-band, as it suffered heavy foreground extinction and observations of its $V$-band light curve are limited.}
\label{fig:lcpub2}
\end{figure*}

Radio observation campaigns often last years for a single nova, and over half a century, observation campaigns start and terminate for a range of reasons. The ideal case is where observations begin early in the eruption, bridge the rise from non-detections to luminous emission, and then continue as the nova fades back to non-detections. The number of such thorough campaigns has been increasing,  but many of the radio observations presented in this paper fall short of this ideal. Campaigns may start too late and miss interesting early structure in the radio light curve. Campaigns may terminate too early---because of scheduling difficulties on telescopes or even changes in employment or telescope technology---and miss the evolution of the thermal maximum.

 There are also novae that have a few radio observations, but the data are not sufficient to really constrain the light curve evolution. For completeness, we have sought to collect these observations in the Appendix, but for clarity do not include them in the main discussions of this paper. The observation campaigns relegated to the Appendix tend to be a) only non-detections; b) only one or two epochs featuring detections that are consistent with thermal expansion. 
 
\begin{figure*}
{\includegraphics[width = 0.48\textwidth]{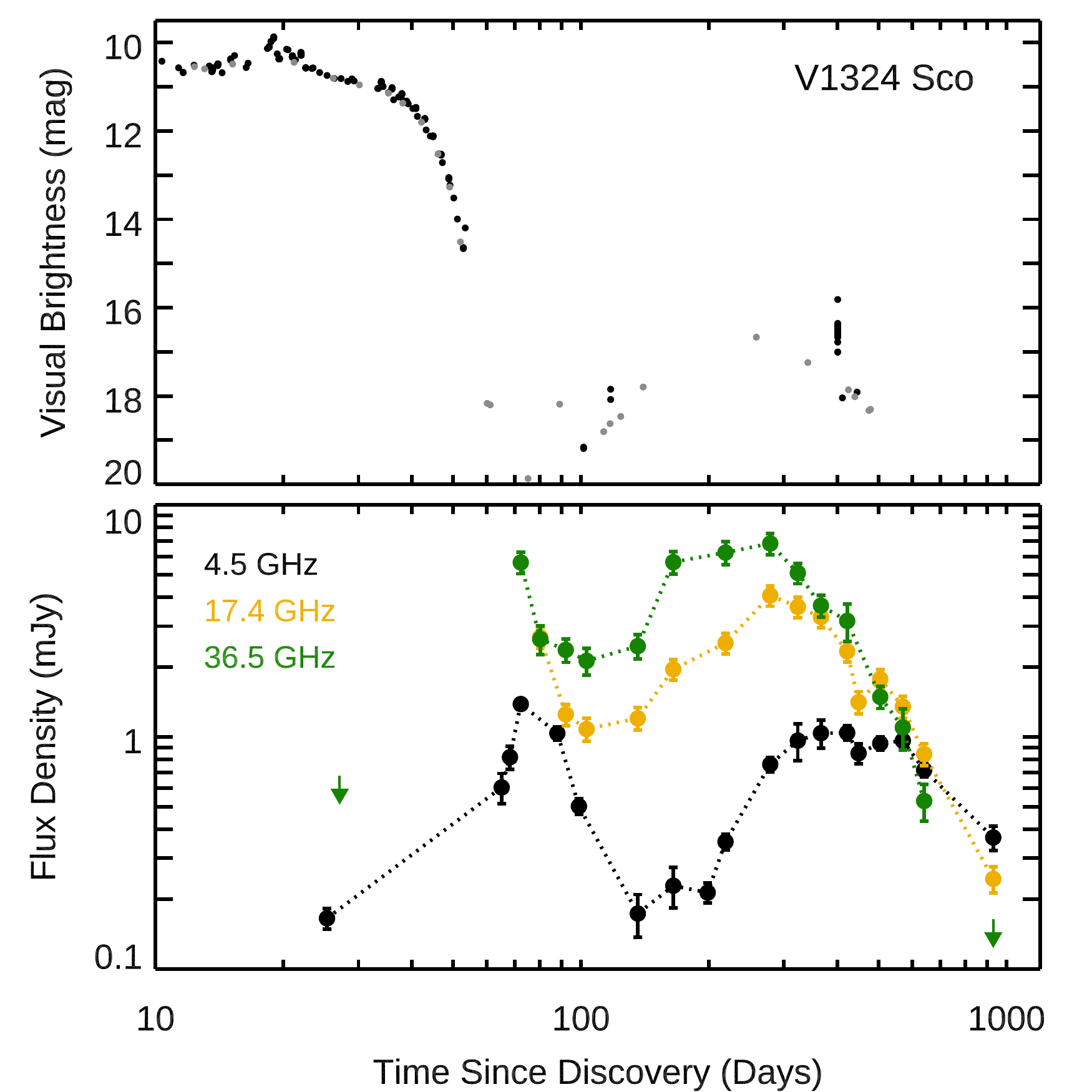}
\includegraphics[width = 0.48\textwidth]{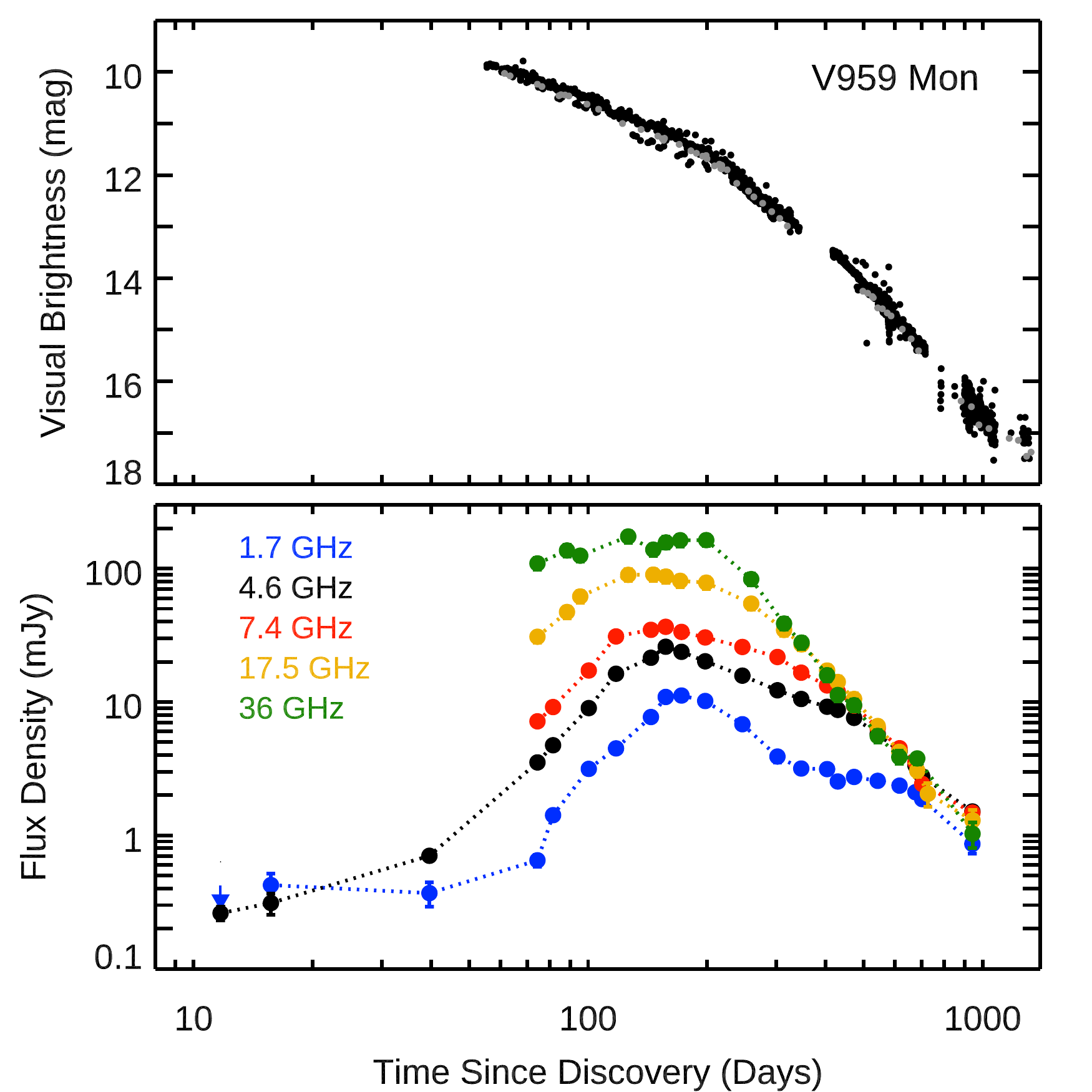}\\
\includegraphics[width = 0.48\textwidth]{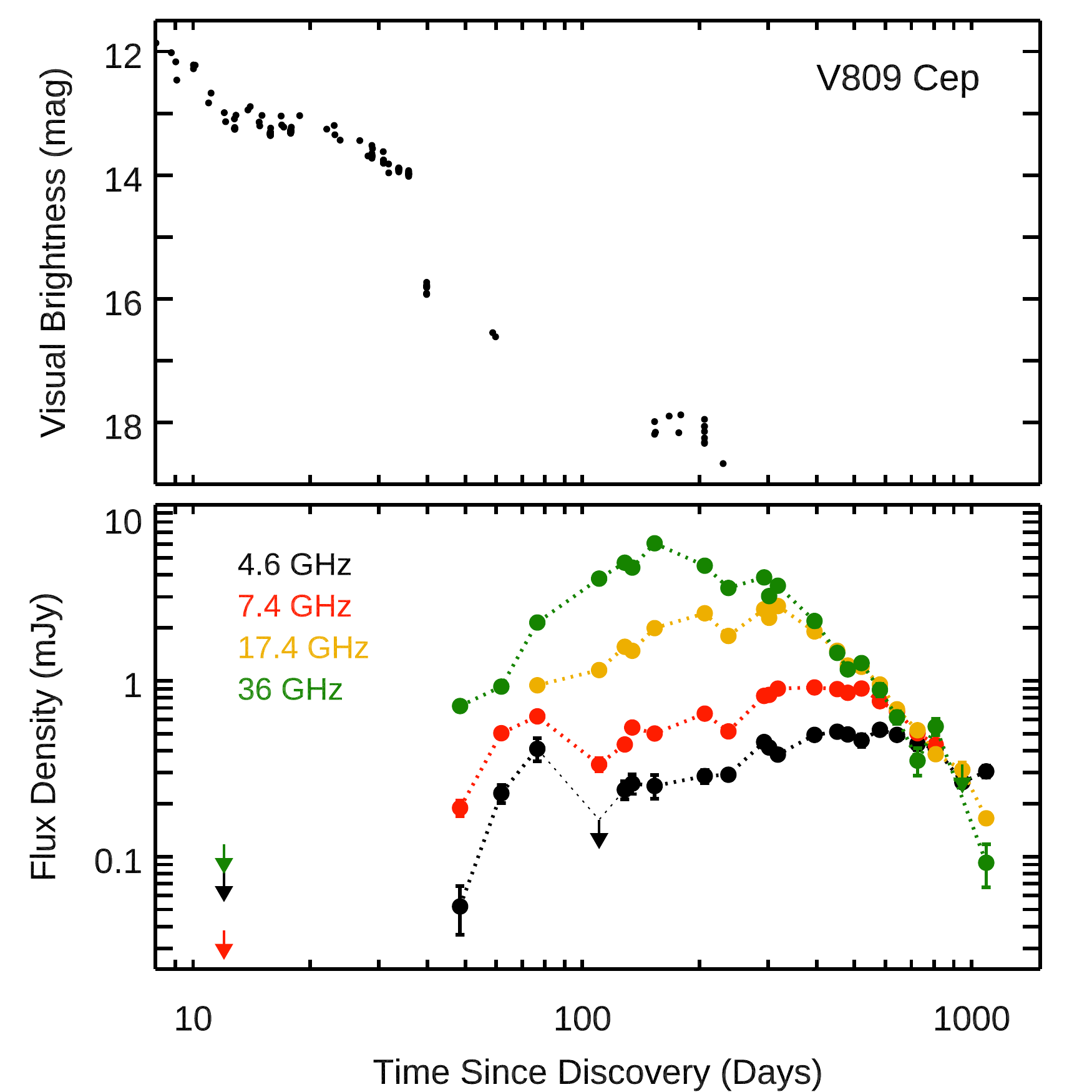}
\includegraphics[width = 0.48\textwidth]{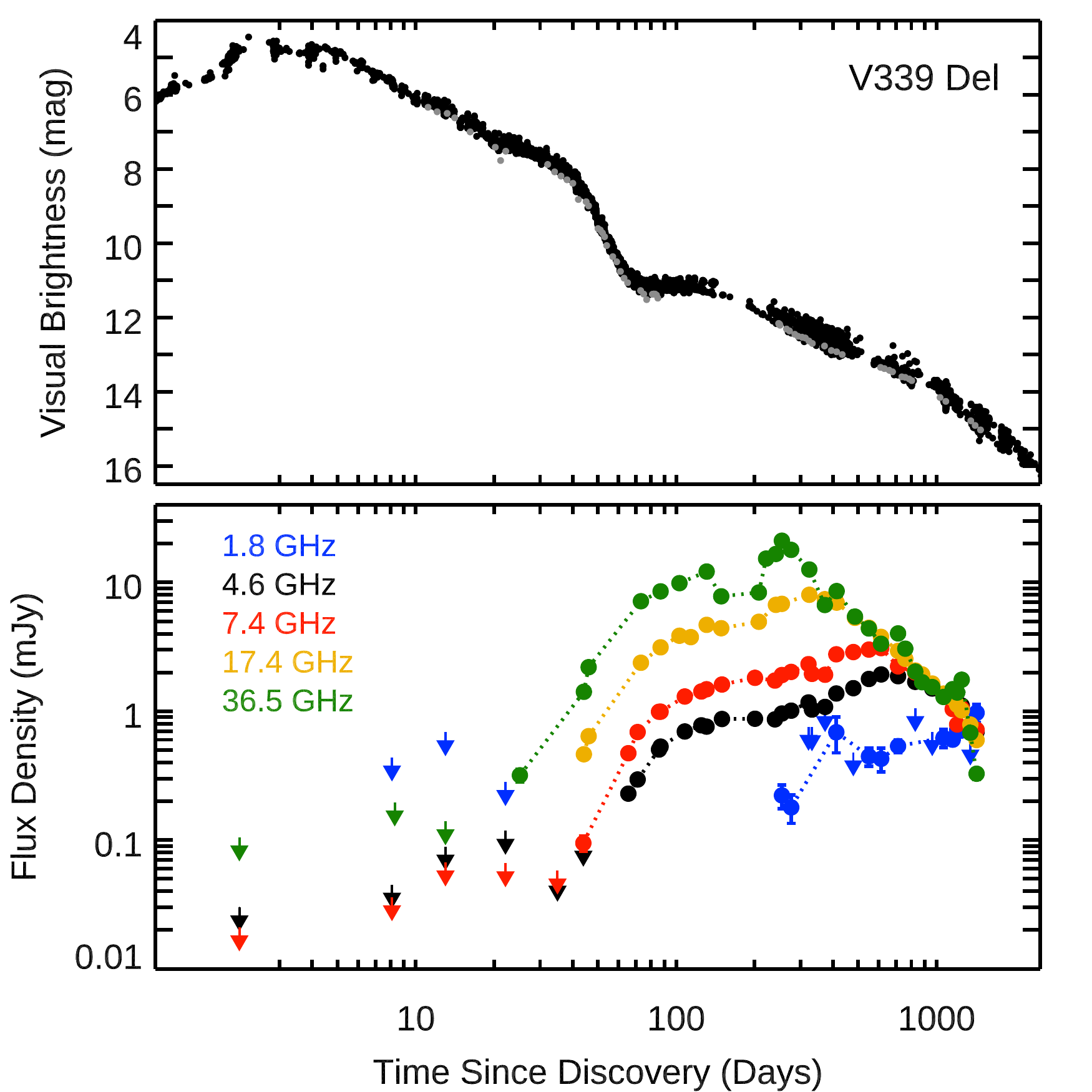}}
\caption{Previously published radio light curves, accompanied by optical light curves, for four novae (clockwise from top left): V1324~Sco (2012), V959~Mon (2012), V339~Del (2013), and V809 Cep (2013). Epochs with non-detections are plotted with downward-facing arrows. Optical light curves are made using AAVSO data (black points; \citealt{Kafka20}) and the Stony Brook/SMARTS data (grey points; \citealt{Walter+12}). V959~Mon lacks optical data for the first $\sim$2 months of its eruption, as it was discovered by \emph{Fermi}-LAT during solar conjunction.}
\label{fig:lcpub3}
\end{figure*}

 In Figures \ref{fig:lcpub1}--\ref{fig:lcpub3}, we plot the previously published radio light curves and place them in context of each nova's optical light curve. Figures \ref{fig:lcnew1}--\ref{fig:lcnew6} plot light curves for 24 novae with new radio observations. This sample is a mix of older archival data collected by diverse nova researchers and newer
 %In the following sub-sections, we discuss details of the radio observations for novae with previously unpublished VLA data. 
 % Our sample also contains a number of more recent novae 
data collected  by our team.
% but not yet published. Radio and optical light curves for these novae are plotted in Figures \ref{fig:lcnew4}--\ref{fig:lcnew6}.

\subsection{Radio Data Reduction} \label{sec:reduction}
For data obtained with the historic VLA (1980--2009), all data were obtained in continuum mode, with two intermediate frequencies each 50 MHz wide. Detailed information about the observations, including calibrators observed, is available in the NRAO archive\footnote{https://archive.nrao.edu/archive/advquery.jsp}. Data were typically obtained at C (4.9 GHz), X (8.4 GHz), U (14.9 GHz), and K (22.4 GHz) bands, with occasional observations at L band (1.4 GHz; novae are usually faintest at low frequencies because their emission is predominantly thermal). 
Data were edited, calibrated, and imaged using standard routines in AIPS \citep{Greisen03}. When possible, we estimated the flux density of the complex gain calibrator by a bootstrapping comparison with a standard absolute flux calibrator (e.g., 3C286). However, many observations did not include an observation of an absolute flux calibrator, and in these cases we are forced to assume a value for the complex gain calibrator, taking the flux estimates determined in adjacent epochs with full calibration.

Observations with the Karl G.\ Jansky VLA (2010--present) yield data with substantially larger bandwidths and volumes. Observations were typically obtained in the C (4--8 GHz), Ku (12--18 GHz), and Ka (26.5--40 GHz) bands, and sometimes at L band (1--2 GHz).
During the first few years of the Jansky VLA (i.e., $\sim$2010--2013), data were  obtained with 8-bit samplers, which yielded two chunks of 1 GHz bandwidth which can be flexibly placed and sampled with 2 MHz-wide channels. Later, we transitioned to using 3-bit samplers, which can yield up to 8 GHz of bandwidth. All observations were obtained in full Stokes mode, although polarization calibration was not typically carried out. As can be seen in Tables \ref{tab:v5666sgr}--\ref{tab:v392per}, sometimes observing frequencies drift over a monitoring campaign,  due to efforts to avoid RFI or changes in instrumental capabilities.  
Jansky VLA data were edited, calibrated, and imaged using standard routines in AIPS, CASA, and Difmap \citep{Greisen03, McMullin+07, Shepherd97}. A bright absolute flux calibrator doubled as the bandpass calibrator. 
As with historic VLA data, details of calibrators and observations are available in the NRAO archive. The flux density of the complex gain calibrator is determined by a bootstrapping comparison with the absolute flux calibrator.

VLA monitoring observations tend to be short ($\sim$5--15 minute) snapshot observations at any particular frequency. However, thanks to the many VLA baselines, uv-coverage is typically sufficient to yield high-quality images. Imaging was performed with a Briggs robust value of 0 or 1, and all images of novae resulted in point-like emission unless otherwise noted. 
Due to the monitoring nature of the observations, data were collected in all VLA configurations and in a range of conditions. Poor weather can decorrelate the signal in interferometric imaging, and thereby lead to underestimates of flux---especially at higher frequencies and in more extended configurations.
A single iteration of phase-only self-calibration was often carried out if there was enough flux in the image ($\geq 10$ mJy), in an attempt to correct for this decorrelation. 

Observations with the 
%Australia Telescope Compact Array (
ATCA made use of the 2 GHz (2048 $\times$ 1 MHz channels) bandwidth CABB system \citep{Wilson+11}, with central frequencies of 34.0, 19.0, 17.0, 9.0, 5.5, and 2.1 GHz. Due to the east--west nature of the ATCA configurations, observations spanned typically 3--6 hours to yield sufficient uv-coverage. A combination of online, automated, and manual flagging of the data was used to excise RFI, then the data were processed and analyzed using tasks within the Miriad package as outlined in the ATCA Users Guide\footnote{\url{https://www.narrabri.atnf.csiro.au/observing/users_guide/html/atug.html}}. The primary flux density and bandpass calibrator was PKS B1934--638, and no self-calibration was applied. Due to the change in beamsize across the 2 GHz band, the data spanning 1.1--3.1 GHz were divided into 4 sub-bands centered on 1.33, 1.84, 2.35, and 2.87 GHz prior to imaging. 
For the imaging, a robust weighting scheme with robustness parameter = 0.5 was used, and the image was cleaned down to a level approaching the theoretical noise. 

%Similar absolute flux calibration errors as for the VLA were assumed. In the absence of any clear detection at the expected position, upper limits of 3 times the measured image noise are given. 

For lower frequency observations ($<$10 GHz), we assumed a calibration error of 5\%, while for higher frequency observations ($>$10 GHz) we assumed 10\% calibration errors. In addition, there were some epochs where no flux calibrator was observed, and we used estimates of the phase calibrator's flux density from adjacent epochs. In these cases, we imposed an additional 5\% calibration error. These calibration errors were ignored in determining which measurements were detections (significant at $>3 \sigma$ level), but are quoted in Tables \ref{tab:v1370aql}--\ref{tab:v392per} as part of the uncertainty in flux density measurements. In cases of non-detection, upper limits on the flux density are estimated by adding three times the rms background noise to the measured flux at the nova position (if it is positive; if a negatively valued noise trough is observed at the nova position, then we just take three times the image noise).

%=================================================

\subsection{Notes on Radio Observations of Individual Sources}  \label{sec:indiv}

\subsubsection{V1370~Aql}
V1370~Aql was discovered on 1982 Jan 27 and was observed in the year following its outburst with the WSRT, primarily at 1.4 GHz \citep{Snijders+87}. The source was revisited on 1984 May 12 by \citet{Bode+87}, who observed it with the VLA at 4.9 GHz and reported a non-detection of $<$0.16 mJy. VLA observations were also obtained on 1985 Jun 14 (VLA program AT59; PI K.\ Turner) and 1987 Jan 24 (VLA program AH254; PI R.\ Hjellming). These observations also yield non-detections. We note a nearby uncatalogued radio source at RA =19h23m22.44s, Dec = $+2^{\circ}29^{\prime}04.3^{\prime\prime}$  (2000; $28^{\prime\prime}$ from the position of V1370~Aql), which is present in all three VLA epochs with a 4.9 GHz flux density of 1--2 mJy and which could cause some confusion if not treated carefully. The WSRT observations and VLA upper limits are plotted in Figure \ref{fig:lcnew1} and are listed in Table \ref{tab:v1370aql}.

\subsubsection{PW~Vul}
\begin{figure*}
{\includegraphics[width = 0.48\textwidth]{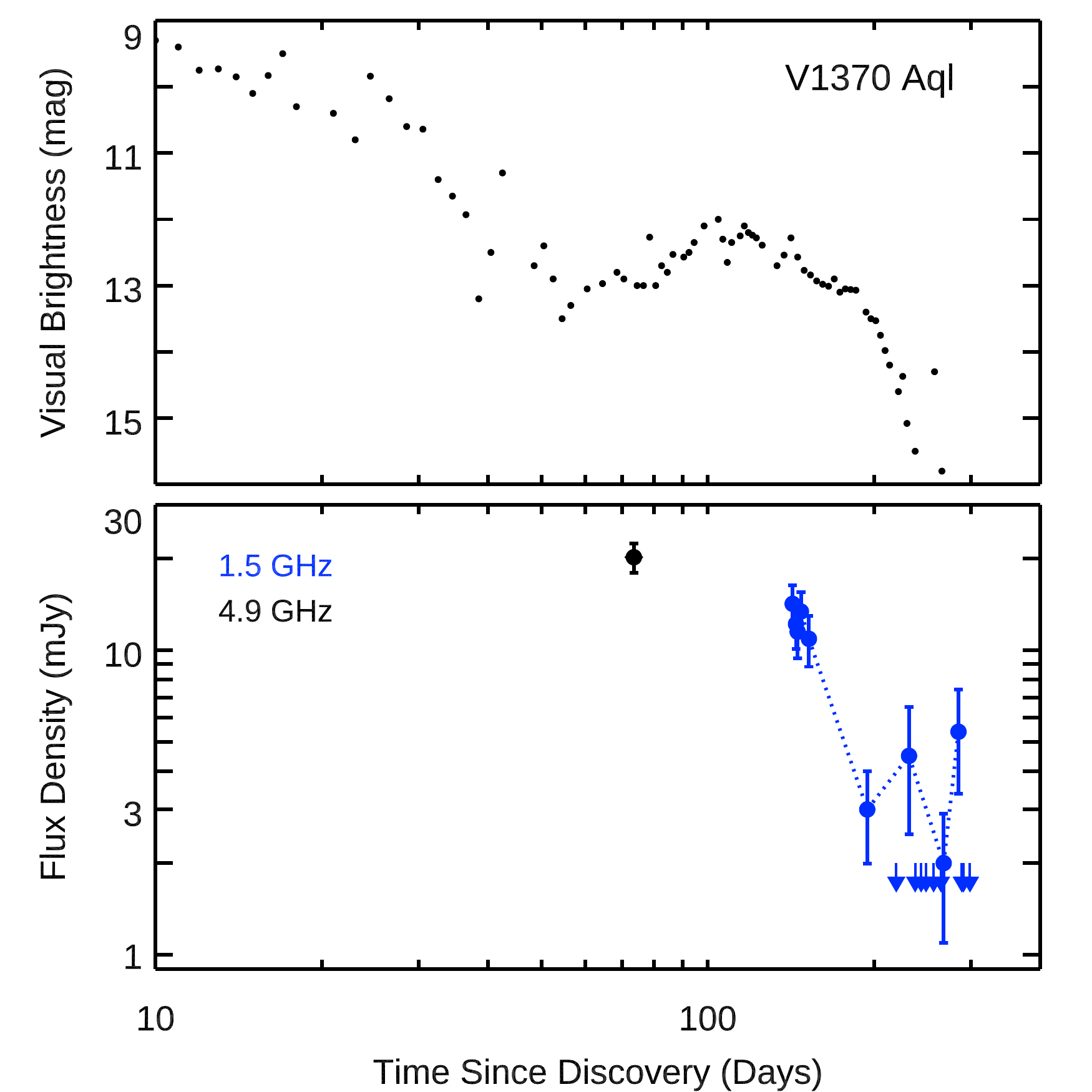}
\includegraphics[width = 0.48\textwidth]{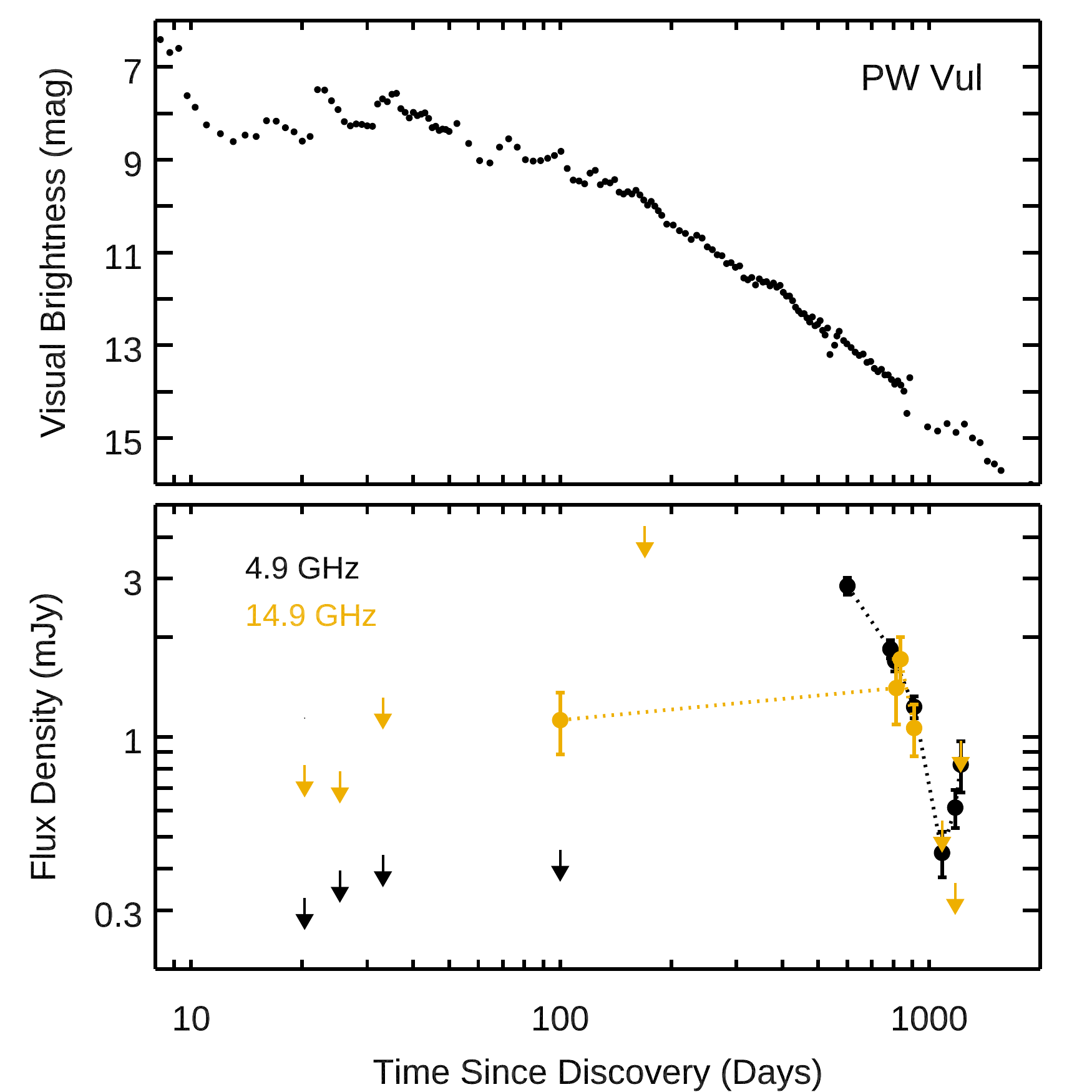}\\
\includegraphics[width = 0.48\textwidth]{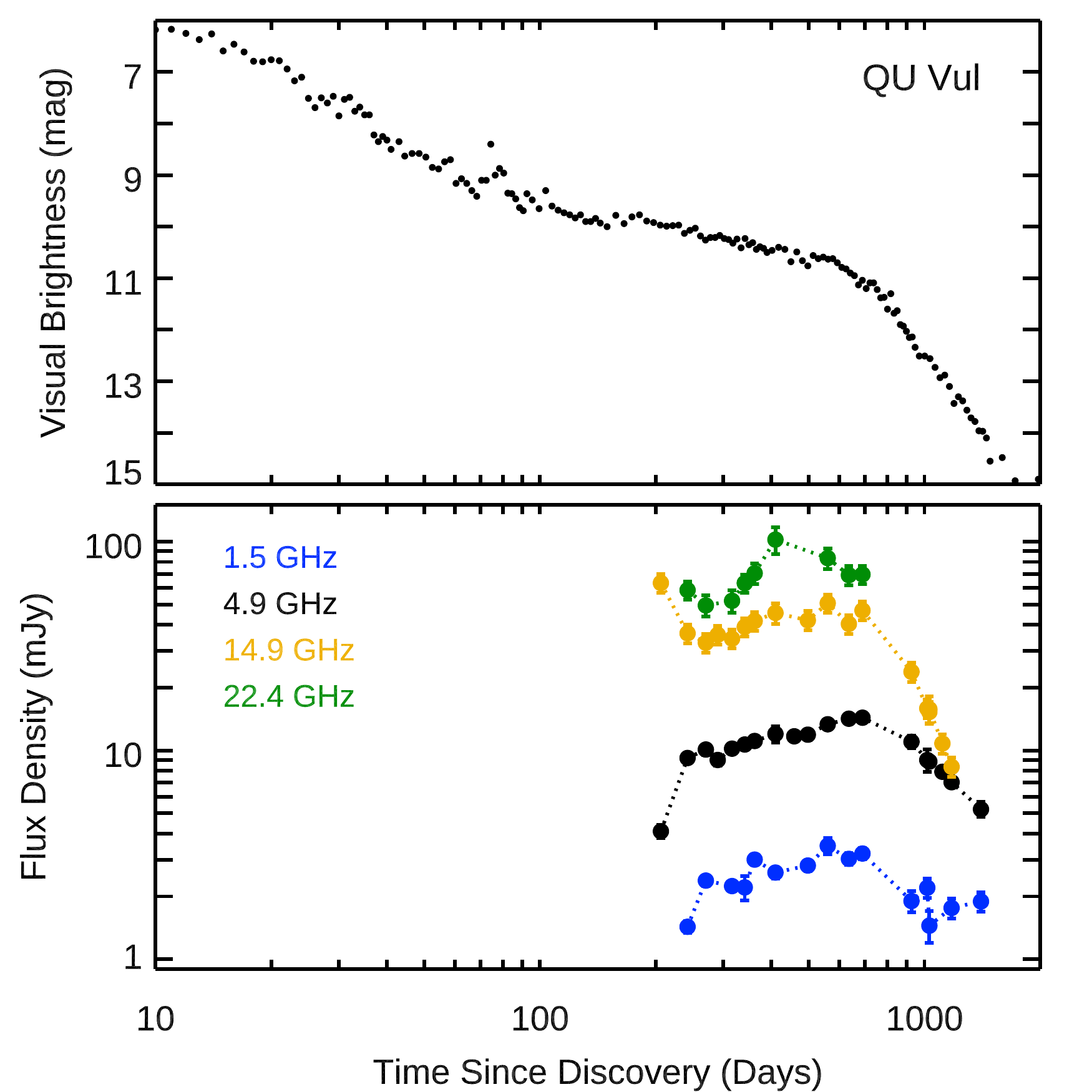}
\includegraphics[width = 0.48\textwidth]{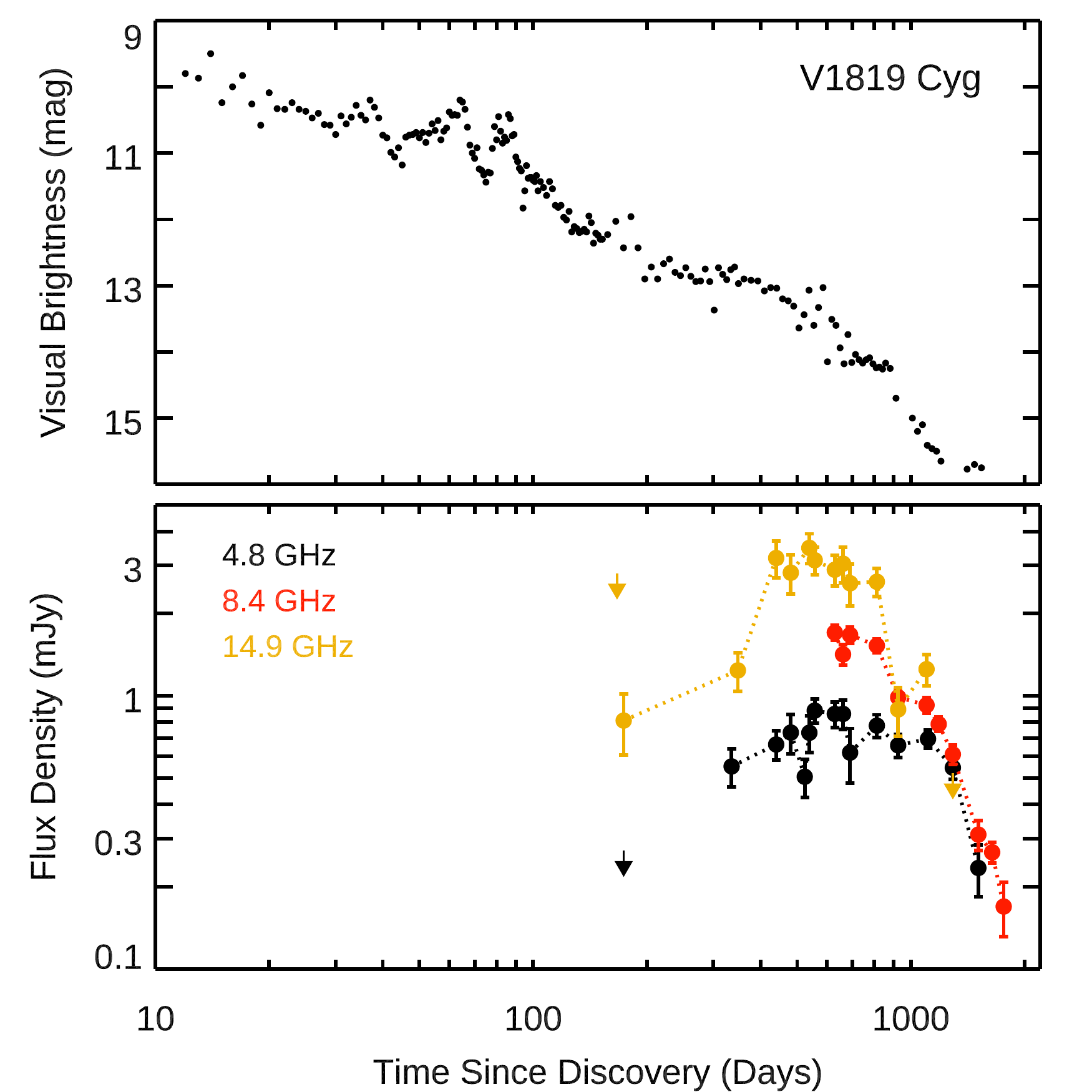}}
\caption{Optical and radio light curves for four novae (clockwise from top left): V1370~Aql (1982), PW~Vul (1984), V1819~Cyg (1986), and QU~Vul (1984). For radio epochs with non-detections, 3$\sigma$ upper limits are plotted with arrows; there are additional non-detections not plotted here, for clarity. The visual-band light curves are binned AAVSO data published by \citet{Strope+10}. }
\label{fig:lcnew1}
\end{figure*}

PW~Vul was observed with the VLA under program IDs AH179, AH195,  AH227, AH254, AH279 (PI R.\ Hjellming),
AH185 (PI G.\ Hennessy), AH199 (PI E.\ Hummel),  AT60,  AT69, and AT76 (PI A.~R.\ Taylor). It was observed during 1984 Aug 17--1987 Nov 28, encompassing 20--1218 days after discovery.  
%The complex gain calibrator was $1923+210$ for all observations except 1986 Nov 11, when the relatively distant calibrator $1741-038$ was used instead. 
%It is possible that this observation was therefore poorly calibrated, and the flux measurement from this date should be treated as a lower limit. 
PW~Vul is consistent with a point source in all epochs and frequencies, although it may be marginally resolved in A-configuration observations from 1987 Jul/Oct.
Flux measurements can be found in Table \ref{tab:pwvul}, and the light curve is plotted in Figure \ref{fig:lcnew1}. 
% radio position?

\clearpage
%\begin{turnpage}
\def\arraystretch{1.4}
\begin{deluxetable}{lllllccccccccccc}
\tablewidth{0 pt}
\tabletypesize{\footnotesize}
\setlength{\tabcolsep}{0.025in}
\tablecaption{ \label{tab:props}
Properties of Radio-Observed Novae (in Chronological Order of Discovery)}
\tablehead{Name & RA (J2000) & Dec (J2000) & Discovery Date & $V_{\rm peak}$ & t$_2$ &  E($B-V$) & $v_1$& $v_2$  & Distance\tablenotemark{$\dagger$} & P$_{\rm orb}$\tablenotemark{b}  \\
 & (h:m:s) & (d:$^{\prime}$:$^{\prime\prime}$) & (UT) & (mag) & (days) & (mag) & (km s$^{-1}$) & (km s$^{-1}$) & (kpc) & (hr)}
\startdata
HR~Del & 20 42 20.344 &	$+$19 09	39.16 & 1967 Jul 8\tablenotemark{g} & 3.6\tablenotemark{d} & 167\tablenotemark{a} &  $0.17\pm0.02$\tablenotemark{e} & 800\tablenotemark{h} & 1300\tablenotemark{h} & $0.87\pm0.02$ & 5.1\tablenotemark{f}   \\
 FH~Ser & 18 30 47.042 & $+$02 36 51.98 & 1970 Feb 13\tablenotemark{i} &  4.5\tablenotemark{d} & 49\tablenotemark{d} & $0.6\pm0.1$\tablenotemark{e} & 800\tablenotemark{j} & 1750\tablenotemark{j} & $1.00_{-0.06}^{+0.07}$ & MS\tablenotemark{k}  \\
 V1500~Cyg & 21 11	36.571 & $+$48 09 01.86 & 1975 Aug 29\tablenotemark{l} & 1.9\tablenotemark{d} & 2\tablenotemark{d} & $0.45\pm0.07$\tablenotemark{e} & ? & 4000\tablenotemark{m}  & $1.71_{-0.28}^{+0.40}$ & 3.4\tablenotemark{f}  \\
V1370~Aql & 19 23 21.241 & $+$02 29 26.20 & 1982 Jan 27\tablenotemark{n}   & 7.7\tablenotemark{d} & 15\tablenotemark{d} & $0.35\pm0.05$\tablenotemark{e} & ? & 2200\tablenotemark{o}  & $2.8_{-0.8}^{+1.7}$ & MS\tablenotemark{p}  \\
PW~Vul & 19 26 05.050 & $+$27 21 58.10 & 1984 Jul 28\tablenotemark{q}  & 6.4\tablenotemark{d} & 44\tablenotemark{d} & $0.6\pm0.1$\tablenotemark{e} & 800\tablenotemark{r} & 1650\tablenotemark{r} & $2.16_{-0.32}^{+0.45}$ & 5.1\tablenotemark{f}  \\
QU~Vul & 20 26 46.021 & $+$27 50 43.11  & 1984 Dec 22\tablenotemark{s} & 5.3\tablenotemark{d} & 20\tablenotemark{d} &  $0.55\pm0.05$\tablenotemark{e} & 700\tablenotemark{t} & 1400\tablenotemark{t} & $1.4_{-0.4}^{1.1}$ & 2.7\tablenotemark{f} \\
V1819~Cyg & 19 54 37.545 & $+$35 42 15.40 & 1986 Aug 4\tablenotemark{u}  & 9.3\tablenotemark{d} & 95\tablenotemark{d} & $0.35\pm0.15$\tablenotemark{e} & 650\tablenotemark{v} & 1450\tablenotemark{v} & $7.4\pm1.4$\tablenotemark{e}& ? \\
V827~Her & 18 43 42.506	 & $+$15 19 18.22 & 1987 Jan 25\tablenotemark{w} & 7.5\tablenotemark{d} & 21\tablenotemark{d} & 0.1\tablenotemark{e} & ? & 2000\tablenotemark{x}  & $2.1_{-0.7}^{+1.4}$ &? \\
V838~Her & 18 46 31.468	 & $+$12 14 02.00 & 1991 Mar 24\tablenotemark{y} & 5.3\tablenotemark{d} & 1\tablenotemark{d} & $0.5\pm0.1$\tablenotemark{e} & ? & 4500\tablenotemark{z} & $3.2_{-1.6}^{+2.5}$ & 7.1\tablenotemark{f} \\
V351~Pup & 08 11 38.391 & $-$35 07 30.27 & 1991 Dec 27\tablenotemark{aa} & 6.4\tablenotemark{d} & 9\tablenotemark{d} & $0.5\pm0.1$\tablenotemark{e} & ? & 3000\tablenotemark{ab} & $5\pm1.5$\tablenotemark{ac} & 2.8\tablenotemark{f} \\
V1974~Cyg & 20 30 31.651 & $+$52 37 50.74  & 1992 Feb 19\tablenotemark{ad}  & 4.3\tablenotemark{d} & 19\tablenotemark{d} & $0.26\pm0.03$\tablenotemark{e} & 1350\tablenotemark{ae}  & 2300\tablenotemark{ae}  & $1.58_{-0.13}^{+0.14}$ & 2.0\tablenotemark{f} \\
V705~Cas & 23 41 47.230 & $+$57 31 00.79  & 1993 Dec 7\tablenotemark{af} & 5.7\tablenotemark{d} & 33\tablenotemark{d} &$0.41\pm0.06$\tablenotemark{e} & ? & 1700\tablenotemark{ag}  &  $2.44_{-0.33}^{+0.44}$ & 5.5\tablenotemark{f}  \\
V1419 Aql & 19	 13 06.791 & $+$01 34 23.23 & 1993 May 14\tablenotemark{ah} & 7.6\tablenotemark{d} & 25\tablenotemark{d} & $0.50\pm0.05$\tablenotemark{e} & 900\tablenotemark{ai} & 1400\tablenotemark{ai} & $5.0_{-2.6}^{+4.0}$ &? \\
V723~Cas & 01 05 05.354 & $+$54 00 40.23 & 1995 Aug 24\tablenotemark{aj}  & 7.1\tablenotemark{d}  & 263\tablenotemark{d} & 0.45\tablenotemark{e} & ? & 1750\tablenotemark{ak} & $4.6	_{-0.6}^{+0.8}$ & 16.6\tablenotemark{f}    \\
U~Sco & 16 22 30.779 & $-$17 52 43.29 & 1999 Feb 24\tablenotemark{al}  & 7.5\tablenotemark{d} & 1\tablenotemark{d} & $0.14\pm0.12$\tablenotemark{e} & ? & 5000\tablenotemark{am} & $12\pm2$\tablenotemark{an} & 29.5\tablenotemark{f}   \\
V4743~Sgr & 19 01 09.339 & $-$22 00 06.12 & 2002 Sep 20\tablenotemark{ao} & 5.0\tablenotemark{d} & 6\tablenotemark{d} & 0.25\tablenotemark{e} & ? & 2700\tablenotemark{ap} & $3.7_{-0.7}^{+0.9}$ & 6.7\tablenotemark{f}  \\
V598~Pup & 07 05 42.501 & $-$38 14 39.32 & 2007 Jun 5\tablenotemark{aq} & 4\tablenotemark{aq} & 9\tablenotemark{aq} & $0.09\pm0.08$\tablenotemark{e} & ? & 2100\tablenotemark{aq} & $1.7\pm0.1$ & ?\\
 V2491~Cyg & 19 43 01.973 & $+$32 19 13.46 & 2008 Apr 10\tablenotemark{ar} & 7.5\tablenotemark{d} & 4\tablenotemark{d} & $0.23\pm0.01$\tablenotemark{e} & ? & 4500\tablenotemark{as} & $7.8_{-2.3}^{+3.5}$ & 2.6\tablenotemark{f} \\
V2672~Oph & 17 38 19.710 & $-$26 44 13.58  & 2009 Aug 16\tablenotemark{at} &11.4\tablenotemark{at} & 2\tablenotemark{au} & $1.6\pm0.1$\tablenotemark{e} & ? & 5000\tablenotemark{au} & $9.4_{-4.2}^{+6.7}$ & MS or SG\tablenotemark{au}  \\
V1723~Aql & 18 47 38.38\tablenotemark{a}  & $-$03 47 14.1 & 2010 Sep 11\tablenotemark{av} & 16.0\tablenotemark{aw} & 12\tablenotemark{aw}  & 4.3\tablenotemark{c}& ? & 1500\tablenotemark{av} & $5.7\pm0.4$\tablenotemark{ax} & ? \\
T~Pyx & 09 04 41.503 & $-$32 22 47.50  & 2011 Apr 14\tablenotemark{ay} & 6.4\tablenotemark{d} & 32\tablenotemark{d} & $0.25\pm0.02$\tablenotemark{d} & ? & 2000\tablenotemark{az} & $2.60_{-0.18}^{+0.20}$ & 1.8\tablenotemark{ba} \\
V5589~Sgr & 17 45 28.033 & $-$23 05 22.80 & 2012 Apr 21\tablenotemark{bz}  & 8.8\tablenotemark{e} & 4.5\tablenotemark{e} & $0.8\pm0.2$\tablenotemark{e} & ? & 4800\tablenotemark{bc}   & $7.7_{-3.2}^{+5.2}$ & 38.2\tablenotemark{bb}  \\ 
V1324~Sco & 17 50 53.94\tablenotemark{a}  & $-$32 37 20.5 & 2012 Jun 1\tablenotemark{bd}  & 10.1\tablenotemark{e} & 25\tablenotemark{e} & $1.2\pm0.1$\tablenotemark{e} & 1300 & 3200  & $>$6.5\tablenotemark{be} & 3.2\tablenotemark{f} \\
V959~Mon & 06 39 38.600 & $+$05 53 52.84  & 2012 Jun 19\tablenotemark{bf} & ? &? & $0.4\pm0.1$\tablenotemark{e} & 2100 & ? & $1.4\pm0.4$\tablenotemark{bg} & 7.1\tablenotemark{bh}  \\
V809 Cep & 23 08 04.722	 & $+$60 46 51.75  & 2013 Feb 2\tablenotemark{ac} & 11.2\tablenotemark{bi} & 16\tablenotemark{bi} & 1.7\tablenotemark{bi} & 800 & 1600 & $>$6.0\tablenotemark{bi} & MS or SG\tablenotemark{bi}  \\
 V339~Del & 20 23 30.682 & $+$20 46 03.64  & 2013 Aug 14  & 4.4\tablenotemark{bj} & 10\tablenotemark{bj}  & $0.18\pm0.04$\tablenotemark{bj} & 1000 & 2600 & $4.9\pm1.0$\tablenotemark{bk} & 3.1\tablenotemark{bl}  \\ 
V1369~Cen & 13 54 45.323 & $-$59 09 04.30 & 2013 Dec 2 & 3.3\tablenotemark{bk} & 40\tablenotemark{bk} & $0.06\pm0.01$\tablenotemark{bk} & 1150 & 1700  & $1.0\pm0.1$\tablenotemark{bk} & 3.8\tablenotemark{bm} \\
V5666~Sgr & 18 25 08.769 & $-$22 36 03.12 & 2014 Jan 26\tablenotemark{bn} & 8.7\tablenotemark{c} & 90\tablenotemark{c} & ? & 700 & 1200  & $6.1_{-1.5}^{+2.4}$ & ?  \\
V2659~Cyg & 20 21 42.321 & $+$31 03 29.29 & 2014 Mar 30\tablenotemark{bo} & 10.9\tablenotemark{c} & 115\tablenotemark{c} & 0.63\tablenotemark{bp} & 700 & 1300 & $6.8_{-1.7}^{+2.7}$ & ? \\
V5667~Sgr & 18 14 25.159 & $-$25	54 34.57 & 2015 Feb 12\tablenotemark{bq} & 9.0\tablenotemark{c} & 55\tablenotemark{c} &  ? & 1500& ? & $5.6_{-1.0}^{+1.5}$ & ? \\
V5668~Sgr & 18 36 56.84\tablenotemark{a}  & $-$28 55 40.1 & 2015 Mar 15\tablenotemark{br} & 4.3\tablenotemark{bs} & 100\tablenotemark{bs} & $0.7\pm0.1$\tablenotemark{bk} & 1100 & 1600  & $2.8\pm0.5$\tablenotemark{bk} & ? \\
V5855~Sgr & 18 10 28.29\tablenotemark{a}  & $-$27 29 59.4 & 2016 Oct 20\tablenotemark{bt} &  7.5\tablenotemark{e} & 18\tablenotemark{e} & 0.5\tablenotemark{e} & 800 & 2700 & $3.9\pm0.5$ & ? \\
V5856~Sgr & 18 20 52.25\tablenotemark{a}  & $-$28 22 12.2 & 2016 Oct 25\tablenotemark{bu}   & 5.9\tablenotemark{e} & 11\tablenotemark{e}  & 1.02\tablenotemark{e} & 900 & 2800 & $2.5\pm0.5$\tablenotemark{bk} &? \\
V357~Mus  & 11 26 15.003 & $-$65 31 24.21 & 2018 Jan 14\tablenotemark{bv} & 7.0\tablenotemark{bk} & 40\tablenotemark{bk} & $0.5\pm0.1$\tablenotemark{bk}  & 750 & 2500  & $3.3_{-0.7}^{+1.2}$ & ? \\
V906~Car & 10	 36 15.413 & $-$59 35 53.64   & 2018 Mar 16\tablenotemark{bw} & 5.9\tablenotemark{bw} & 44\tablenotemark{bw} & $0.35\pm0.05$\tablenotemark{bw}  & 350 & 2500 & $4\pm1.5$\tablenotemark{bw} & 1.6/3.3\tablenotemark{bw}   \\
V392 Per & 04 53 21.370 & $+$47 21 25.84 & 2018 Apr 29\tablenotemark{bx} & 5.6\tablenotemark{bk} & 3\tablenotemark{bk} & $0.40\pm0.05$\tablenotemark{bk} & 2800 & 4100 & $3.5_{-0.5}^{+0.7}$ & 81.9\tablenotemark{by} \\

\enddata
\tablenotetext{a}{Position was measured using VLA radio data presented here.}
\tablenotetext{$\dagger$}{Distances are determined from \emph{Gaia} EDR3 parallaxes, unless otherwise noted.}
\tablenotetext{b}{In some cases, the orbital period is not known, but the nature of the companion can be determined as main sequence (MS), sub-giant (SG), or red giant (RG).}
\tablenotetext{}{References: $^{c}$this work; $^{d}$\citet{Strope+10}; $^{e}$\citet{Ozdonmez+18}; $^{f}$\citet{Ritter&Kolb03}; $^{g}$\citet{Candy+67}; $^{h}$\citet{Hutchings70}; $^{i}$\citet{Seki+70}; $^{j}$\citet{Rosino+86}; $^{k}$\citet{Darnley+12}; $^{l}$\citet{Honda+75};   $^{m}$\citet{Hutchings+78}; $^{n}$\citet{Kosai+82};  $^{o}$\citet{Rosino+83};  $^{p}$\citet{Tappert+14}; $^{q}$\citet{Kosai+84}; $^{r}$\citet{Rosino&Iijima87}; $^{s}$\citet{Collins+84}; $^{t}$\citet{Rosino+92}; $^{u}$\citet{Wakuda&Huruhata86}; $^{v}$\citet{Andrillat&Houziaux89}; $^{w}$\citet{Kosai+87a}; $^{x}$\citet{Andrillat87her}; $^{y}$\citet{Sugano+91}; $^{z}$\citet{Iijima&Cassatella10}; $^{aa}$\citet{Camilleri+92};  $^{ab}$\citet{dellaValle+92}; $^{ac}$\citet{Wendeln+17}; $^{ad}$\citet{Collins+92};  $^{ae}$\citet{Chochol+93}; $^{af}$\citet{Nakano+93}; $^{ag}$\citet{Hauschildt+94a}; $^{ah}$\citet{Hirayama+93}; $^{ai}$\citet{Arkhipova+94};  $^{aj}$\citet{Hirosawa+95}; $^{ak}$\citet{Iijima06}; $^{al}$\citet{Schmeer+99}; $^{am}$\citet{Anupama+13}; $^{an}$\citet{Schaefer10}; $^{ao}$\citet{Kato+02}; $^{ap}$\citet{Morgan+03}; $^{aq}$\citet{Read+08}; $^{ar}$\citet{Nakano+08}; $^{as}$\citet{Munari+11b};  $^{at}$\citet{Nakano+09}; $^{au}$\citet{Munari+11}; $^{av}$\citet{Yamanaka+10} and \citet{Balam+10}; $^{aw}$\citet{Nagashima+13}; $^{ax}$\citet{Weston+16a}; $^{ay}$\citet{Waagan+11}; $^{az}$\citet{Pavana+19}; $^{ba}$\citet{Uthas+10}; $^{bb}$\citet{Mroz+15}; $^{bc}$\citet{Weston+16b};  $^{bd}$\citet{Finzell+18};  $^{be}$\citet{Finzell+15}; $^{bf}$\citet{Ackermann+14}; $^{bg}$\citet{Linford+15}; $^{bh}$\citet{Page+13}; $^{bi}$\citet{Munari+14}; $^{bj}$\citet{Chochol+14}; $^{bk}$\citet{Gordon+21}; $^{bl}$\citet{Chochol+15}; $^{bm}$\citet{Mason+21}; $^{bn}$\citet{Furuyama14}; $^{bo}$\citet{Nishiyama14}; $^{bp}$\citet{Raj+14}; $^{bq}$\citet{Nishiyama15}; $^{br}$\citet{Seach15}; $^{bs}$\citet{Banerjee+16}; $^{bt}$\citet{Nakano+16}; $^{bu}$\citet{Stanek+16}; $^{bv}$\citet{Kaufman+18}; $^{bw}$\citet{Aydi+20}; $^{bx}$\citet{Endoh+18}; $^{by}$\citet{Munari+20}; ; $^{bz}$\citet{Korotkiy+12}}
\end{deluxetable} 
\def\arraystretch{1.0}
\clearpage

\def\arraystretch{1.4}
\begin{deluxetable}{lccccccccccccc}
\tablewidth{0 pt}
\tabletypesize{\footnotesize}
\setlength{\tabcolsep}{0.025in}
\tablecaption{ \label{tab:radio} 
Overview of Radio Observations of Novae}
\tablehead{Name & Telescope & Date Range & $t_{\rm max}$\tablenotemark{e}  & $S_{\rm max}$\tablenotemark{e}  & Reference\\
 & & & (days) &  (mJy) & }
\startdata
HR~Del & Combination\tablenotemark{a} & 1970 Jun 22--1978 Dec 12 & 1208\tablenotemark{f} & 73\tablenotemark{f}  &  \citet{Hjellming+79}\\ 
FH~Ser & Combination\tablenotemark{b}  & 1970 Jun 22--1978 Dec 12 & 460\tablenotemark{f}  & 53\tablenotemark{f}  & \citet{Hjellming+79}\\
V1500~Cyg & Combination\tablenotemark{d} & 1975 Aug 31--1978 Dec 12 & 105\tablenotemark{f}  & 36\tablenotemark{f}  & \citet{Hjellming+79} \& \citet{Seaquist+80}\\
V1370~Aql & WSRT/VLA & 1982 Mar 27--1987 Jan 24 & $\leq$142\tablenotemark{g}  & $\geq$14.2\tablenotemark{g}  & \citet{Snijders+87}, \citet{Bode+87} \& This work\\
PW~Vul & VLA & 1984 Aug 17--1987 Nov 28 & 600.6 & 2.85 & This work\\
QU~Vul & VLA & 1985 Jul 16--1988 Oct 24 & 690.0 & 14.35 & \citet{Taylor+87, Taylor+88} \& This work\\
V1819~Cyg & VLA & 1987 Jan 17--1991 May 31 & 556.5 & 0.88 &This work \\
V827~Her & VLA & 1987 Jul 2--1989 Feb 14 & 307.1 & 2.02 & This work \\
V838~Her & VLA & 1991 Mar 28--1991 Oct 31 & 25.5 & 4.8 & This work \\
V351~Pup & VLA & 1992 Mar 7--1995 Jan 7 & 391.0 & 16.6 & \citet{Wendeln+17}\\
V1974~Cyg & VLA & 1992 Feb 25--1996 Feb 6 & 344.6 & 24.59 & \citet{Hjellming96} \& This work\\
V705~Cas & VLA/Merlin & 1993 Dec 24--1998 Dec 8& 108.5 & 3.37 &\citet{Eyres+00} \& This work \\
V1419 Aql & VLA & 1993 Jun 19--1995 Jul 5 & 214.8 & 0.83 & This work \\
V723~Cas & Merlin & 1996 Dec 13--2001 Oct 26 & 1697.5 & 13.5 & \citet{Heywood+05} \\
U~Sco (1987) & VLA & 1987 Jun 2--1987 Jul 5 & & &This work \\
 U~Sco (1999) & VLA & 1999 Mar 4--1999 Apr 25 & 33.4\tablenotemark{f}  & 0.14\tablenotemark{f}  &This work \\
U~Sco (2010) & GMRT & 2010 Jan 29--2010 Mar 2 & & & \citet{Anupama+13} \\
 V4743 Sgr & VLA & 2002 Oct 11--2003 Feb 21 & $\geq$154.6\tablenotemark{f}  & $\geq5.10$\tablenotemark{f}  &  This work \\
 V598 Pup & VLA & 2007 Nov 18--2008 Aug 3 & 205.0\tablenotemark{f}  & 11.1\tablenotemark{f}  & This work \\
 V2491 Cyg & VLA & 2008 Apr 28--2008 Sep 19 & 41.4\tablenotemark{f}  & 1.46\tablenotemark{f}  & This work\\
 V2672~Oph & VLA & 2009 Sep 1 --2009 Nov 7 & $\leq$16.1\tablenotemark{f}  & $\geq$0.55\tablenotemark{f}  & This work \\
V1723~Aql & Jansky VLA & 2010 Sep 25--2014 Mar 15 & 57.3 & 4.08 & \citet{Krauss+11} \& \citet{Weston+16a}\\
T~Pyx & Jansky VLA & 2011 Apr 22--2012 Sep 23 & 265.9 & 4.35 & \citet{Nelson+14} \\
V5589~Sgr & Jansky VLA & 2012 Apr 23--2013 Aug 26 & 62.3 & 3.42 & \citet{Weston+16b}\\
V1324~Sco & Jansky VLA & 2012 Jun 26--2014 Dec 18 & 72.2 & 1.39 & \citet{Finzell+18}\\
V959~Mon & Jansky VLA & 2012 Jun 30--2014 Feb 25 & 157.3 & 25.94 & \citet{Chomiuk+14}\\
V809 Cep & Jansky VLA & 2013 Feb 14--2016 Jan 28 & 581.2 & 0.53 & Babul et al.\ in prep\\
V339 Del & Jansky VLA & 2013 Aug 16--2017 Jul 8 & 612.7 & 1.93 & Nyamai et al.\ in prep\\
V1369~Cen & ATCA & 2013 Dec 5--2014 Apr 1 & $\geq176.0$\tablenotemark{f} & $\geq29.9$\tablenotemark{f} & This work\\
V5666~Sgr & Jansky VLA & 2014 Feb 19--2017 Aug 22 & 929.6 & 0.34 & This work\\
V2659~Cyg & Jansky VLA & 2014 Apr 5--2018 Sep 27 & 1024.6 & 0.97 & This work\\
V5667~Sgr & Jansky VLA & 2015 Feb 19--2019 Dec 22 & 1307.0 & 1.62 &This work\\
V5668~Sgr & Jansky VLA & 2015 Mar 17--2019 Dec 22 & 385.5 & 15.51 &This work\\
V5855~Sgr & Jansky VLA & 2016 Nov 4--2021 Feb 9 & 692.0 & 0.62 & This work\\
V5856~Sgr & Jansky VLA & 2016 Nov 11--2021 Feb 9 & 1148.8 & 2.36 & This work \\
V357 Mus & ATCA & 2018 Jan 18--2020 Sep 12 & 64.0 & 31.0 & This work\\
V906~Car & ATCA & 2018 Apr 3--2020 Sep 12 & 911.0 & 11.5 & This work\\
V392 Per & Jansky VLA & 2018 Apr 30--2020 May 15 & 32.6 & 5.23 & This work\\

\enddata
\tablenotetext{a}{NRAO Interferometer/NRAO 11m/NRAO 140ft/Bonn 100m/VLA}
\tablenotetext{b}{NRAO Interferometer/NRAO 11m/NRAO 140ft/VLA}
\tablenotetext{c}{NRAO Interferometer/NRAO 11m}
\tablenotetext{d}{WSRT/NRAO Interferometer/ARO 46m/NRAO 11m/VLA}
\tablenotetext{e}{Measured at C band (4--6 GHz) unless it is not available and otherwise noted in this column}
\tablenotetext{f}{Measured at X band (8--9 GHz)}
\tablenotetext{g}{Measured at 1.5 GHz}
\end{deluxetable} 
\def\arraystretch{1.0}
\clearpage

\subsubsection{QU~Vul}
Radio data covering the first 500 days of outburst were published by \citet{Taylor+87}. Multi-frequency light curves reveal a double-peaked structure, with the first radio maximum taking place before day 206 (and only sampled on the decline), and the second radio maximum sampled only on the rise (out to day 497).  Radio monitoring of this nova continued for several more years after the publications of \citet{Taylor+87, Taylor+88}, and these archival data are reduced and published for the first time as part of this work in Table \ref{tab:quvul} and Figure \ref{fig:lcnew1}. Overall, the light curve spans 206--1400 days after outburst, and was obtained using VLA programs AT64 (PI A.~R.\ Taylor), AH185 (PI G.\ Hennessy), AH254, AH301, AH573 (PI R.\ Hjellming), and AJ154 (PI K.\ Johnston). Flux measurements can be found in Table \ref{tab:quvul}, and the light curve is plotted in Figure \ref{fig:lcnew1}. 

During the VLA's 1986 and 1987 A configurations, QU~Vul was spatially resolved as discussed in \citet{Taylor+88}. The next time the VLA was in A configuration, in 1988 Oct, QU~Vul was again resolved. In fact, the nova remnant is so extended in the 14.9 GHz image from this time that it starts to be ``resolved out" (i.e., the image is very low S/N).
%show images? more details?
% radio position?

\subsubsection{V1819~Cyg}
V1819~Cyg was observed with the VLA under program IDs AH185 (PI G.~Hennessy), AH254, AH279, AH301, AH316, AH366, AH390 (PI R.~Hjellming) and AH385 (PI X.~Han). It was observed over 1987--1991, encompassing 166--1761 days after discovery. Flux measurements can be found in Table \ref{tab:v1819cyg}, and the light curve of V1819~Cyg is plotted in Figure \ref{fig:lcnew1}. The X-band receivers (8.4 GHz) were commissioned partway through the evolution of V1819~Cyg's light curve, so there are no 8.4 GHz measurements of the early evolution. V1819~Cyg appears as a point source in all epochs and frequencies. 
% radio position?

 \begin{figure*}
{\includegraphics[width = 0.48\textwidth]{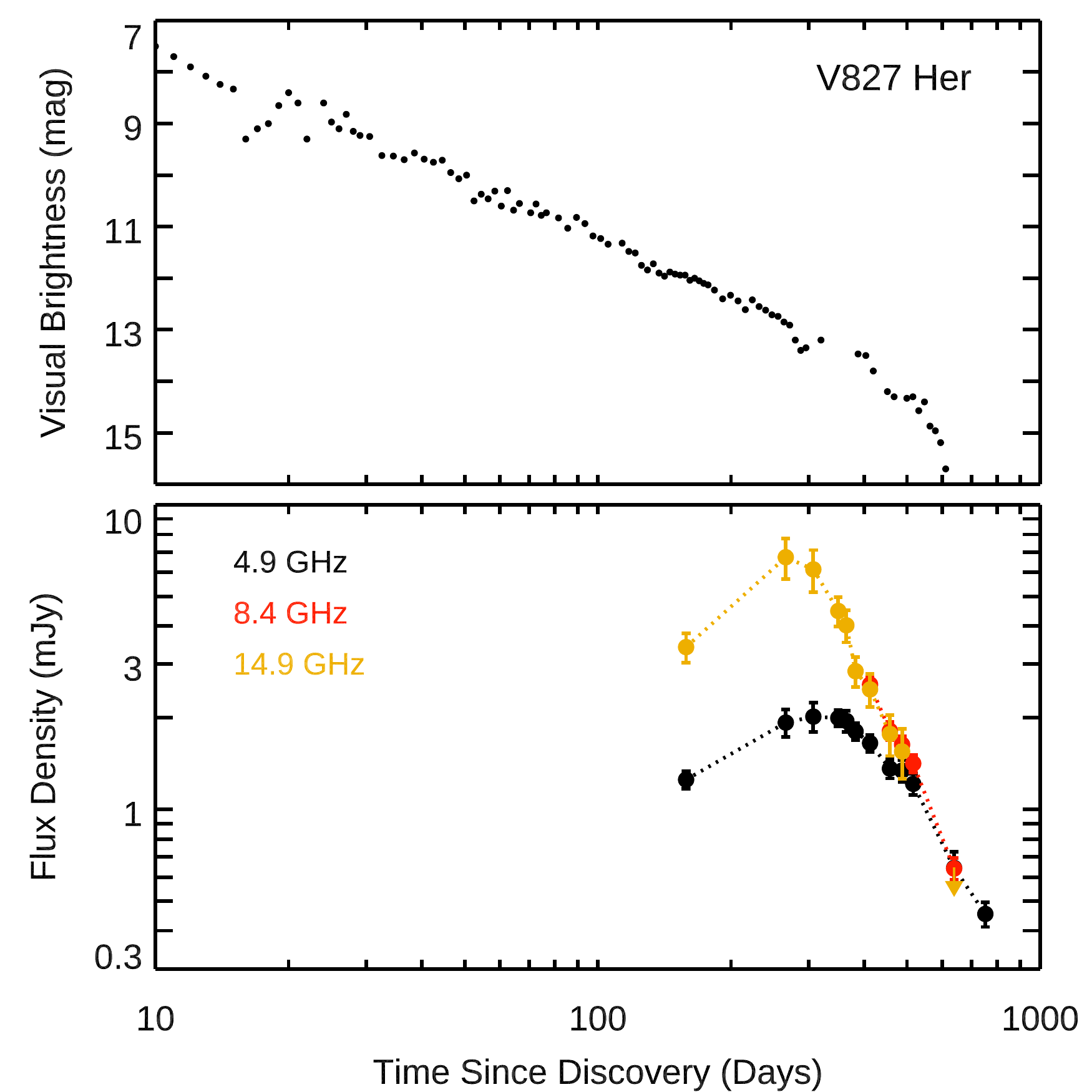}
\includegraphics[width = 0.48\textwidth]{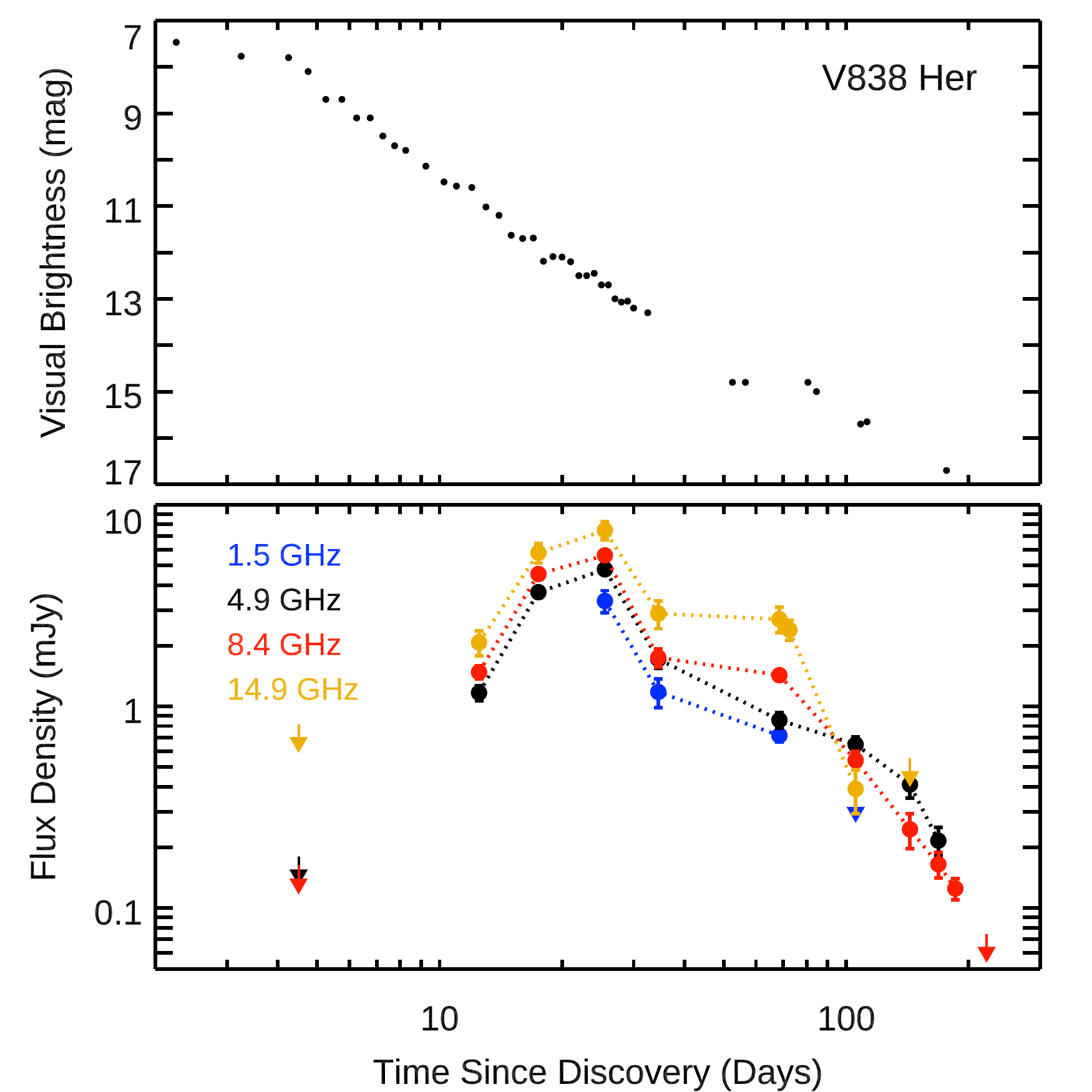}\\
\includegraphics[width = 0.48\textwidth]{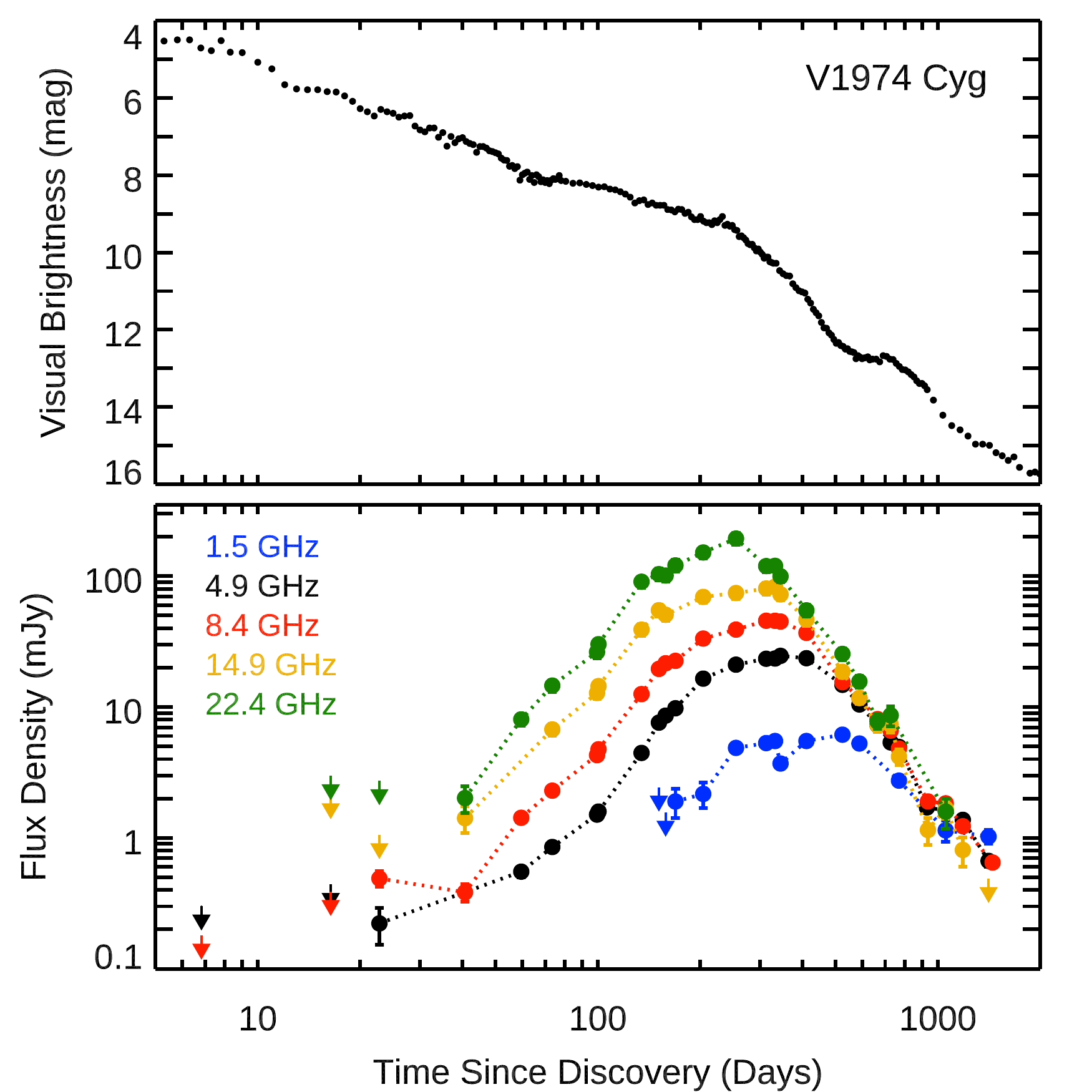}
\includegraphics[width = 0.48\textwidth]{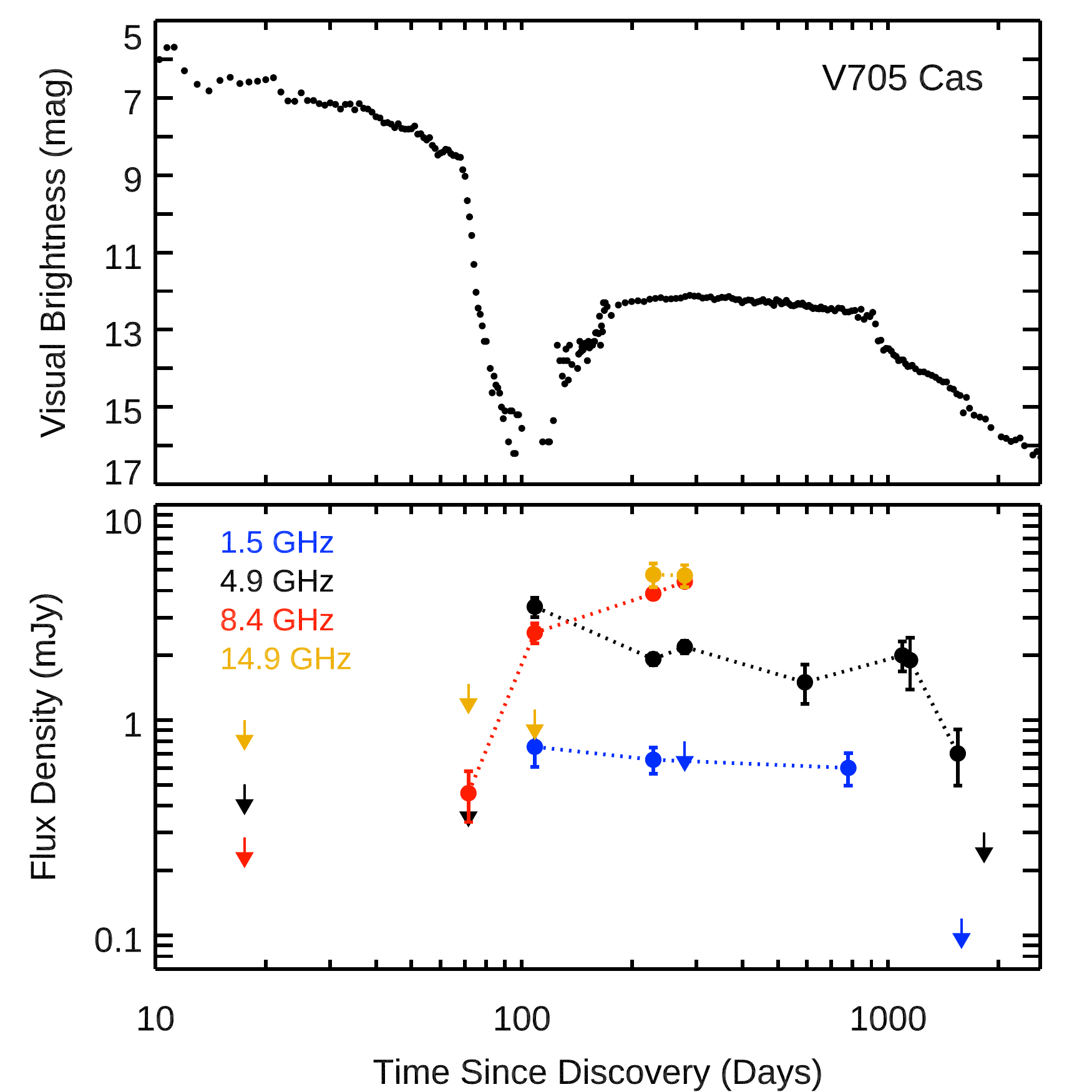}}
\caption{Optical and radio light curves for four novae (clockwise from top left): V827~Her (1987), V838~Her (1991), V705~Cas (1993), and V1974~Cyg (1992). For radio epochs with non-detections, 3$\sigma$ upper limits are plotted with arrows; there are additional non-detections not plotted here, for clarity.. The visual-band light curves  are binned AAVSO data published by \citet{Strope+10}.  }
\label{fig:lcnew2}
\end{figure*}

\subsubsection{V827~Her}
VLA observations of V827~Her were obtained under program codes AH254, AH301, AH316 (PI R.~Hjellming) and AJ154 (PI K.~Johnston). Observations took place during 1987--1989, or 158--752 days after discovery.  Flux measurements can be found in Table \ref{tab:v827her}, and the light curve is plotted in Figure \ref{fig:lcnew2}. The X-band receivers (8.4 GHz) were commissioned partway through the evolution of V827~Her's light curve, so there are no 8.4 GHz measurements of the early evolution. V827~Her appears as a point source in all epochs and frequencies. 
% radio position?

\subsubsection{V838~Her}
VLA observations of V838~Her were obtained under program codes  AB601 (PI D.~A.\ Brown), AH390, AH428, AH444 (PI R.~Hjellming), and AH424 (PI X.~H.\ Han). Observations took place during 1991, covering 4--222 days after discovery.  Flux measurements can be found in Table \ref{tab:v838her}, and the light curve is plotted in Figure \ref{fig:lcnew2}. V838~Her appears as a point source in all epochs and frequencies. 
% radio position?

\subsubsection{V1974 Cyg}
VLA observations of V1974~Cyg were tentatively presented in \citet{Hjellming96}, but never fully catalogued in the literature, so we present them in full detail here. High-resolution MERLIN observations were presented in \citet{Pavelin+93} and  \citet{Eyres+96}, while monitoring at millimeter wavelengths is presented in \citet{Ivison+93}.

The VLA observations were obtained under program codes  AF211 (PI R.~L.\ Fiedler), AH390, AH492, AH573 (PI R.\ Hjellming), and AL314 (PI E.~P.\ Liang). Observations took place during 1992 Feb 25 to 1996 Feb 6, or 6--1448 days after discovery.  Flux measurements can be found in Table \ref{tab:v1974cyg}, and the light curve is plotted in Figure \ref{fig:lcnew2}. 

During the 1992 A configuration, V1974~Cyg was already strongly resolved at U- and K-bands. Images were broadly consistent with a circularly symmetric ring, and total flux densities were estimated by integrating over the emitting region with AIPS/\verb|tvstat|. Meanwhile, at the lower frequencies in A configuration, fluxes were determined with Gaussian fitting, but we allowed the width to vary. By the time of the 2014 A configuration, V1974~Cyg's 22.4 GHz emission was resolved out and went non-detected. Total fluxes in X and U bands were integrated with \verb|tvstat|, while Gaussian fits were performed at L and C bands.

%\iffalse
 \begin{figure*}
{\includegraphics[width = 0.48\textwidth]{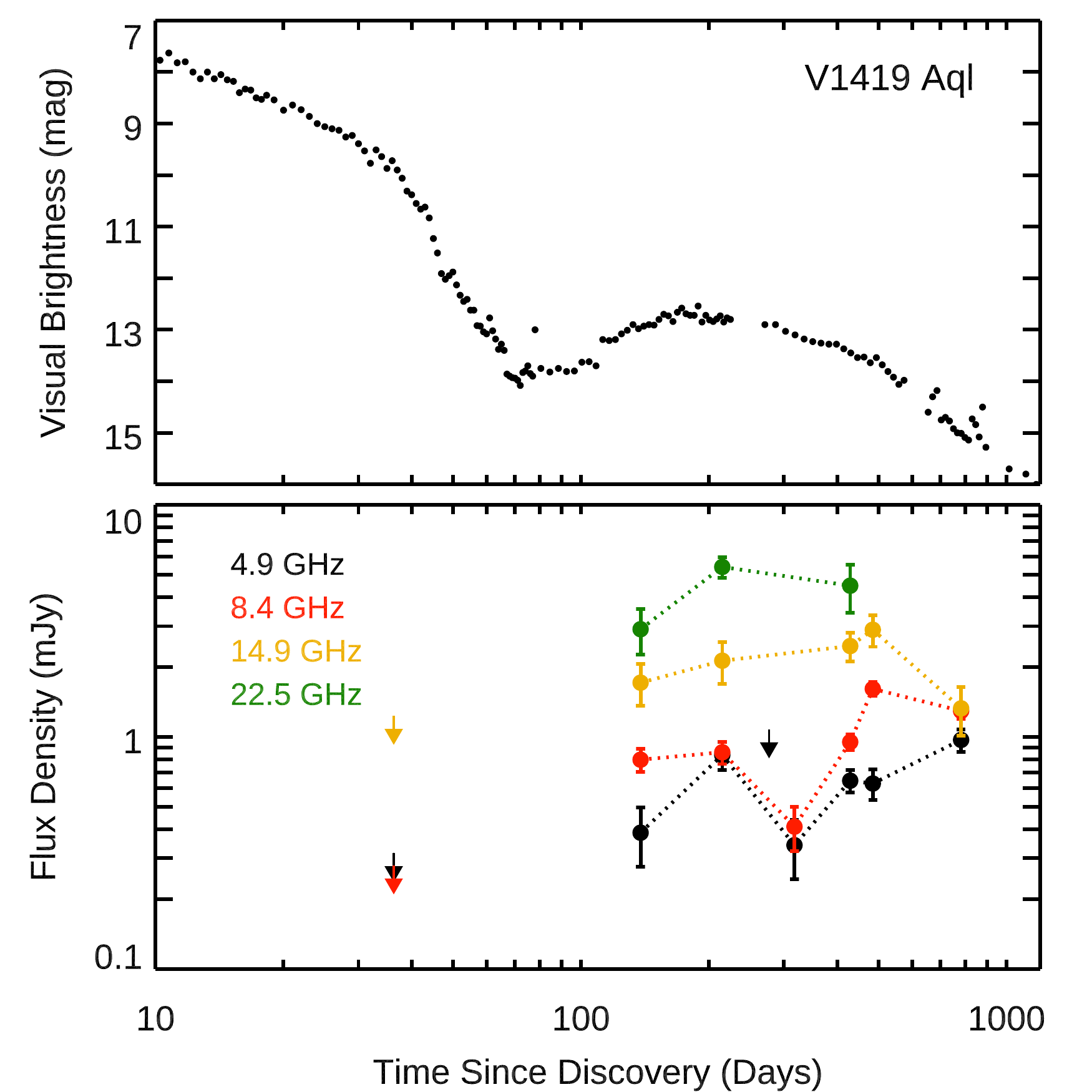}
\includegraphics[width = 0.48\textwidth]{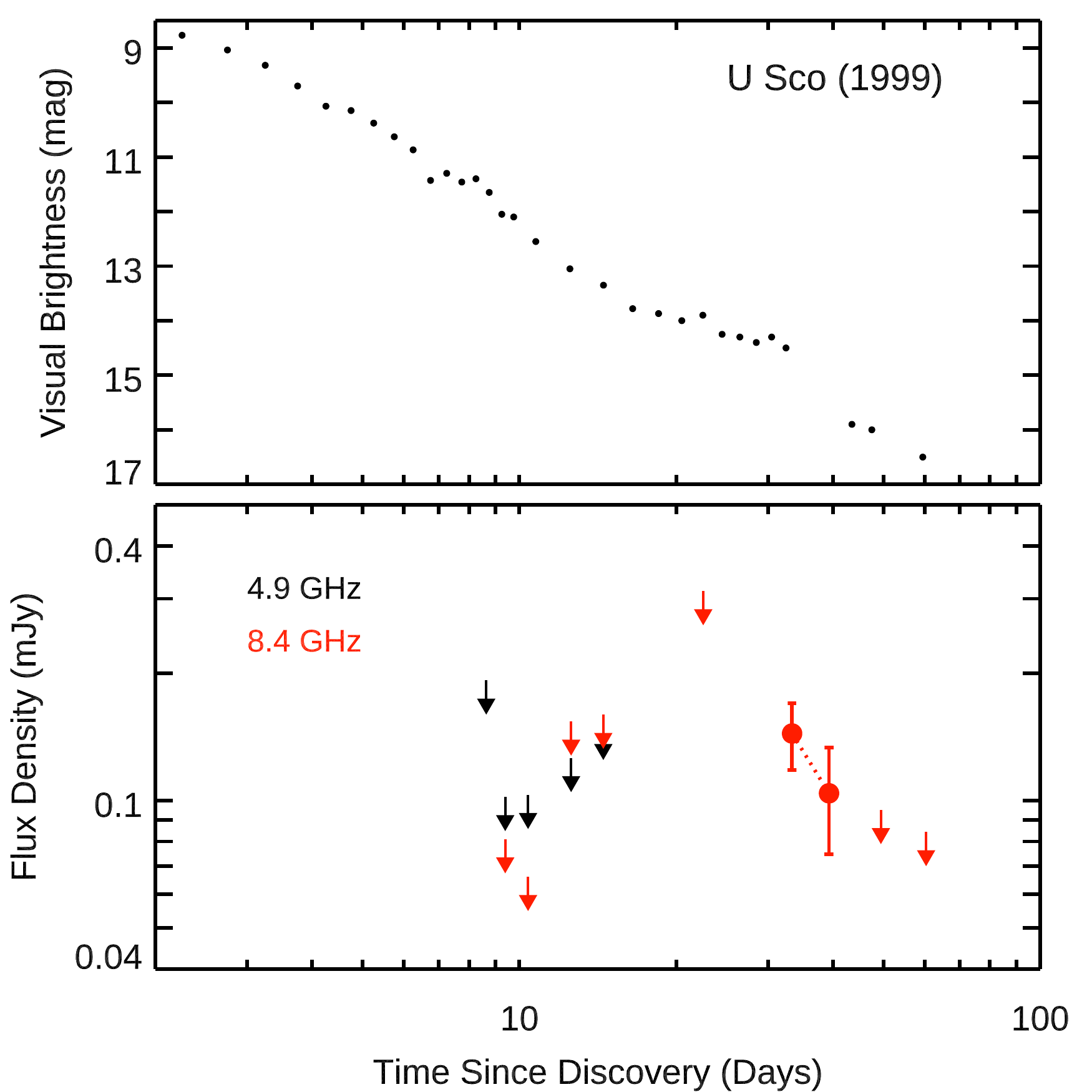}\\
\includegraphics[width = 0.48\textwidth]{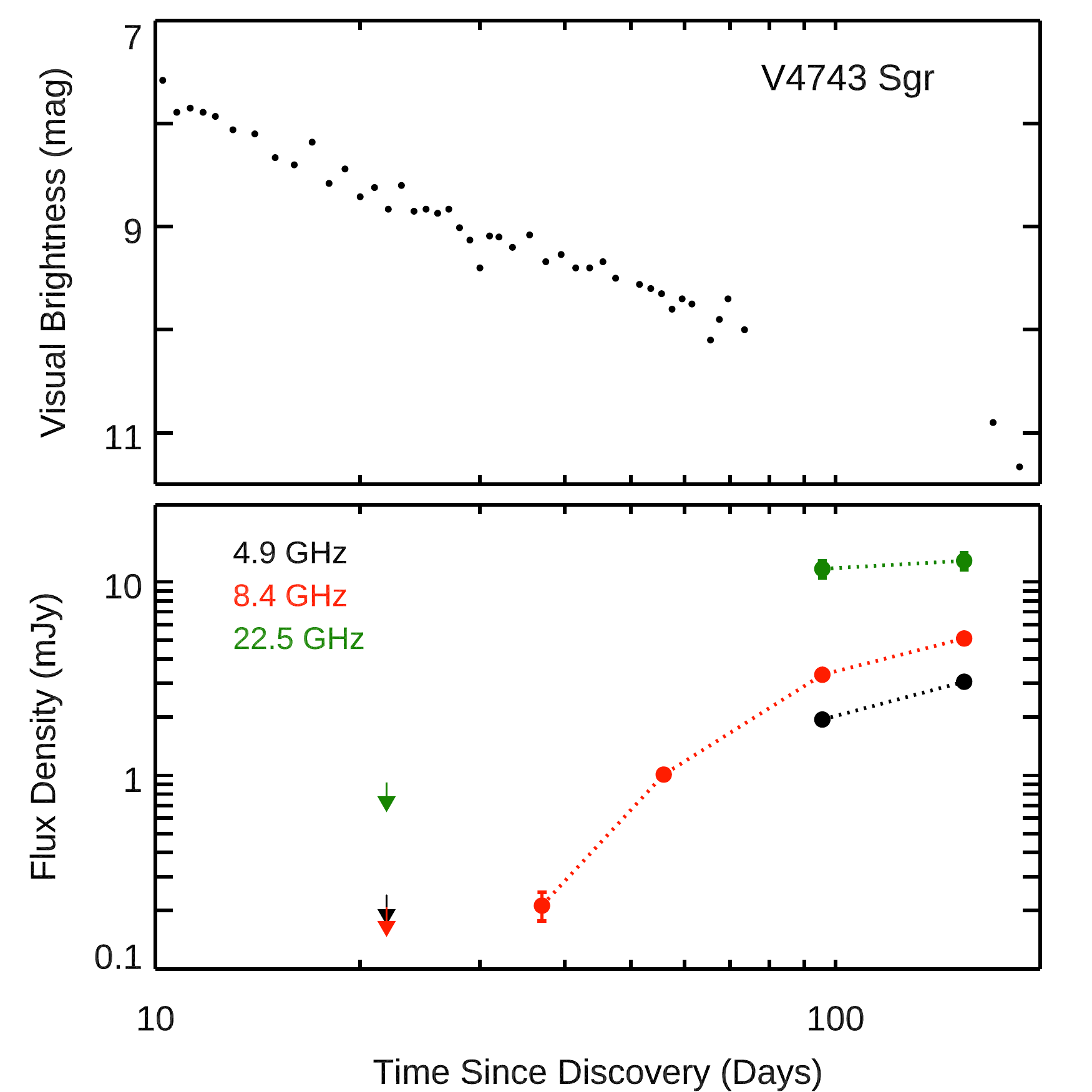}
\includegraphics[width = 0.48\textwidth]{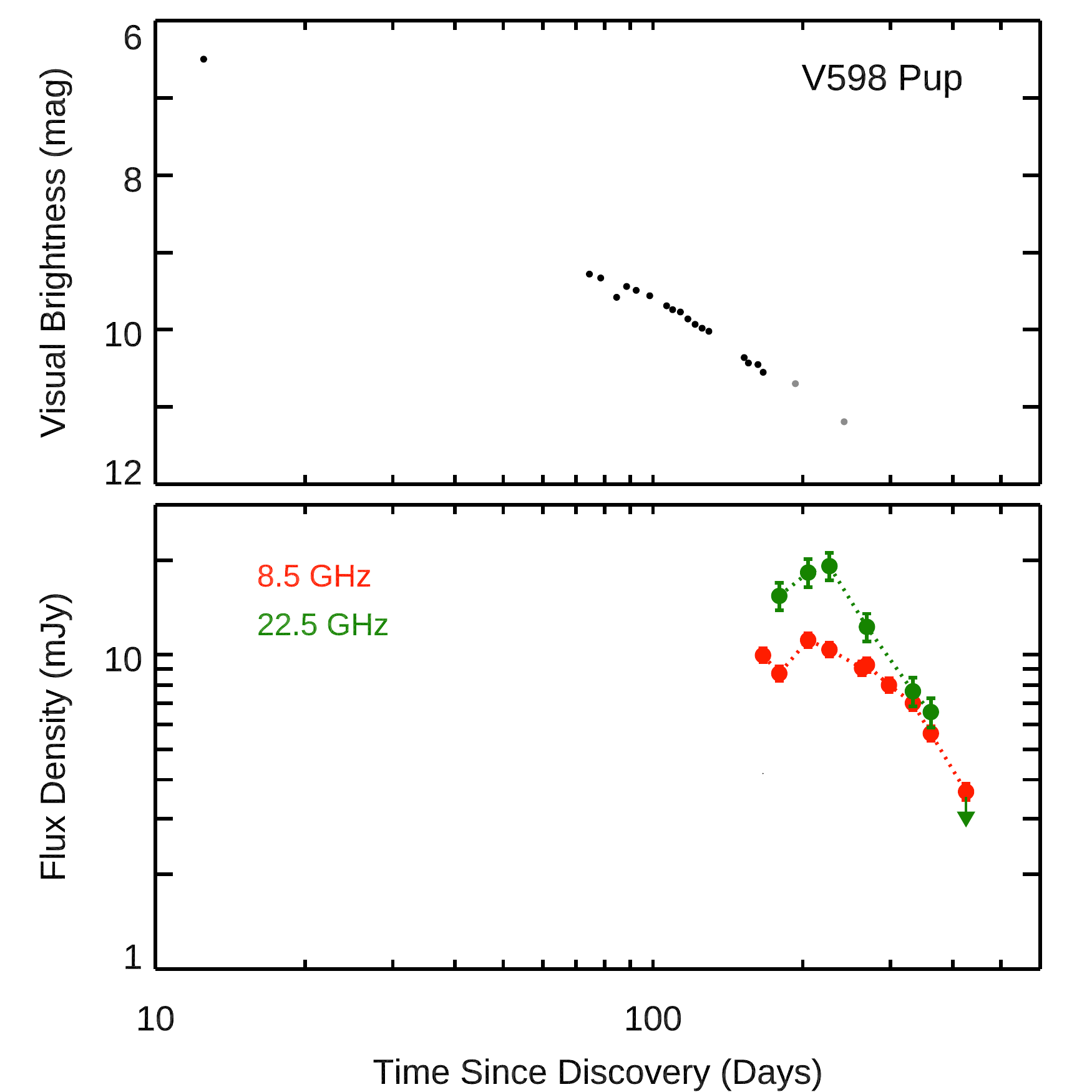}}
\caption{Optical and radio light curves for four novae (clockwise from top left): V1419~Aql (1993), U~Sco (1999), V598~Pup (2007), and V4743~Sgr (2002).  For epochs with non-detections, 3$\sigma$ upper limits are plotted with arrows. For the recurrent nova U~Sco, we only plot data from its 1999 eruption and the first radio detections ever reported for this source; only two of its radio points are $>3\sigma$ detections, both at 8.4 GHz. The visual-band light curves for V1419~Aql, U~Sco, and V4743~Sgr are binned AAVSO data published by \citet{Strope+10}, while the V598~Pup $V$-band light curve is from \citet{Pojmanski+07}. }
\label{fig:lcnew3}
\end{figure*}
%\fi

\subsubsection{V705 Cas}

VLA and MERLIN observations of the eruption of V705~Cas were presented by \citet{Eyres+00} starting on 1994 Jul 23 (day 228), which they claim is the first radio detection of the nova. There are several earlier epochs of VLA observations, obtained under program  AH492 (PI R.\ Hjelliming) 17--108 days after discovery, which Eyres et al.\ present as non-detections. We re-reduce these data here, and find detections at lower frequencies ($\leq$8.4 GHz) on 1994 Feb 16 (day 71) and 1994 Mar 25 (day 108).
The flux measurements derived from these early epochs should be considered skeptically as they are afflicted by several issues, including very distant ($\gg 10^{\circ}$) phase calibrators that may lead to data being decorrelated and flux densities being underestimated. The AH492 epoch from 1994 Jul 16 is unsalvageable and is omitted here, but luckily observations were obtained just seven days later under a different program; this 1994 Jul 23 epoch is presented by \citet{Eyres+00}. While our analysis of the $\geq$4.9 GHz data is consistent with theirs, Eyres et al.\ report a non-detection at 1.4 GHz while we detect the nova. Flux measurements can be found in Table \ref{tab:v705cas}, and the light curve is plotted in Figure \ref{fig:lcnew2}. 

V705 Cas was resolved by MERLIN on days 592--1548. Here we use Eyres et al.'s ``visibilities" method to estimate the integrated flux densities; note that \citet{Eyres+00} warn caution in interpreting these measurements, as some flux may be resolved out by MERLIN baselines. Flux measurements from these later epochs (listed in Table \ref{tab:v705cas}) should be treated as lower limits.
%There is some indication of an early non-thermal bump, but this light curve is a mess.

\subsubsection{V1419~Aql}
VLA observations of V1419~Aql were obtained under program codes  AH492 (PI R.~Hjellming) and AL314  (PI E.~P.\ Liang).
 Observations took place during 1991, covering 4--222 days after discovery.  Flux measurements can be found in Table \ref{tab:v1419aql}, and the light curve is plotted in Figure \ref{fig:lcnew3}. V1419~Aql appears as a point source in all epochs and frequencies. 
% radio position?

\subsubsection{U Sco}
There are archival VLA data covering the 1987 and 1999 outbursts of the recurrent nova U~Sco. 
The 1987 outburst was discovered on May 15, after light curve maximum \citep{Overbeek+87, Schaefer10}. 
Radio observations of the 1987 eruption were obtained under programs AF138 (PI E.\ Fomalont) and AH185 (PI G.\ Hennessy), and covered 18 and 51 days after discovery. Both VLA observations from this outburst yielded non-detections (Table \ref{tab:usco}).

The 1999 outburst was discovered on Feb 24, and peaked on 1999 Feb 25 \citep{Schmeer+99, Schaefer10}. VLA data for this outburst were obtained by PI S.\ Eyres as part of VLA program AE124, and covered 8--60 days after discovery.
These  data are published for the first time here, and remarkably, U~Sco was detected at 8.4 GHz in two VLA epochs during the 1999 outburst, on days 33 and 39 (Table \ref{tab:usco}). These two detections are surrounded by a substantial number of non-detections, so it is clear that radio emission from U~Sco is faint and fleeting. U~Sco appears as a point source in the few epochs when it was detected. The 1999 light curve is plotted in Figure \ref{fig:lcnew3}.  

The 2010 outburst of U~Sco was was discovered and peaked on Jan 28. It was observed at longer radio wavelengths by \citet{Anupama+13}, using the GMRT. They report three non-detections spanning days 1--33, as listed in Table \ref{tab:usco}.

\subsubsection{V4743~Sgr}
VLA observations of V4743~Sgr were obtained under program codes AE134 (PI S.\ Eyres) and AI104 (PI R.\ Ivison) during Oct 2002--Feb 2003, covering 21--154 days after discovery.   
 Flux measurements can be found in Table \ref{tab:v4743sgr}, and the light curve is plotted in Figure \ref{fig:lcnew3}. V4743~Sgr appears as a point source in all epochs and frequencies.

\subsubsection{V598~Pup}
VLA observations of V598~Pup were obtained under program code  AR642 (PI M.\ Rupen) during 2007--2008, covering 166-425 days after discovery.  In many epochs, no standard absolute gain calibrator was observed. Instead, 0713+438 was observed as the flux calibrator, which had ongoing NRAO flux monitoring on a fortnightly basis as part of their polarization calibration strategy. We do not include the Q-band observations here, as many of these are severely decorrelated (presumably due to atmospheric conditions), and they do not have sufficient S/N for self calibration.
 Flux measurements can be found in Table \ref{tab:v598pup}, and the light curve is plotted in Figure \ref{fig:lcnew3}. V598~Pup appears as a point source in all epochs and frequencies.

%\iffalse
 \begin{figure*}
{\includegraphics[width = 0.48\textwidth]{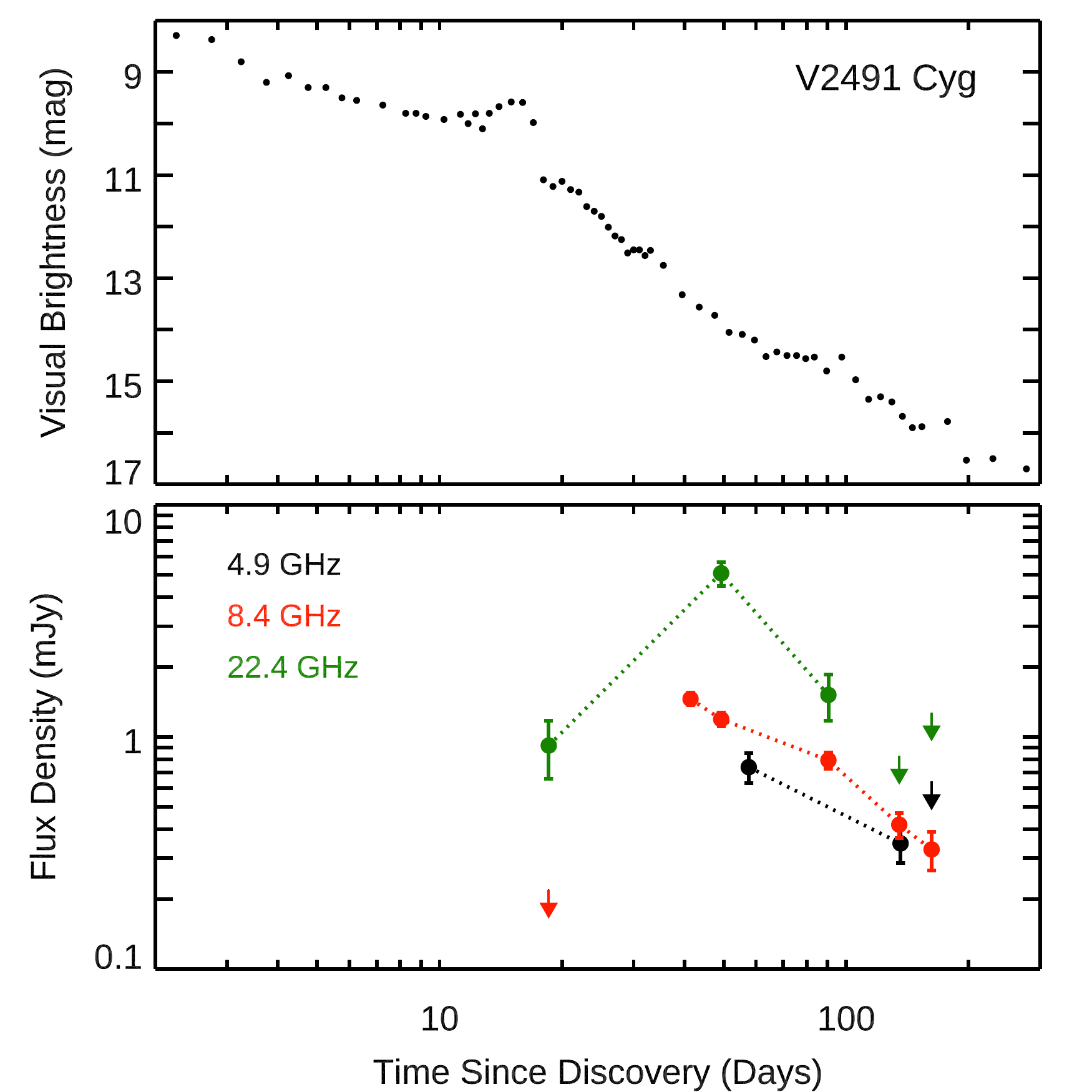}
\includegraphics[width = 0.48\textwidth]{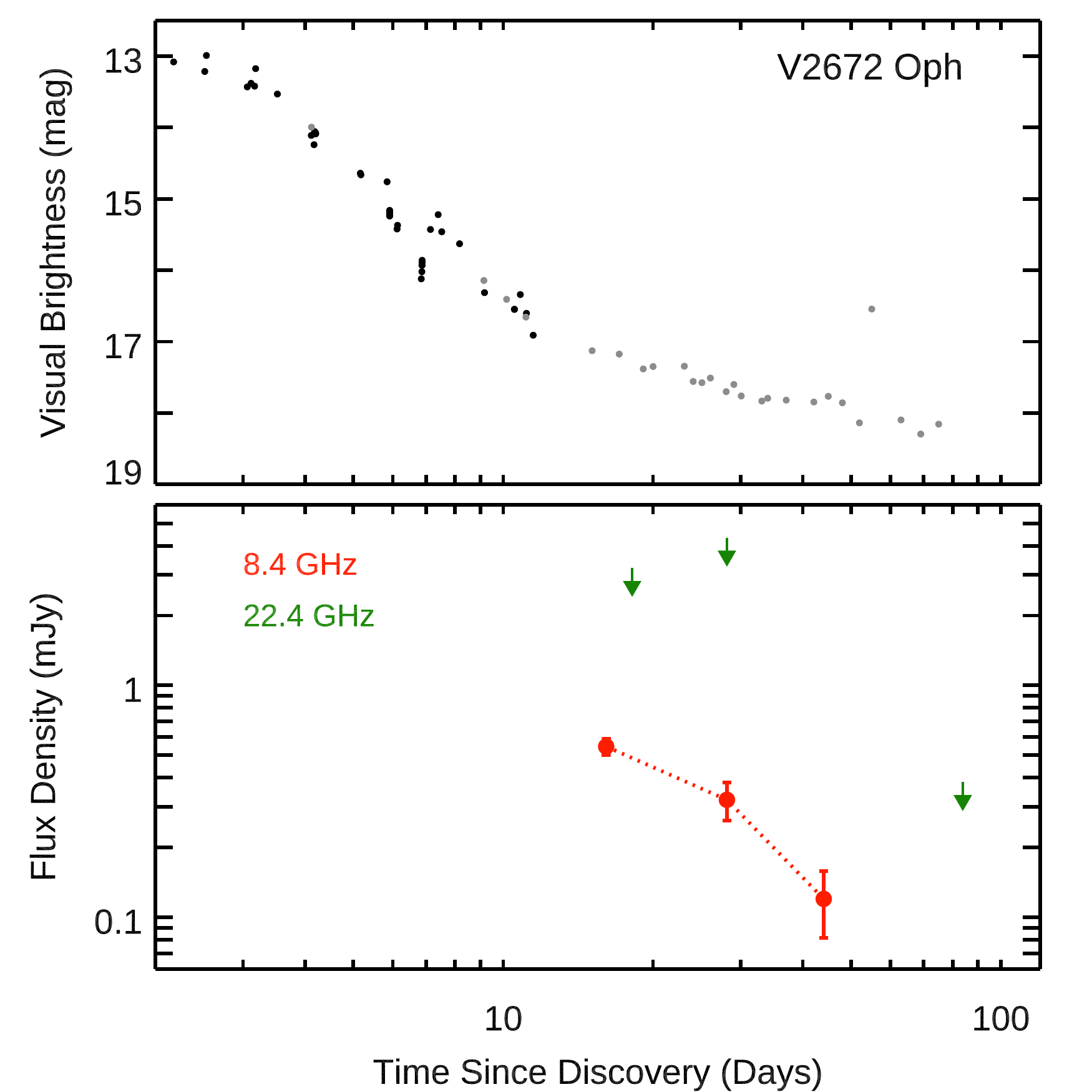}\\
\includegraphics[width = 0.48\textwidth]{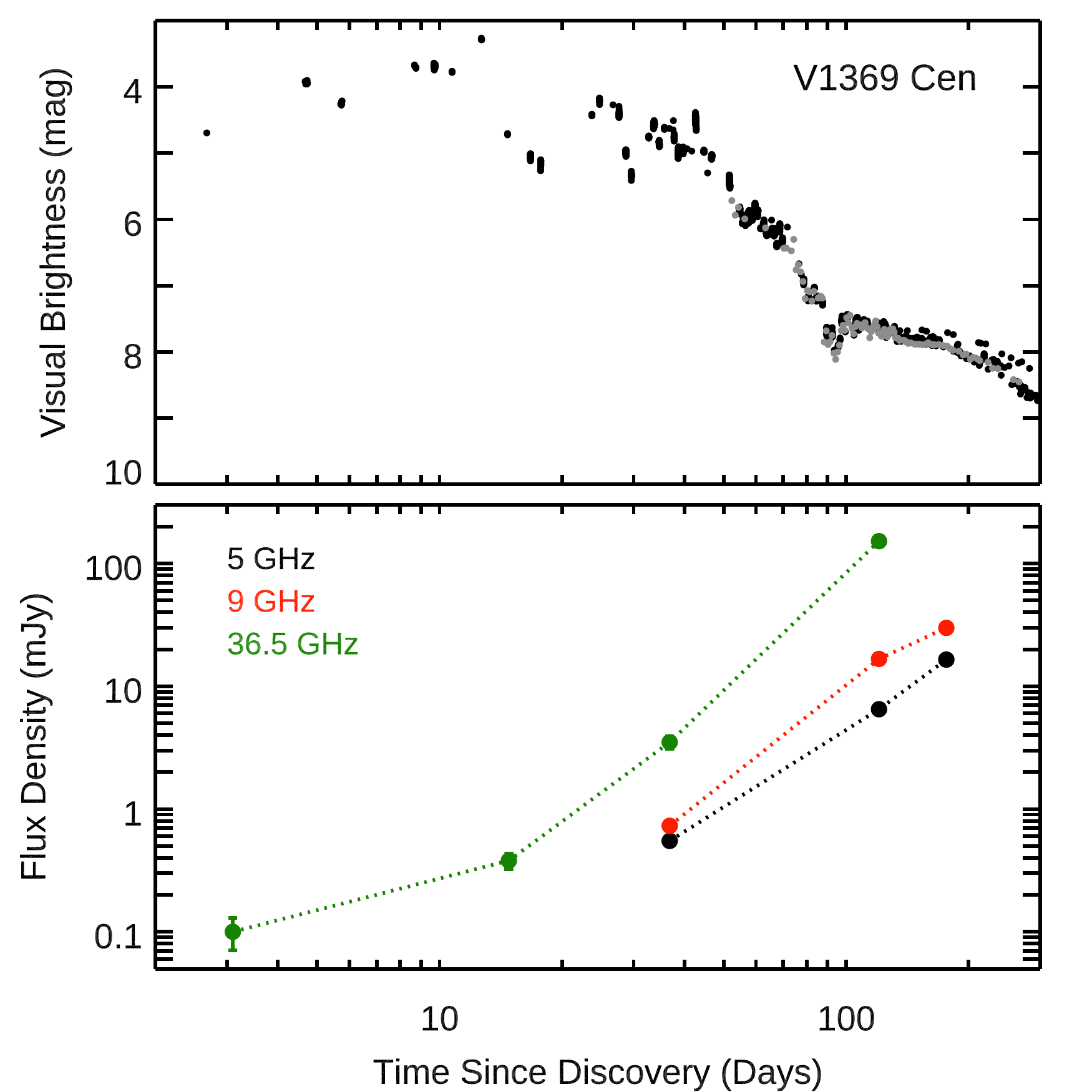}
\includegraphics[width = 0.48\textwidth]{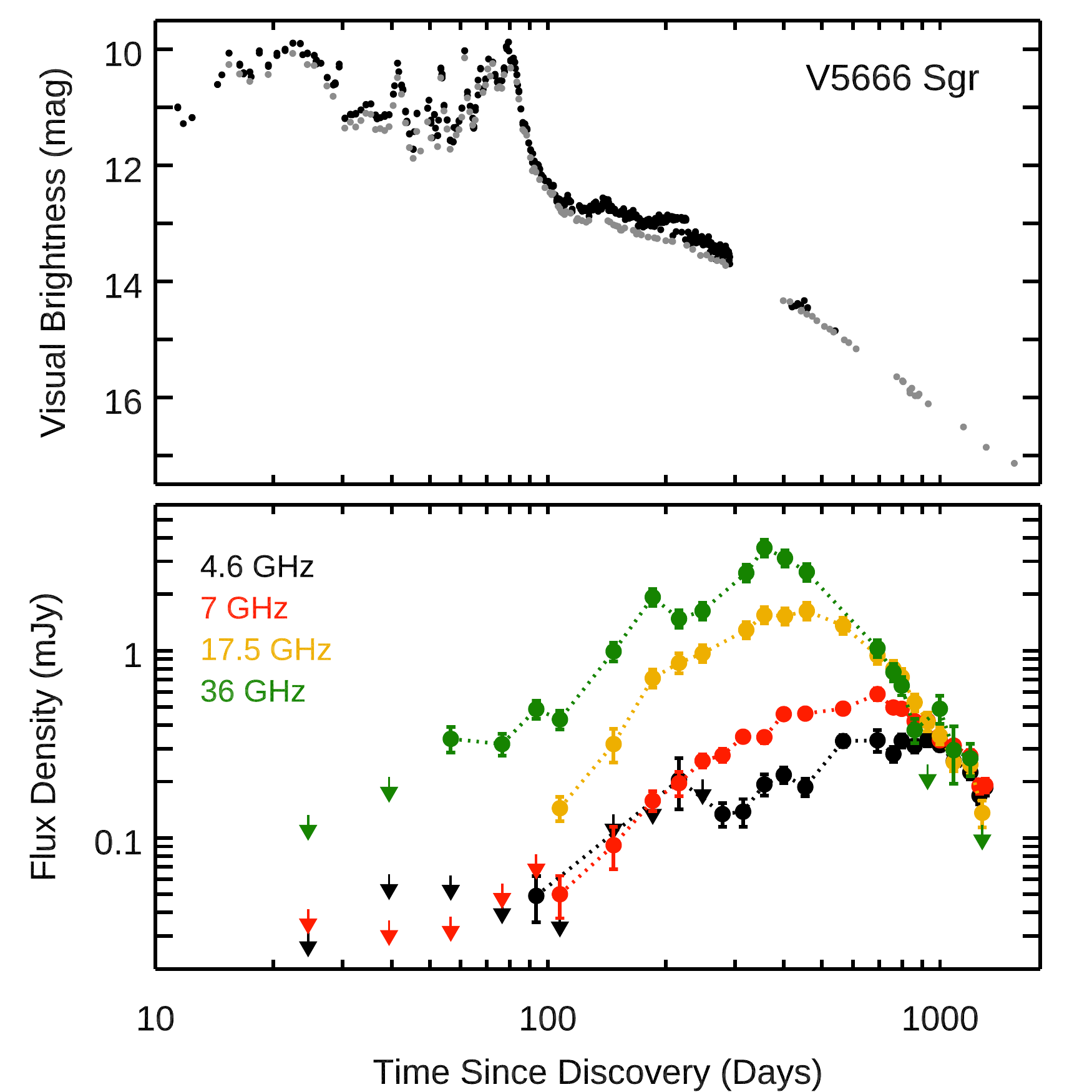}}
\caption{Optical and radio light curves for four novae (clockwise from top left): V2491~Cyg (2008), V2672~Oph (2009), V5666~Sgr (2014), and V1369~Cen (2013).  For radio epochs with non-detections, 3$\sigma$ upper limits are plotted with arrows, and some non-detections are omitted for clarity. The visual-band light curve for V2491~Cyg is binned AAVSO data published by \citet{Strope+10}, while we plot individual $V$-band AAVSO measurements (black points) and SMARTS/Stonybrook photometry (grey points) for the other three novae. }
\label{fig:lcnew4}
\end{figure*}
%\fi

\subsubsection{V2491~Cyg}
VLA observations of V2491~Cyg were obtained under program code  AS946 (PI J.\ Sokoloski) during 2008, covering 18--162 days after discovery.   
 Flux measurements can be found in Table \ref{tab:v2491cyg}, and the light curve is plotted in Figure \ref{fig:lcnew4}. V2491~Cyg appears as a point source in all epochs and frequencies.

\subsubsection{V2672~Oph}
VLA observations of V2672~Oph were obtained under program codes  AK720 and AK722 (PI M.~Krauss) during 2009, covering 16--83 days after discovery.   
 Flux measurements can be found in Table \ref{tab:v2672oph}, and the light curve is plotted in Figure \ref{fig:lcnew4}. V2672~Oph appears as a point source in all epochs and frequencies.

%\iffalse
 \begin{figure*}
{\includegraphics[width = 0.48\textwidth]{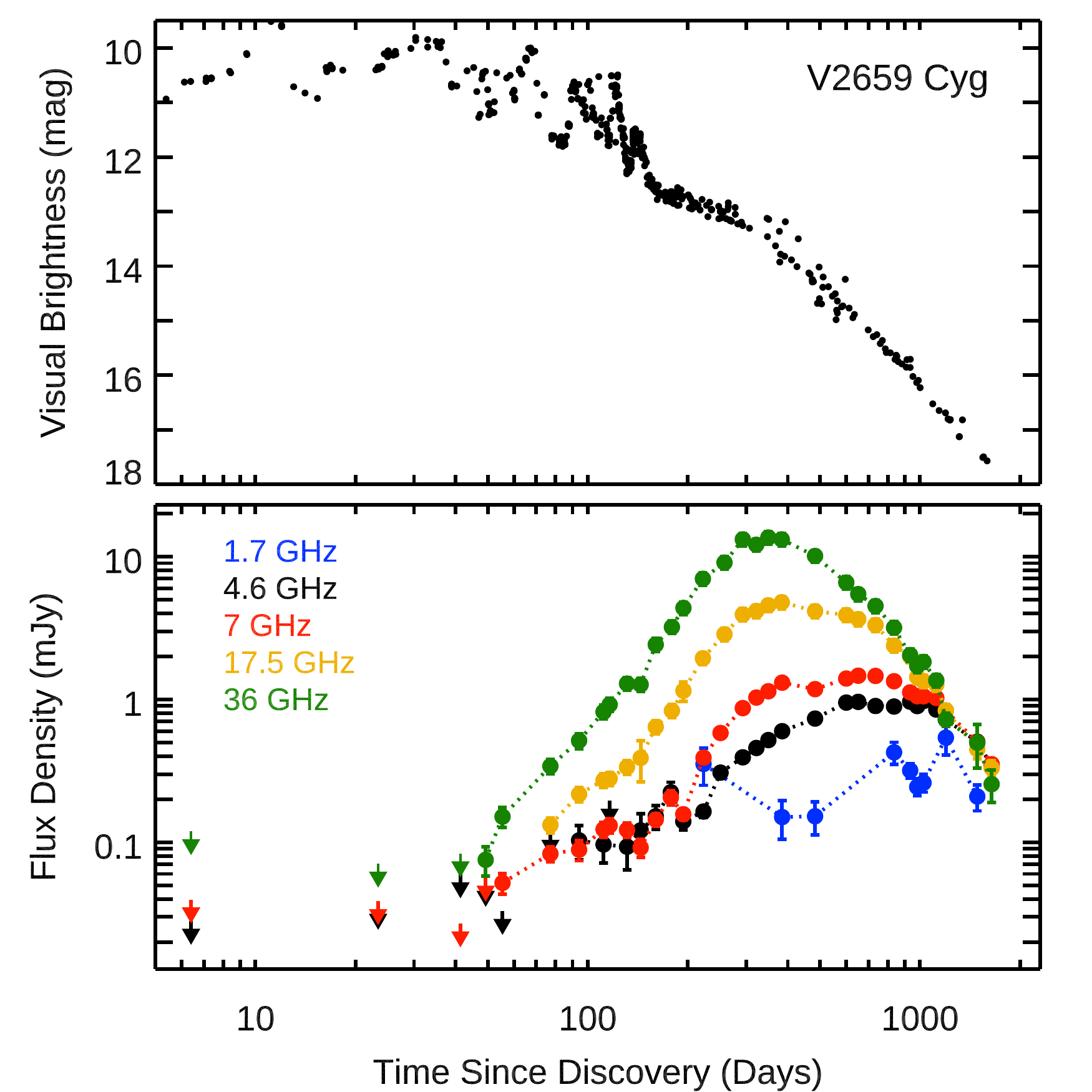}
\includegraphics[width = 0.48\textwidth]{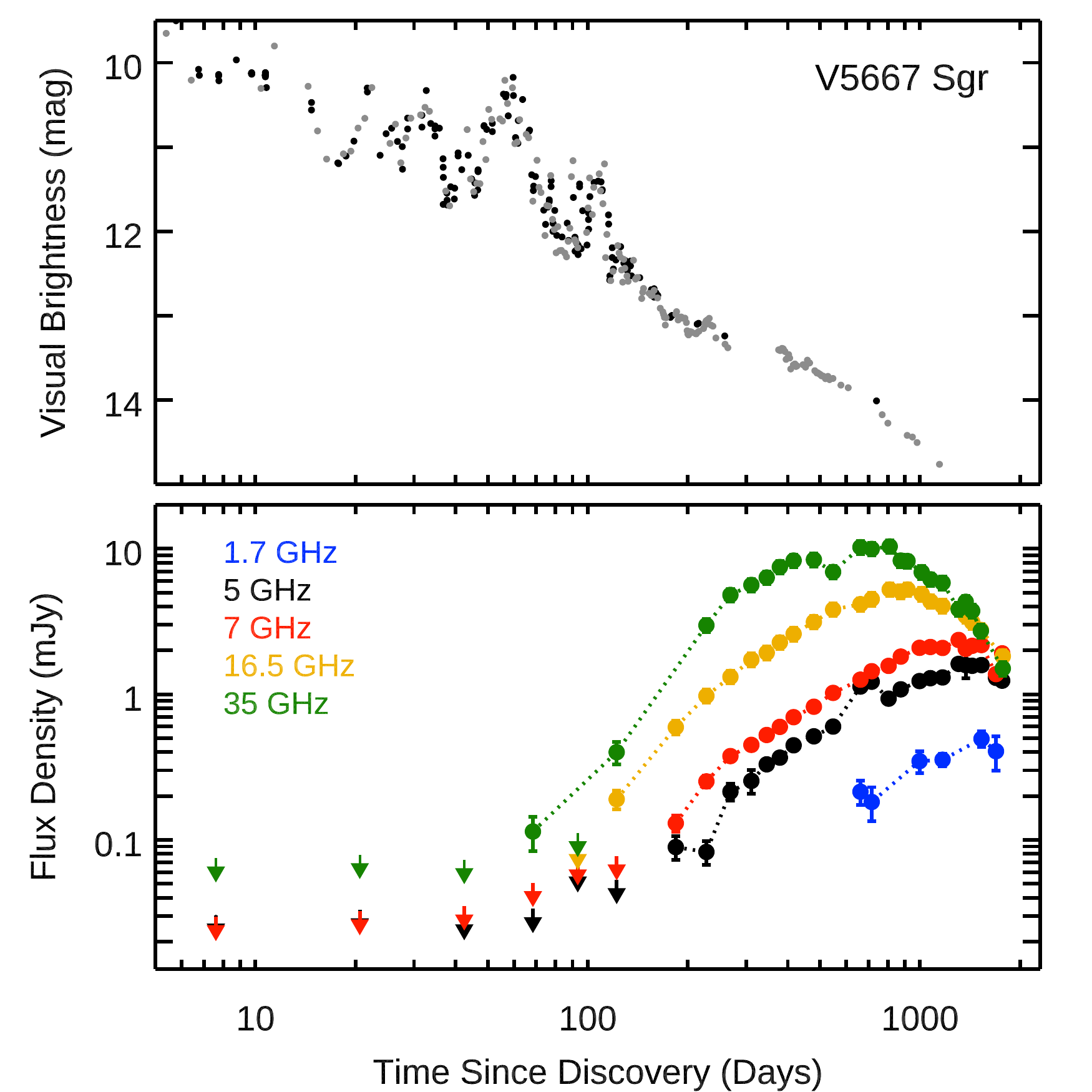}\\
\includegraphics[width = 0.48\textwidth]{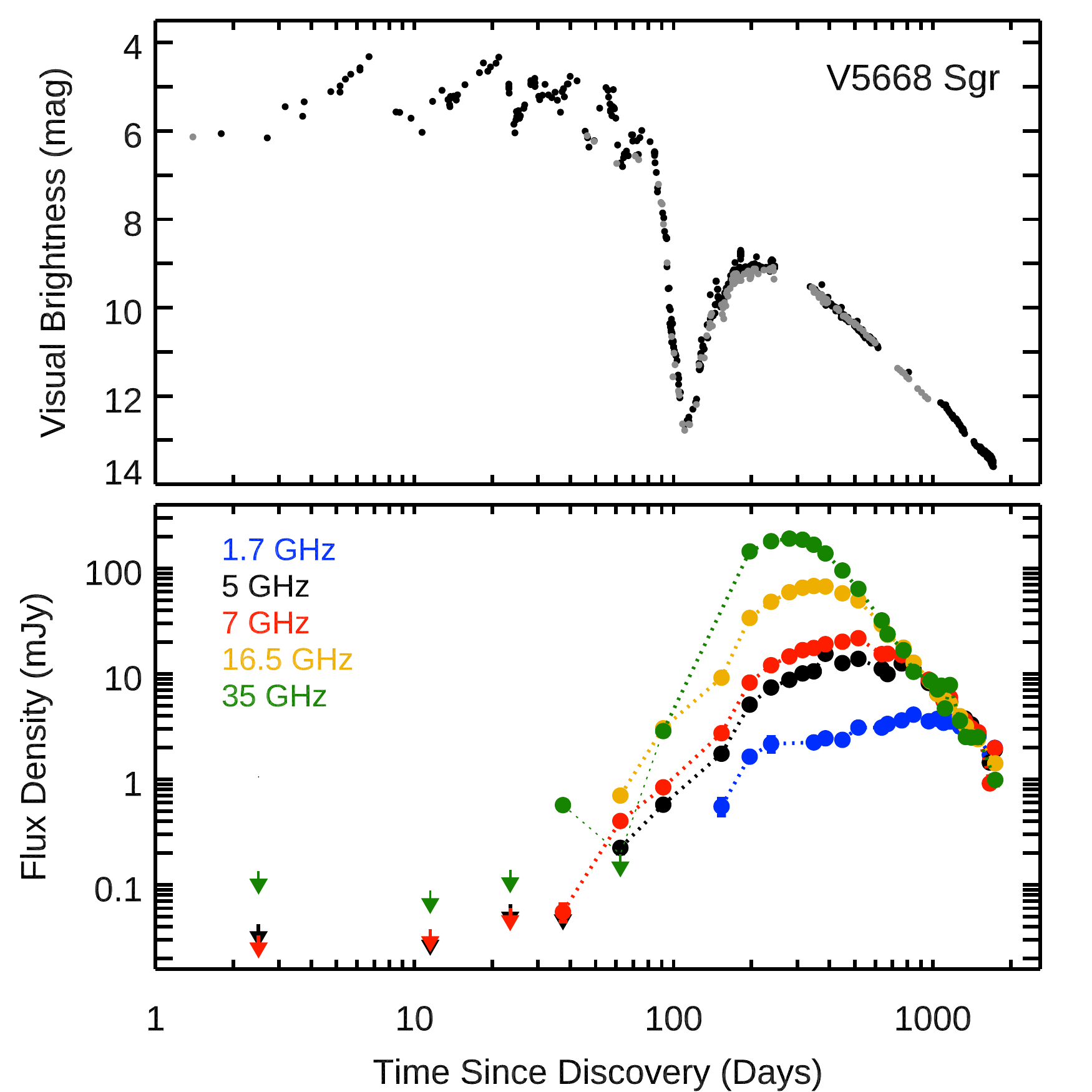}
\includegraphics[width = 0.48\textwidth]{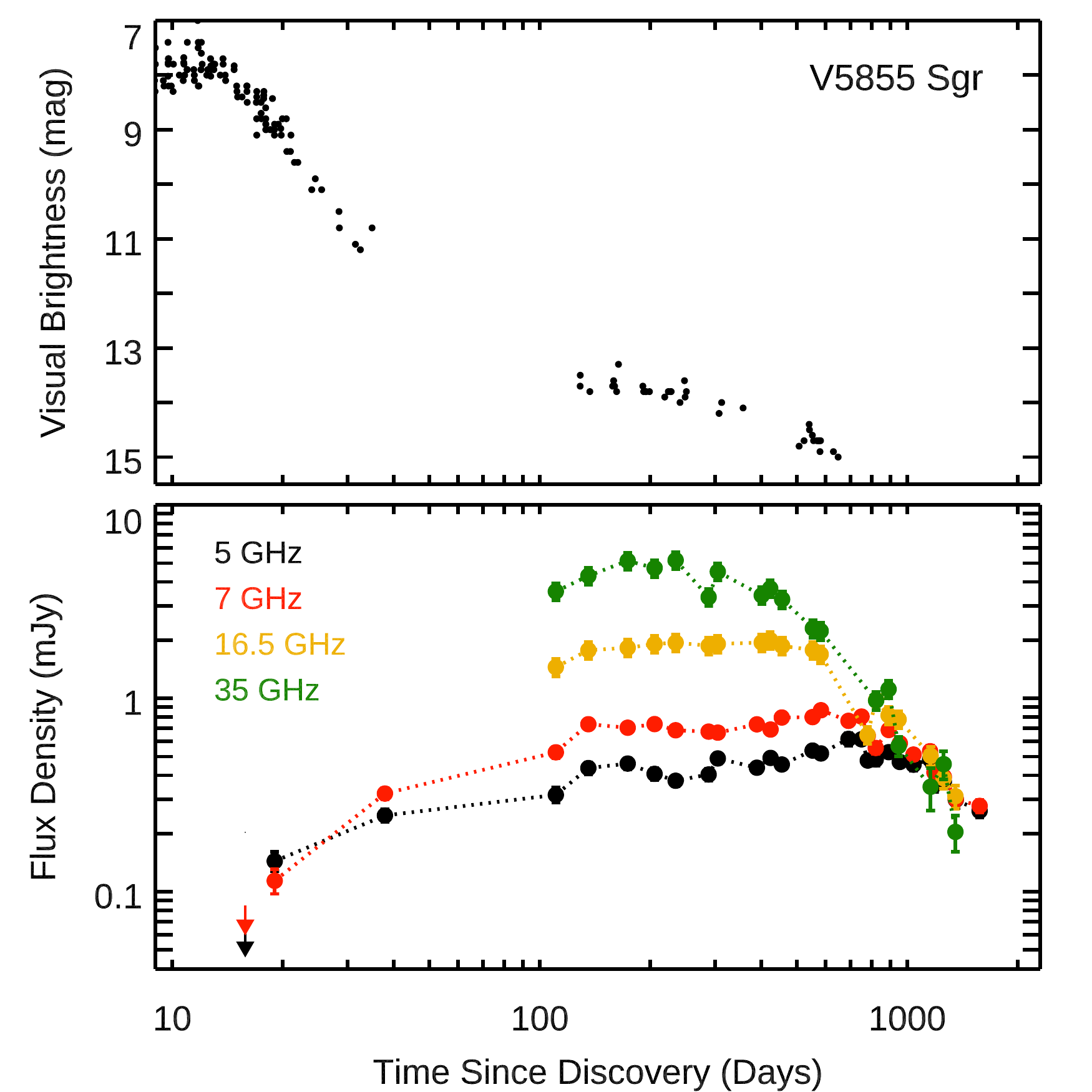}}
\caption{Optical and radio light curves for four novae (clockwise from top left): V2659~Cyg (2014), V5667~Sgr (2015), V5855~Sgr (2016) and V5668~Sgr (2015).  For epochs with non-detections, 3$\sigma$ upper limits are plotted with arrows. Optical light curves are made using $V$-band or ``Vis." AAVSO measurements (black points; \citealt{Kafka20}) and Stony Brook/SMARTS data (grey points; \citealt{Walter+12}). Note that V5668~Sgr has an early 35 GHz detection on day 37, but then faded to a non-detection on day 62; these points are connected with lighter dotted lines. We also do not plot 1.7 GHz upper limits, as their large quantity and relatively high fluxes lead to confusion with other frequencies. }
\label{fig:lcnew5}
\end{figure*}
%\fi

\subsubsection{V1369~Cen}

We obtained ATCA observations of the southern nova V1369~Cen under programs CX282 and VX21 (PI K.\ Bannister). Observations took place during the first seven months of outburst, while the nova dramatically brightened at radio  wavelengths. Unfortunately, due to scheduling difficulties, monitoring efforts did not continue past May 2014 for this nova.  Flux measurements can be found in Table \ref{tab:v1369cen}, and the light curve is plotted in Figure \ref{fig:lcnew4}. V1369~Cen appears as a point source in all epochs and frequencies. 
%Mention VLBI non-detection??? - only if keith wants to write it up

%{\bf Ryan is still working on later epochs}

\subsubsection{V5666 Sgr}
Jansky VLA observations of V5666~Sgr were obtained under program codes 13B-057, 16A-258 (PI L.\ Chomiuk), 15B-343, and 17A-335 (PI J.\ Linford). Observations took place during Feb 2014--Aug 2017, covering 24--1304 days after discovery.  Flux measurements can be found in Table \ref{tab:v5666sgr}, and the light curve is plotted in Figure \ref{fig:lcnew4}. During the 2016--2017 A configuration, V5666~Sgr appears marginally resolved in our Ka-band observations; in these cases we let the width of the gaussian vary and fit for the integrated flux density.

%\iffalse
 \begin{figure*}
{\includegraphics[width = 0.48\textwidth]{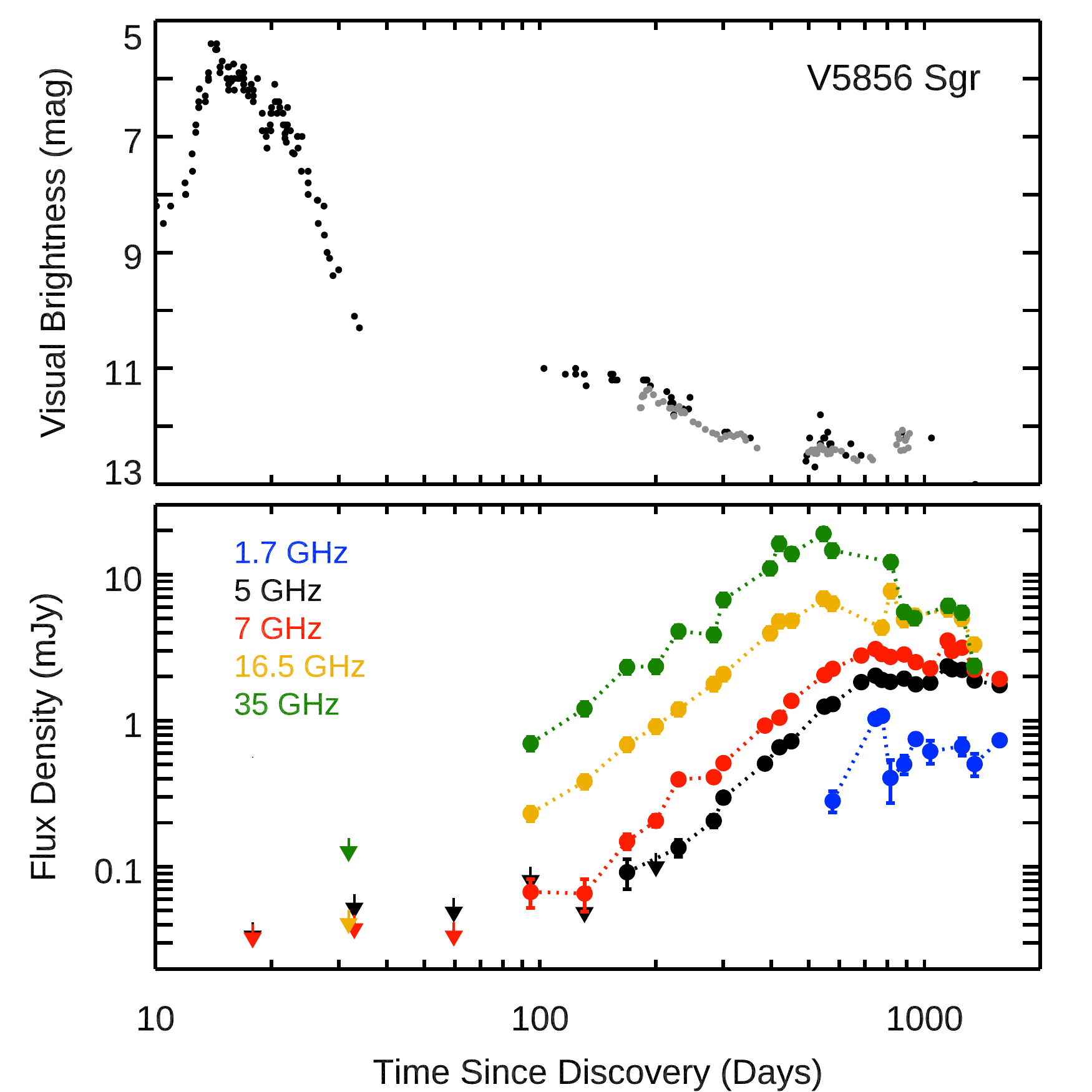}
\includegraphics[width = 0.48\textwidth]{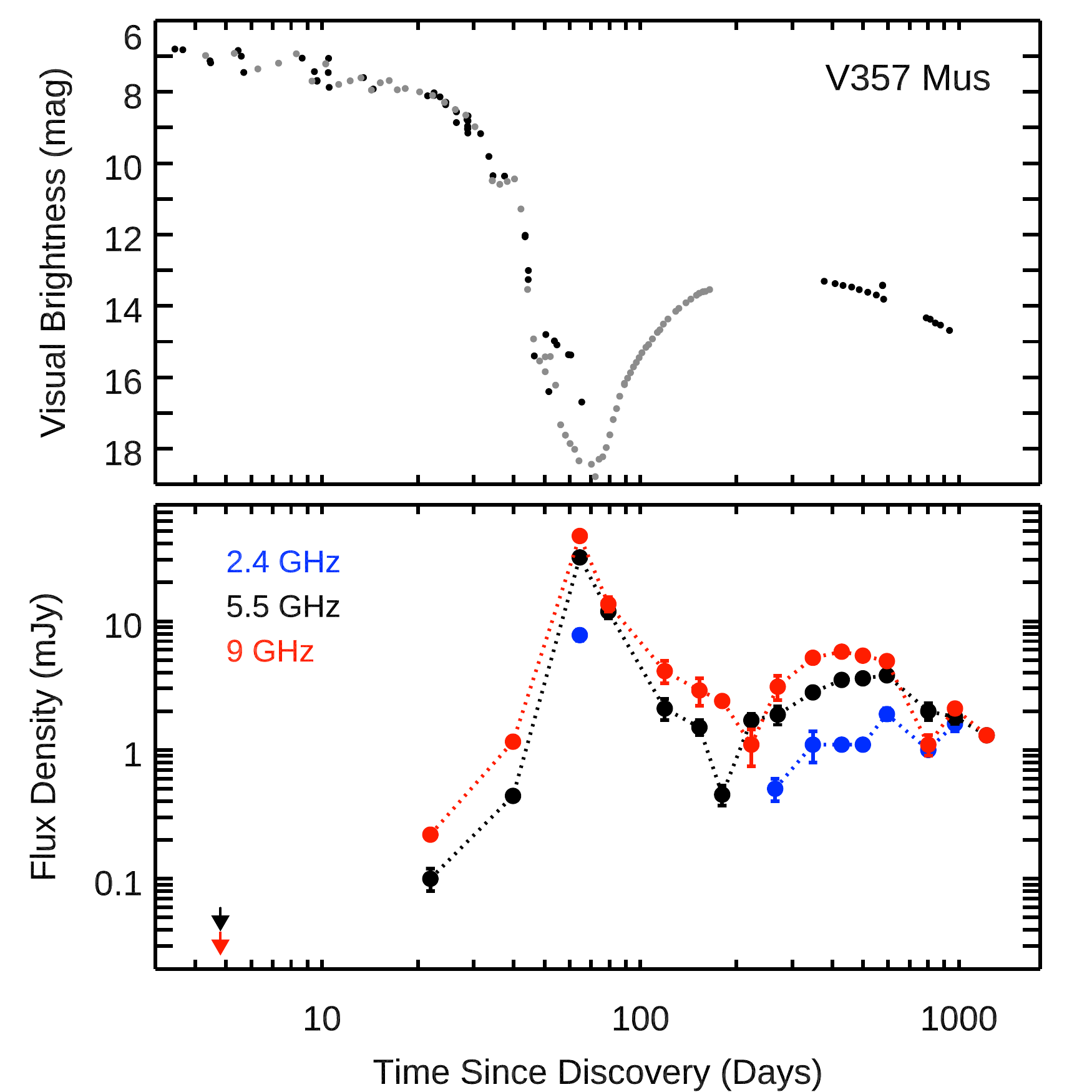}\\
\includegraphics[width = 0.48\textwidth]{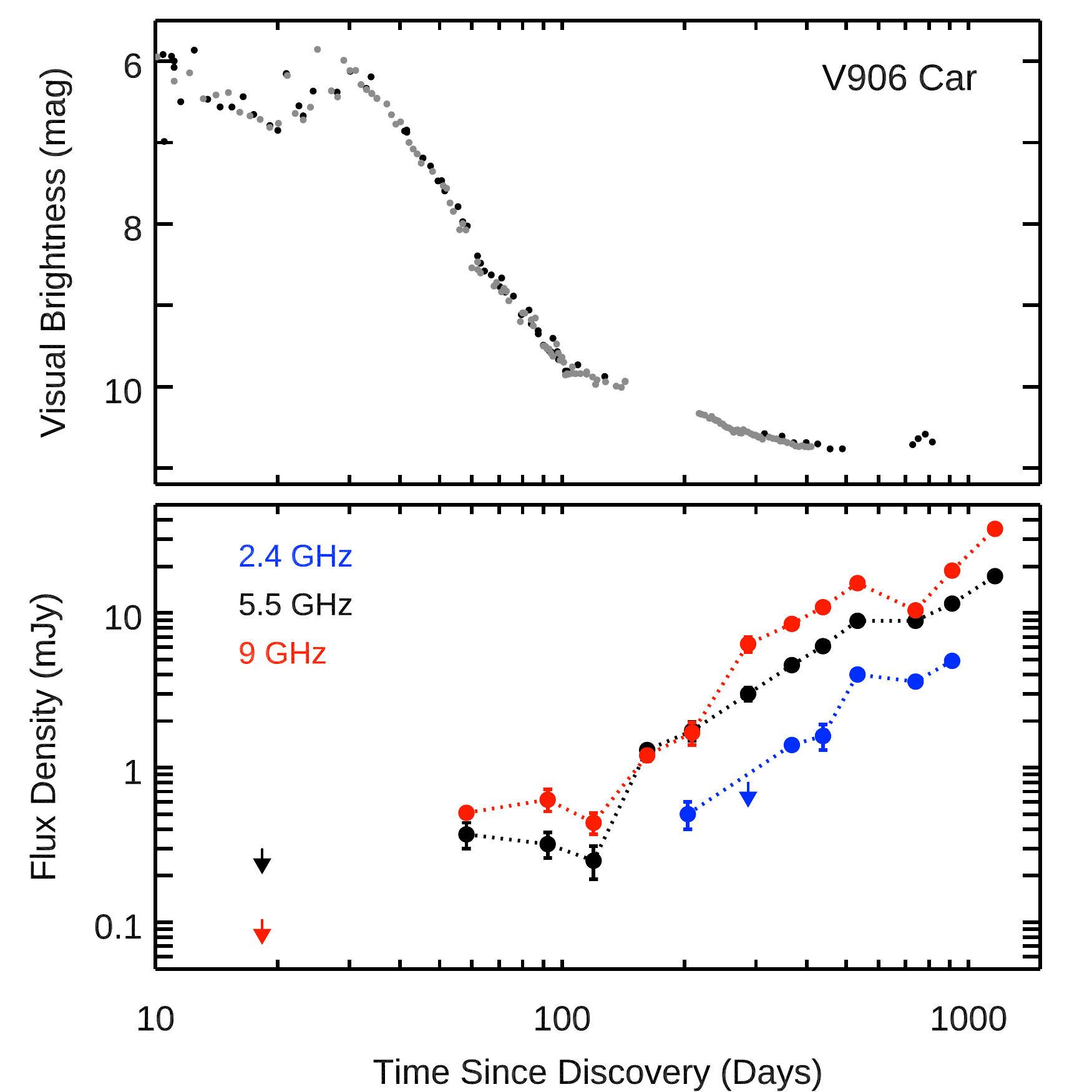}
\includegraphics[width = 0.48\textwidth]{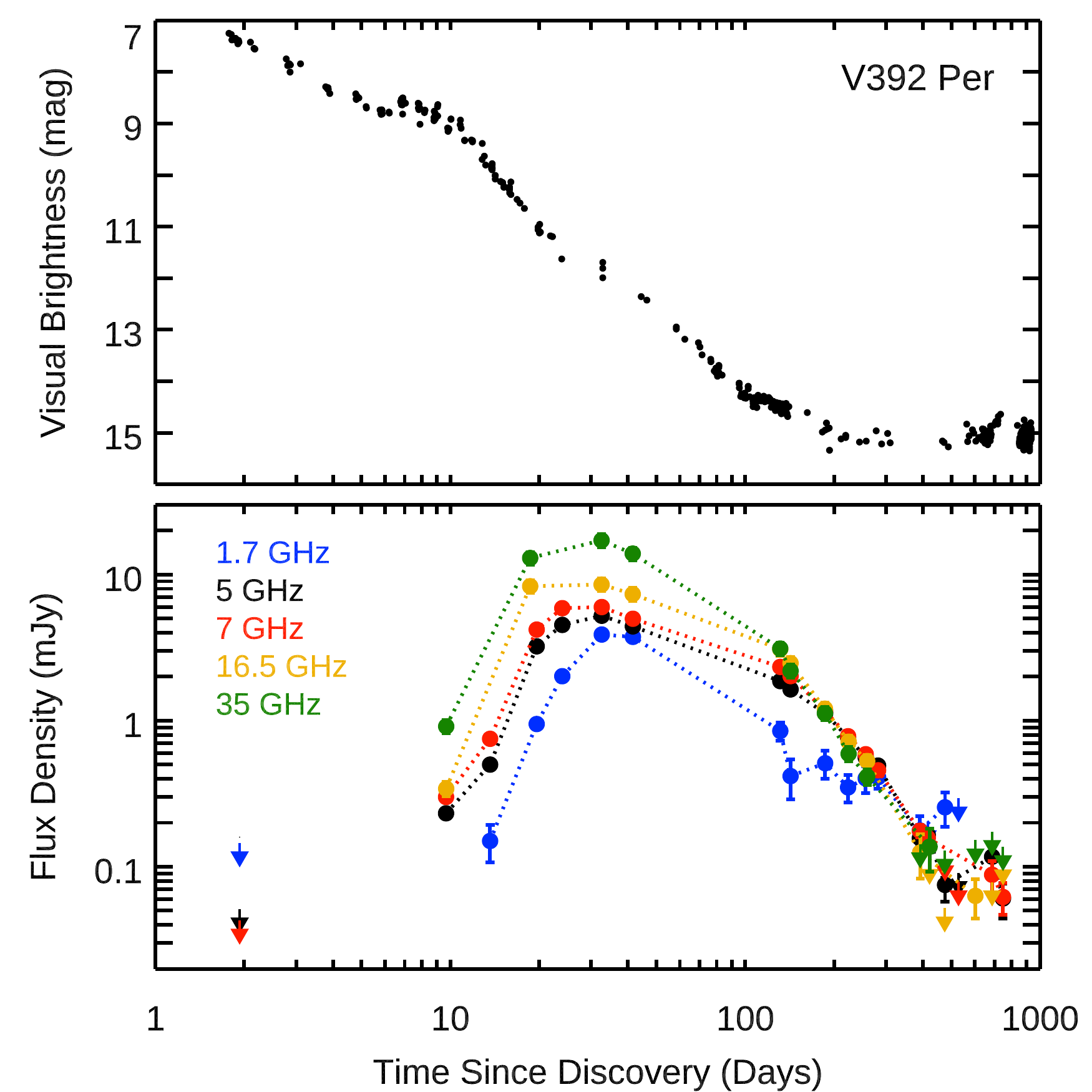}}
\caption{Optical and radio light curves for four novae (clockwise from top left): V5856~Sgr (2016), V357~Mus (2018), V392~Per (2018),  and V906~Car (2018).  For epochs with non-detections, 3$\sigma$ upper limits are plotted with arrows.
Optical light curves are made using $V$-band or ``Vis." AAVSO measurements (black points; \citealt{Kafka20}) and Stony Brook/SMARTS data (grey points; \citealt{Walter+12}). For V5856 Sgr, we do not plot 1.7 GHz upper limits, as their large quantity and relatively high fluxes lead to confusion with other frequencies.}
\label{fig:lcnew6}
\end{figure*}
%\fi

\subsubsection{V2659 Cyg}
Jansky VLA observations of V2659 Cyg were obtained under program codes 13B-057,  16A-258 (PI L.\ Chomiuk), 15B-343, 16B-330, 17A-335 (PI J.\ Linford), and 18A-415 (PI J.\ Sokoloski).
Observations took place during Apr 2014--Sep 2018, covering 6--1642 days after discovery. 
%Early-time observations were obtained with unusually high cadence, in order to look for relatively rapid fluctuations that might result from shocks. aka xmm monitoring
Flux measurements can be found in Table \ref{tab:v2659cyg}, and the light curve is plotted in Figure \ref{fig:lcnew5}. 

During the 2015 A configuration, V2659~Cyg appears marginally resolved in our Ka-band observations, and we let the width of the gaussian vary in JMFIT and fit for the integrated flux density. In the 2016--2017 and 2018 A configurations, V2659~Cyg was resolved at $\geq$16.5 GHz, and we estimated integrated flux densities by fitting gaussian components in the uv-plane with Difmap.

\subsubsection{V5667 Sgr}
Jansky VLA observations of V5667~Sgr were obtained under program codes S61420, 13B-057, 16A-258 (PI L.\ Chomiuk), 15B-343, 17A-335, 17B-313, 18B-273, 19A-298 (PI J.\ Linford), 18A-415 (PI J.\ Sokoloski), and 19B-244 (PI K.\ Sokolovsky). Observations took place during Feb 2015--Dec 2019, covering 7--1774 days after discovery.  Flux measurements can be found in Table \ref{tab:v5667sgr}, and the light curve is plotted in Figure \ref{fig:lcnew5}. 
During the 2016 and 2018 A configurations, V5667~Sgr appears marginally resolved in our Ka-band observations, and we let the width of the gaussian vary and fit for the integrated flux density.

\subsubsection{V5668 Sgr}
Jansky VLA observations of V5668~Sgr were obtained under program codes S61420, SA1159, 13B-057, 16A-258 (PI L.\ Chomiuk), 15B-343, 16B-330, 17A-335, 17B-313, 18A-365, 18B-273, 19A-298 (PI J.\ Linford), and 19B-244 (PI K.\ Sokolovsky). Observations took place during Mar 2015--Dec 2019, covering 2--1744 days after discovery.  Flux measurements can be found in Table \ref{tab:v5668sgr}, and the light curve is plotted in Figure \ref{fig:lcnew5}. 

As reported in \citet{Diaz+18}, V5668~Sgr is an excellent target for interferometric imaging. It was resolved in the A configurations of 2016, 2018, and 2019. For our purposes here, integrated flux densities were estimated by intergrating over the emission area in Difmap. The images will be the subject of Takeda  et al.\ 2021, in preparation.

\subsubsection{V5855 Sgr}
Jansky VLA observations of V5855~Sgr were obtained under program codes SA1159, 16A-318, 17B-313, 18A-365, 18B-273, 19A-298,  20B-302 (PI J.\ Linford),19B-244,  and 20A-395,  (PI K.\ Sokolovsky). Observations took place during Nov 2016--Feb 2021, covering 15--1573 days after discovery.  Flux measurements can be found in Table \ref{tab:v5855sgr}, and the light curve is plotted in Figure \ref{fig:lcnew5}. 
During the 2018 A configuration, V5855~Sgr appears marginally resolved in our Ka-band observations, and we let the width of the gaussian vary and fit for the integrated flux density.

\subsubsection{V5856 Sgr}
Jansky VLA observations of V5856~Sgr were obtained under program codes  SA1159, 16A-318, 17B-313, 18A-365, 18B-273, 19A-298,  20B-302 (PI J.\ Linford), 19B-244,  and 20A-395,  (PI K.\ Sokolovsky). Observations took place during Nov 2016--Feb 2021, covering 18--1568 days after discovery.  Flux measurements can be found in Table \ref{tab:v5856sgr}, and the light curve is plotted in Figure \ref{fig:lcnew6}. 
During the 2018 A configuration, V5855~Sgr appears marginally resolved in our Ka-band observations, and we let the width of the gaussian vary and fit for the integrated flux density.

\subsubsection{V357~Mus}
We obtained ATCA observations of V357~Mus under programs CX371 and C3279 during Jan 2018--Sep 2020, covering 4--972 days after discovery. Most epochs in the first few months of eruption only covered C band (5.5 and 9 GHz), with the exception of the epoch on 2018 Mar 19.0, when we also obtain data at 1--3 GHz, and which also just happens to be the peak of the first radio maximum.
 Flux measurements can be found in Table \ref{tab:v357mus}, and the light curve is plotted in Figure \ref{fig:lcnew6}. V357~Mus appears as a point source in all epochs and frequencies.

\subsubsection{V906~Car}
ATCA radio observations of V906~Car are published by \citet{Aydi+20}. However, that publication occurred while the radio light curve was still evolving, and we publish three additional, later epochs here (in fact, the radio light curve continues to rise up until present day). 
These ATCA observations were obtained under program codes CX371 and C3279. ATCA monitoring took place during Apr 2018--Sep 2020, covering 18--911 days after discovery, although observations continue to the present day. Flux measurements can be found in Table \ref{tab:v906car}, and the light curve is plotted in Figure \ref{fig:lcnew6}.  V906~Car appears as a point source in all ATCA images.

\subsubsection{V392~Per}
Jansky VLA observations of V392~Per were obtained under program codes  17B-352, 19A-298  (PI J.\ Linford), 19B-244, and 20A-395 (PI K.\ Sokolovsky). Observations took place during Apr 2018--May 2020, covering 1--748 days after discovery.  Flux measurements can be found in Table \ref{tab:v392per}, and the light curve is plotted in Figure \ref{fig:lcnew6}. During the 2018 A configuration, V392~Per is unresolved. During the 2019 A configuration, V392 Per was undetected at higher frequencies, and consistent with a low S/N point source at lower frequencies.

We note that radio observations of the eruption of V392 Per were also obtained with the Arcminute Microkelvin Imager Large Array (AMI-LA), at higher cadence than the VLA observations but only at a single frequency (15.5 GHz; \citealt{Linford+18}). These observations will be the subject of a future paper exploring the source in more detail.

\subsection{Multi-wavelength nova properties}  \label{sec:mwprop}

Some basic properties of the novae in our sample are listed in Table \ref{tab:props}. Sources were investigated using \emph{Gaia} EDR3 \citep{Gaia16,Gaia21}. 
%In order to determine reliable equatorial coordinates and the \emph{Gaia} counterpart, we started with coordinates reported in the the International Variable Star Index (VSX) database.
 If there was an unambiguous \emph{Gaia} match to the nova, we took the \emph{Gaia} equatorial coordinates and list them in Table \ref{tab:props}. In cases with more than one potential \emph{Gaia} match to the nova position, we used radio observations to hone the position and determine the \emph{Gaia} counterpart. Several novae have no \emph{Gaia} counterpart, and in these cases we report a position measured from radio observations (these cases are noted in Table \ref{tab:props}).
 
 In cases where \emph{Gaia} parallaxes are measured at $\geq2.5 \sigma$ significance in EDR3, distances are estimated taking the prior suggested by \citet{Schaefer18}: 150 pc /sin($l$), where $l$ is Galactic latitude (the prior is set to zero beyond a maximum distance of 8 kpc). Lower significance \emph{Gaia} measurements were carefully considered; in some cases there is essentially no useful information in \emph{Gaia} EDR3, as the nova was in outburst and highly variable during \emph{Gaia} data acquisition. In these cases, we take distances from \citet{Gordon+21} or other sources in the literature. In other cases, where there is some, albeit low significance, \emph{Gaia} constraint, we consider both the literature and \emph{Gaia} estimates and take whichever is most constraining and reliable. If no superscript is listed in the Distance column of Table \ref{tab:props}, this implies that the distance is from a \emph{Gaia} parallax measurement, which should be revisited after future \emph{Gaia} data releases. Uncertainties correspond to roughly 1$\sigma$ (68\%) confidence intervals.

The date of discovery is taken as $t_0$, the time of start of eruption (in all cases, this is bound to be after the time of the true eruption start). All novae were discovered with optical surveys except V598~Pup and V959~Mon.
V598~Pup was discovered months into eruption at X-ray wavelengths by the XMM-Newton slew survey \citep{Saxton+08}. The discovery date listed in Table \ref{tab:props} is actually the time V598~Pup brightened in archival optical observations obtained with the All Sky Automated Survey \citep{Pojmanski02}, and is remarkably well constrained to within three days \citep{Read+08}. V959~Mon was discovered at GeV $\gamma$-ray wavelengths with the Large Area Telescope  on the \emph{Fermi Gamma Ray Space Telescope} while the source was in solar conjunction, and was not optically identified as a nova until months later \citep{Ackermann+14}. Here we take the first $\gamma$-ray detection of V959~Mon as its time of discovery.

\begin{figure*}[t]
\begin{center}
\includegraphics[height=3.5in,angle=0]{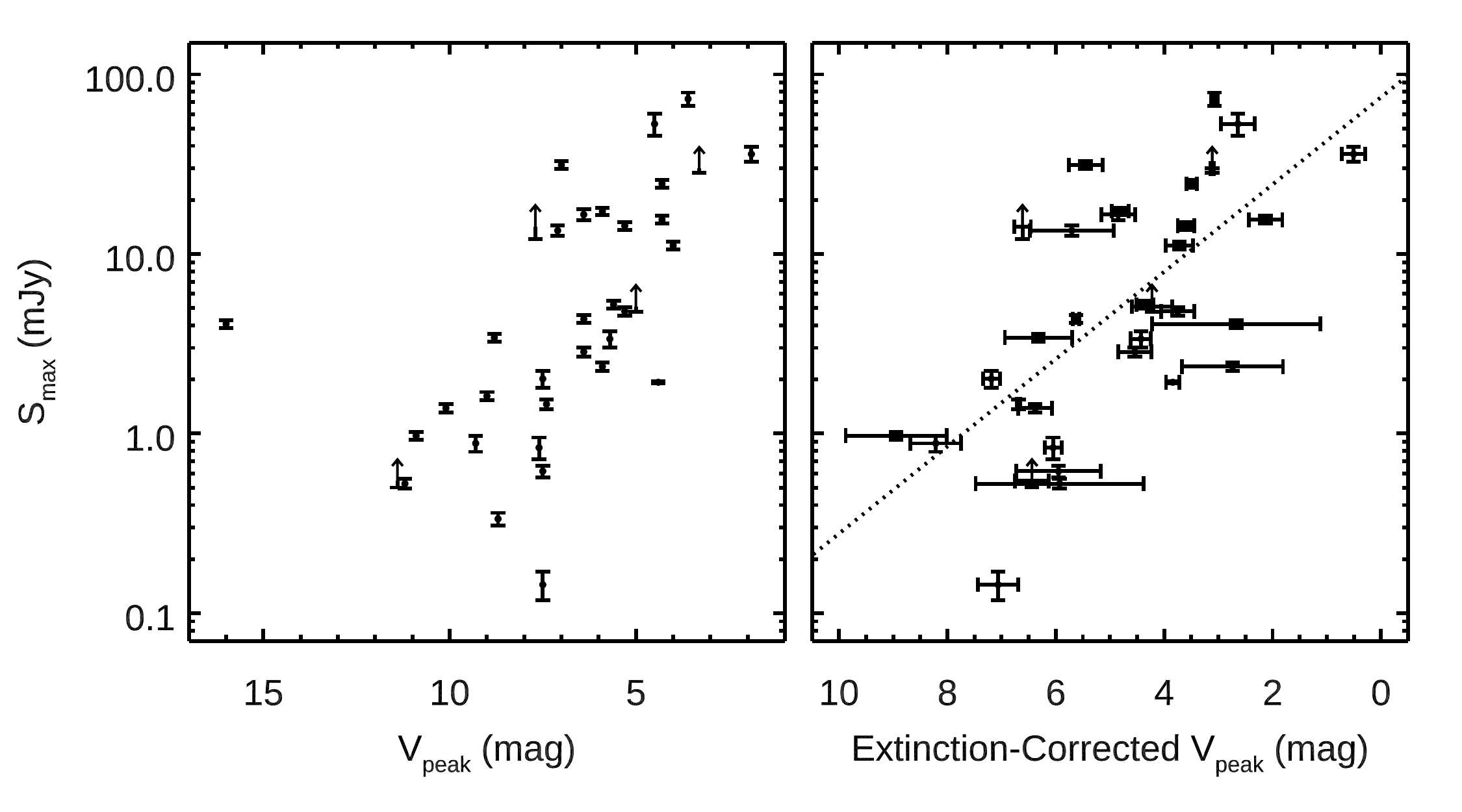}
\caption{The flux density at radio light curve maximum ($S_{\rm max}$) compared with the observed peak optical apparent magnitude ($V_{\rm peak}$; left panel) and the extinction-corrected peak optical apparent magnitude (right panel). The dotted line in the right panel represents a linear fit to the data (see \S \ref{sec:optvrad} for details). Radio flux densities are measured at 5 GHz, or when 5 GHz is not available, 8.5 GHz; optical brightnesses are measured in the $V$ band.}
\label{fig:vmax}
\end{center}
\end{figure*}

\begin{figure*}[t]
\begin{center}
\includegraphics[height=3.5in,angle=0]{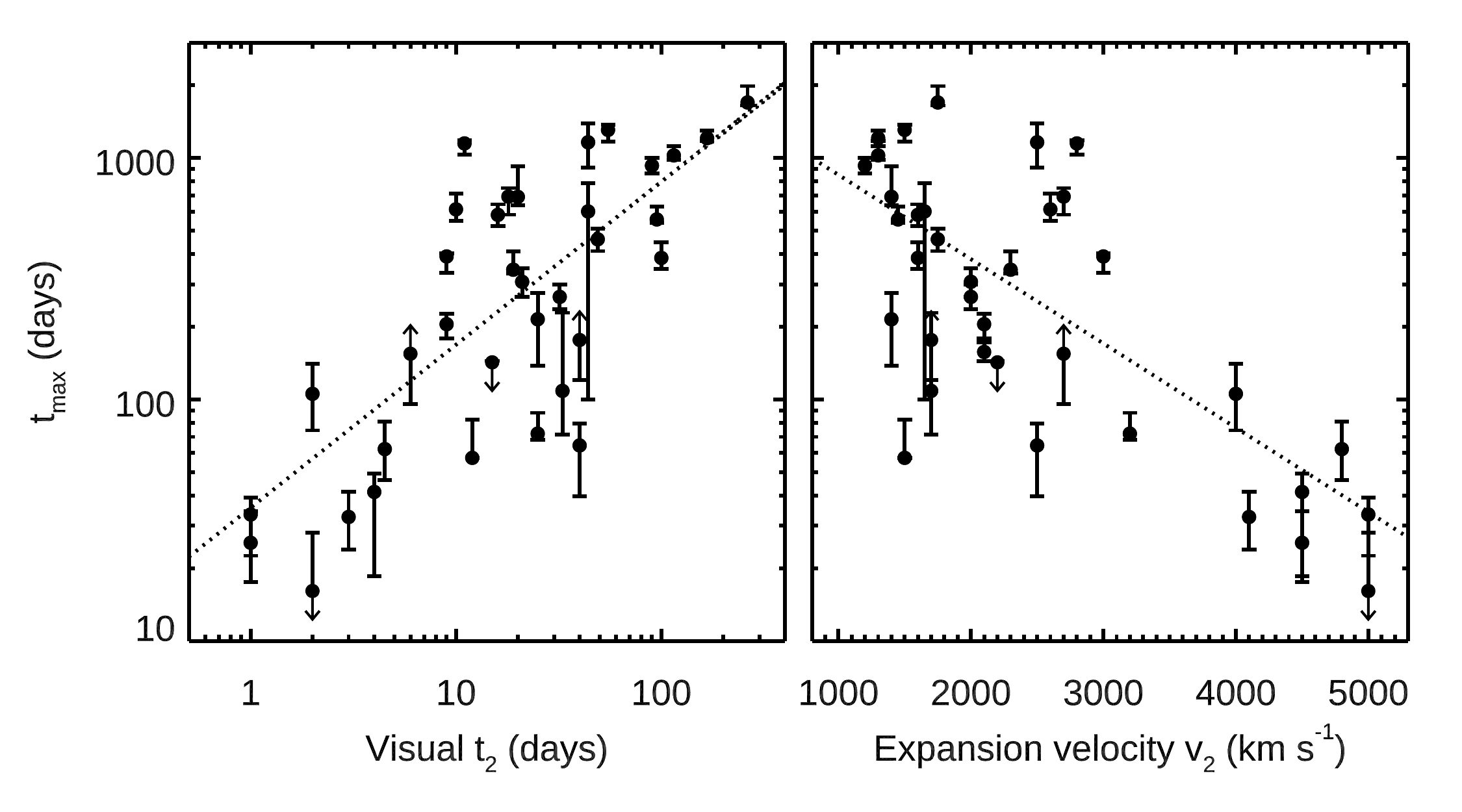}
\caption{The time to radio light curve maximum ($t_{\rm max}$) is compared, in the left panel, with the time for the optical light curve to decline by two magnitudes from peak ($t_2$). In the right panel, $t_{\rm max}$ is compared with the expansion velocity of the ejecta $v_2$. The dotted lines represent linear correlations fit to the data (see \S \ref{sec:optvrad} for details).
%The dashed line in the left panel is not a fit, but simply a fiducial representing $t_{\rm max} = 10 t_2$. 
As in Figure \ref{fig:vmax}, $t_{\rm max}$ is measured at 5 GHz or 8.5 GHz. The plotted uncertainties on $t_{\rm max}$ show the time interval between the last observation before $t_{\rm max}$ and the first observation after $t_{\rm max}$.}
\label{fig:tmax} 
\end{center}
\end{figure*}

The optical light curves are measured in the $V$ band, and the peak observed magnitude is listed as $V_{\rm peak}$ in Table \ref{tab:props} (also Figure \ref{fig:vmax}).
We parameterize how rapidly the optical light curve of a nova evolves with $t_2$, the time for the light curve to decline by two magnitudes from visual peak (Figure \ref{fig:tmax}). In cases where the optical light curve shows variations and ``jitters", we measure $t_2$ as the last time the light curve crossed the $V_{\rm peak} - 2$ mag threshold, as performed by \citet{Strope+10}.

Ejecta expansion velocities are estimated and interpreted within the framework put forward by \citet{Aydi+20b}.
$v_1$ is the velocity of the ``intermediate component/principal component" in the H Balmer lines, measured from the absorption trough of the P Cygni profiles (on top of the broad emission) a few days after optical peak.
$v_2$ is the FWZI/2 of the broad H Balmer emission measured several days after optical peak. When spectra were available, we self-consistently measure these quantities (if no reference is provided in Table \ref{tab:props}, velocities are measured in this current work using spectra published by e.g., \citealt{Aydi+20b}). However, for older novae we are forced to estimate $v_1$ and $v_2$ from figures or values quoted in the literature. In cases where only a FWHM is available, we quote this value as $v_2$, as we have empirically estimated that FWZI/2$ \approx$ FWHM for typical novae around light curve maximum.

We obtained other parameters in Table \ref{tab:props} from the compilations of \citet{Ritter&Kolb03}, \citet{Strope+10}, \citet{Ozdonmez+18}, and \citet{Gordon+21}. 
%For objects missing from these compilations, we searched the literature and references are as noted in Table \ref{tab:props}. In the ``Dust?" column, we report a ``Y" if there is evidence of dust formation in the nova eruption. A ``N" is reported if dust is searched for (using e.g., IR data) and not found; a ``?" is listed if no searches for dust are reported in the literature. In some cases like V4743~Sgr, there is evidence for old cold dust that must pre-date the eruption, but no evidence of dust newly formed in the eruption \citep{Nielbock&Schmidtobreick03}; such targets are listed as ``N" in the `Dust?" column. Ideally, the inclination column of Table \ref{tab:props} would represent the inclination of the binary orbit, but in many cases it is indirectly estimated from the expanding nova ejecta, assuming they have bipolar morphology (e.g., \citealt{Munari+11, Mason+18}).  
 In some cases, we made measurements for the first time. Light curve parameters for V5666~Sgr, V5667~Sgr, and V2659~Cyg were determined from the light curves presented in Figure \ref{fig:lcnew3}. The reddening for V1723~Aql is estimated from the photometry of \citet{Nagashima+13}, who present a color index $V-R \approx 2.5$ around light curve maximum. Assuming an intrinsic color of $(V-R)_0 = 0$ around maximum (e.g., \citealt{dellaValle+02}) yields reddening values of E($V-R$) = 2.5 and E($B-V$) = 4.3 mag.
%NOTE- KVS- back up where $(V-R)_0 = 0$ comes from. best option right now is van den bergh and younger, but they don't have r band.

To constrain the emission mechanism powering nova radio luminosity (i.e., thermal or non-thermal), the brightness temperature (a measure of the radio surface brightness) is a valuable diagnostic.  Brightness temperature is calculated assuming a circular disk-like emitter of diameter $\theta$ using the equation:
\begin{equation}
T_B = 1765.8\, {\rm K}\ \left(\frac{\nu}{\rm GHz}\right)^{-2} \frac{S_{\nu}}{\rm mJy} \left(\frac{\theta}{\rm arcsec}\right)^{-2}
\end{equation}
The most reliable measurements of brightness temperature use high resolution imaging to directly constrain the size of the emitting region, and thereby the surface brightness. Unfortunately, the resolution of ATCA/VLA is not usually high enough for this to be true, but instead we can approximate the angular size of the source if we know its distance, expansion velocity, and expansion time. For each nova, we use the uncertainty on distance listed in Table \ref{tab:props} and the range of velocities spanning $v_1$ to $v_2$ to estimate a plausible range of angular diameters ($\theta$) at a given observation epoch. We then use this range in sizes to estimate a plausible range of brightness temperature, plotted as polygons in Figures \ref{fig:bt_fast}--\ref{fig:bt_slow4}. We note that the brightness temperatures estimated in this way are likely to be lower limits on the true brightness temperature of any non-thermal emission, as high-resolution imaging reveals that synchrotron emission is often arranged as compact knots within the more diffuse thermal ejecta \citep[e.g.,][]{Chomiuk+14}.
The brightness temperatures plotted in Figures \ref{fig:bt_fast}--\ref{fig:bt_slow4} are for the frequency that yields the most constraining (highest) brightness temperature estimates and has reasonable temporal coverage. In most cases, this will be a relatively low frequency, often C band ($\sim$5 GHz)  or, when available, L/S bands (1--3 GHz). 

Radio luminosities are calculated assuming the distances in Table \ref{tab:props}. The luminosities plotted in Figures \ref{fig:bt_fast}--\ref{fig:bt_slow4}  correspond to observing frequency $\sim$5 GHz, or when that is not available, $\sim$8.5 GHz. In the case of V1370~Aql, neither of these frequencies are available, and we instead use the 1.5 GHz light curve. The plotted polygons denote the uncertainty band in luminosity, given measurement uncertainties on flux density and the uncertainties on distance listed in Table \ref{tab:props}.

\section{Discussion}\label{sec:discussion}
The radio light curves of novae are strikingly diverse. Some remain undetected (see the Appendix), while others become some of the brightest transients in the radio sky (at least at high frequencies; e.g., V1974~Cyg). Some remain bright for a few weeks  (e.g., U~Sco), while others are bright for years (e.g., V723 Cas). Many show a single peak in their radio light curves, while others show two distinct peaks (e.g., V1723~Aql). We investigate this diversity in more detail below.

The radio spectra of novae are complex, and often require more than a single power law to describe. Even when the emission mechanism is purely thermal, the radio spectrum turns over first at high frequencies, and then the turn-over proceeds to lower frequencies; intermediate times can require $>$2 power laws to fully describe the spectral shape (see \citealt{Nelson+14} and \citealt{Weston+16a} for some examples). For these reasons, it is difficult to talk about nova spectra in terms of a single spectral index, and discussion of the radio  spectral evolution for 36 novae would rapidly become ungainly and long. We postpone discussion of radio spectral evolution to future papers, where we will investigate the thermal and synchrotron properties of these novae in more detail.

\subsection{Comparison of basic radio and optical properties}\label{sec:optvrad}
Perhaps the diversity in radio light curves is not surprising given the range of optical light curves of novae, which show amplitudes spanning a factor of $\gtrsim$10 magnitudes (factor of $\sim 10^4$) and $t_2$ spanning $>$two orders of magnitude  \citep{Kawash+21}. Both radio and optical light curves start out optically thick and transition to an optically thin state. If both optical and radio emission are thermal, powered by the warm expanding ejecta or a wind, we might expect the radio and optical light curves to track one another.

Very roughly, the peak of each light curve is set by the maximum size of the photosphere.
In Figure \ref{fig:vmax}, we compare the peak radio flux density ($S_{\rm max}$; also listed in Table \ref{tab:radio}) with the peak optical magnitude. The left panel plots peak $V$-band magnitudes as observed, while the right panel corrects for foreground reddening using the E($B-V$) values tabulated in Table \ref{tab:props}. We plot peak radio flux densities as measured at 5 GHz, or when that is not available, as measured at 8.5 GHz (limits are estimated if the radio light curve maximum is not observed). We see that, indeed, the peak brightnesses at optical and radio wavelengths are correlated. V1723~Aql is a strong outlier in the left panel of Figure \ref{fig:vmax} because of the high Galactic extinction along its line of sight, but it falls in line with the rest of the sample in the extinction-corrected right panel.

\begin{figure*}[t]
\begin{center}
\includegraphics[height=3.5in,angle=0]{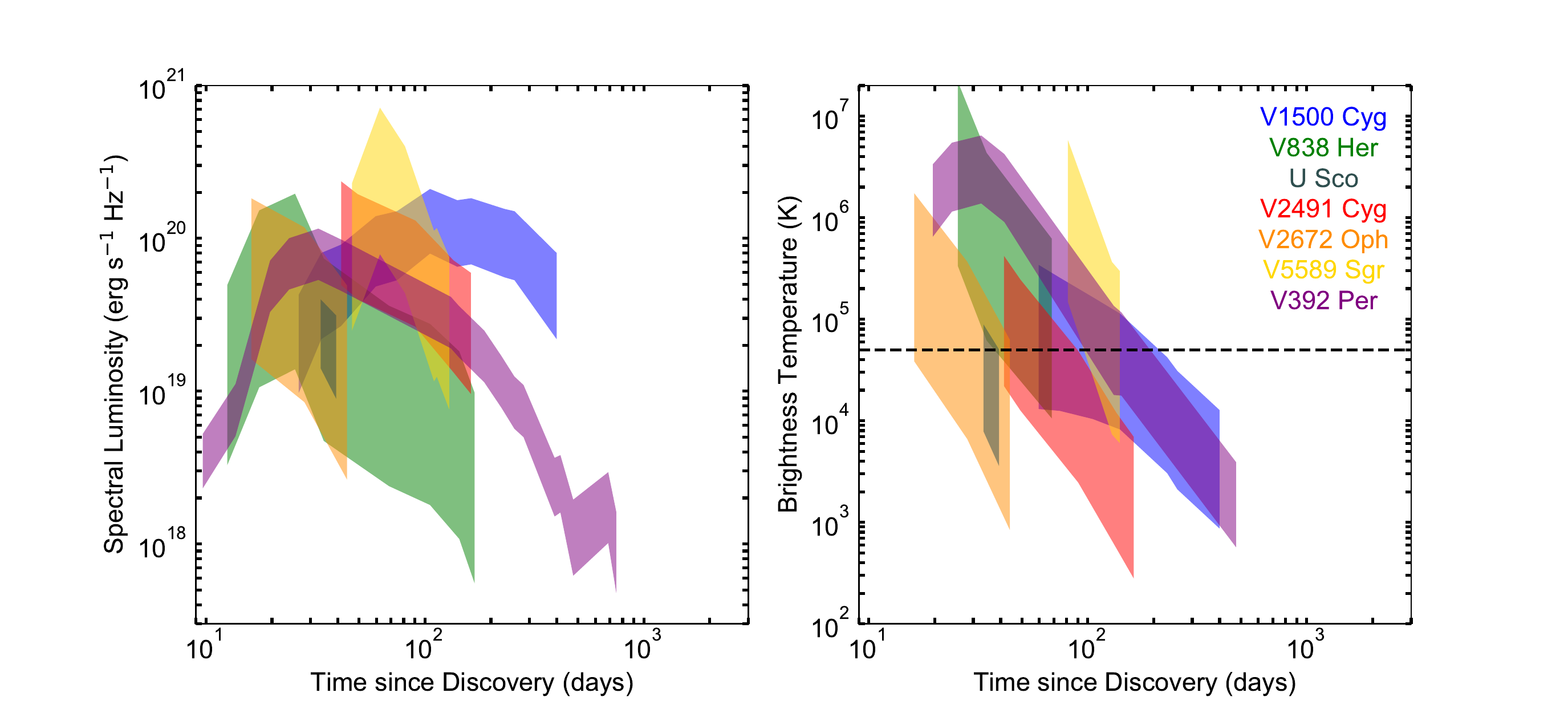}
\caption{Radio luminosity and brightness temperature for seven of the fastest-evolving radio novae. In the left panel, spectral luminosity is plotted as observed at $\sim$5 GHz or $\sim$8.4 GHz; the width of shaded polygons denotes uncertainty on distance and flux density. In the right panel, brightness temperature is plotted as measured at the most constraining frequency. The width of shaded polygons denotes uncertainty in distance, expansion velocity, and flux density. A dashed line is plotted at $T_B = 5 \times 10^4$ K to denote the conservative maximum temperature expected for thermal emission.}
\label{fig:bt_fast}
\end{center}
\end{figure*}

\begin{figure*}[t]
\begin{center}
\includegraphics[height=3.5in,angle=0]{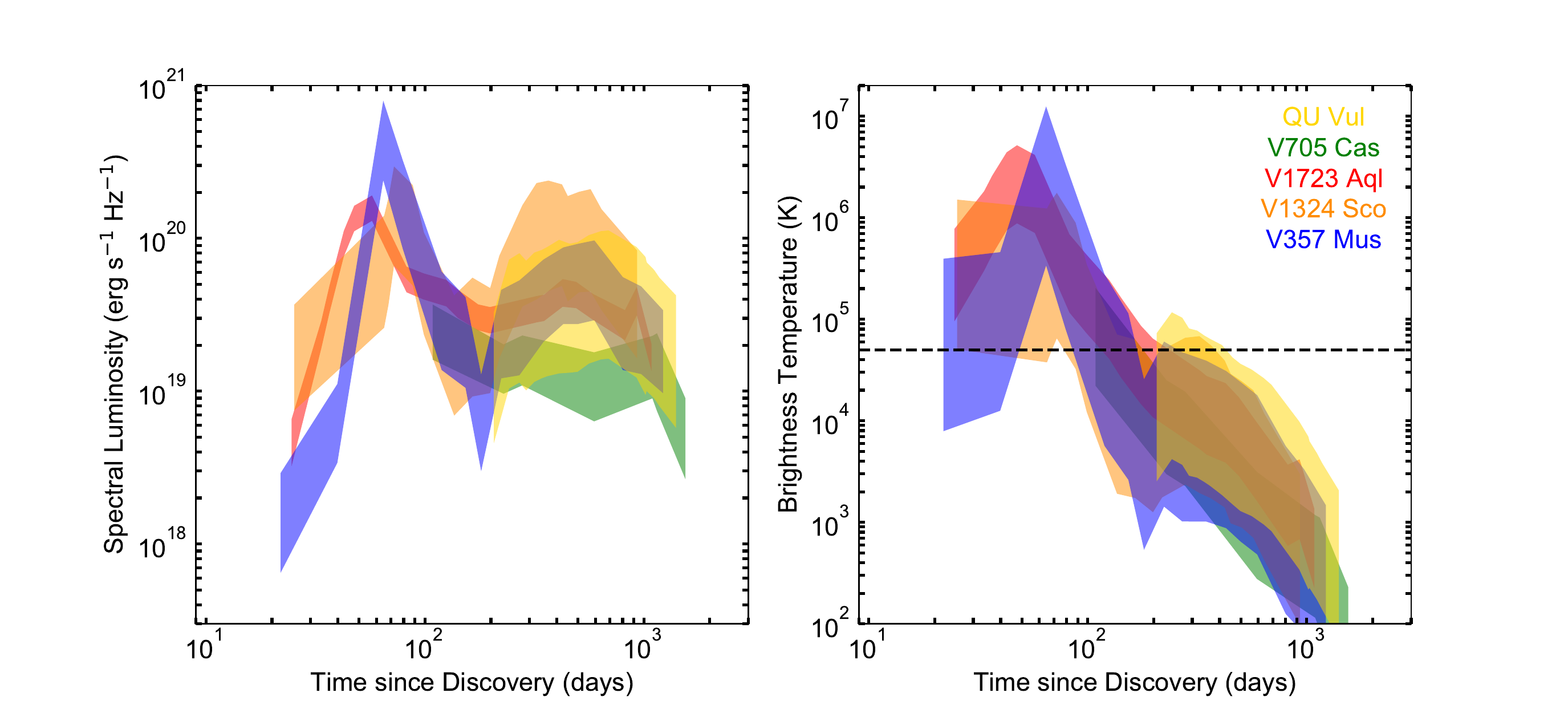}
\caption{As in Figure \ref{fig:bt_fast}, but here we plot five novae that show double-peaked radio light curves. }
\label{fig:bt_double}
\end{center}
\end{figure*}

The correlation between radio and optical peak brightness holds over three orders of magnitude in radio flux density. Fitting a linear relation to the data in the right panel of Figure \ref{fig:vmax},
%\begin{equation}
$log(S_{\rm max}) = \beta (V_{\rm peak,0} - 4.88) + \alpha$,
%\end{equation}
we find  $\alpha = 0.69 \pm 0.09$ and $\beta = -0.24 \pm 0.05$,
with a substantial scatter of $\sim$0.5 dex around the relation ($\sigma = 0.48 \pm 0.07$, to be precise).
We also see that the variation in optical peak brightness is larger than in radio peak brightness (factor of $\sim 10^4$ compared with $10^3$), even after correcting for dust extinction.
In Figures \ref{fig:bt_fast}--\ref{fig:bt_slow4}, we see that peak radio luminosities for the majority of novae span a relatively narrow range of $\sim 10^{19}-10^{20}$ erg s$^{-1}$ Hz$^{-1}$, so much of the spread in peak radio flux density is due to distance.

We might similarly expect the timescales of the radio and optical light curves to be correlated. For example, a low-mass ejection should thin out and fade quickly at both optical and radio wavelengths. 
%If the radio light curve is thermal, the radio light curve is expected to peak at a time post-eruption given by:
%\begin{eqnarray}
%t_{\rm max} & \approx & 2.6\, {\rm yr} \left(\frac{\nu}{\rm GHz}\right)^{-0.42}
%\left(\frac{T_e}{\rm 10^4\, K}\right)^{-0.27}
%\left(\frac{M_{\rm ej}}{10^{-4}\, M_{\odot}}\right)^{0.40} \\
% \notag & & \times \left(\frac{v_{\rm ej}}{\rm 10^3\, km\ s^{-1}}\right)^{-1}
%\end{eqnarray}
%\citep{Seaquist&Bode08}.
In the left panel of Figure \ref{fig:tmax}, we plot  the time to radio maximum $t_{\rm max}$, and compare it with optical $t_2$ (time to optical maximum usually occurs quickly after eruption and is therefore only measured for a few novae; $t_2$ is a more available diagnostic of light curve speed). The novae in our sample span a factor of $\sim$300 in optical $t_2$ and a factor of $\gtrsim$1000 in radio $t_{\rm max}$. We see that radio and optical timescales are indeed positively correlated. 
%The dashed line  represents $t_{\rm max} = 10\, t_2$ to guide the eye, and appears to describe much of the data. 
A linear fit of the form $log(t_{\rm max}) = \beta (log(t_2) - 1.263) + \alpha$ returns $\alpha = 2.40 \pm 0.07$ and $\beta =  0.68 \pm 0.11$, and is plotted as a dotted line in Figure \ref{fig:tmax}. There is a scatter around this line of $\sigma = 0.38 \pm 0.05$.

The right panel of Figure \ref{fig:tmax} plots the time to radio maximum against the expansion velocity $v_2$ (this is expected to be the speed of the faster, more spherical flow as hypothesized by \citealt{Aydi+20b}). Unsurprisingly, novae with faster expansion velocities also show radio light curves that evolve more quickly. We fit a line as $log(t_{\rm max}) = \beta(v_2/1000 - 2.51) + \alpha$, and found $\alpha = 2.40 \pm 0.07$ and $\beta= -0.35 \pm 0.06$. The scatter around this relation is $\sigma = 0.40 \pm 0.05$.

\subsection{The fastest radio novae: synchrotron flares and thermal blips} \label{sec:fastpeak}
%Mostly synchrotron with a sprinkling of thermal emission}

While most novae have radio light curves that peak $>$1 year after eruption, about a quarter show much faster radio light curves with time to radio maximum $\lesssim$100 days (Figure \ref{fig:tmax}). These fast radio novae can, in turn, be divided into two classes---those that show a later, second radio maximum (which are the subject of \S \ref{sec:doublepeak}), and those that rapidly rise to maximum and then simply fade. This second class---novae that show a single rapid radio maximum---is represented here by V1500~Cyg, V838~Her, U~Sco, V2491~Cyg, V2672~Oph, V5589~Sgr, and~V392 Per, and is plotted in Figure \ref{fig:bt_fast}. Note that the temporal coverage on these novae varies greatly, and while in some cases like V5589~Sgr we can exclude the possibility that the nova rose to a second radio maximum, in other cases like V2491~Cyg and V2672~Oph we cannot. In fact, higher cadence observations of V392~Per with AMI-LA imply that it may serve as a bridge between the sources discussed here and the double-peaked sources of \S \ref{sec:doublepeak}, as a second bump is apparent in the 15.5 GHz light curve between days 41 and 131, when the VLA was down for an electrical infrastructure upgrade \citep{Williams+21}.

The right panel of Figure \ref{fig:bt_fast} shows that many of these fastest novae reach brightness temperatures substantially $>10^4$ K at early times. In novae, brightness temperatures in excess of $5 \times 10^4$ K are evidence of non-thermal synchrotron emission for two reasons. The first reason is that the photo-ionized nova ejecta should quickly relax to electron temperatures of $T_e \approx (1-5) \times 10^4$ K, with the value primarily determined by the white dwarf's temperature in the supersoft X-ray phase, and therefore by the white dwarf mass. The temperature should remain near this value for a time set by the ejecta mass and expansion velocity before cooling  \citep{Cunningham+15}. If the gas is optically thick at a given frequency, it should show a brightness temperature equivalent to the electron temperature of the gas; as it transitions to optically thin at that frequency, the corresponding brightness temperature will drop. The implication is that, if photoionization is the only heating source in novae, radio brightness temperatures should not be $>5 \times 10^4$ K. The second reason that brightness temperatures significantly $> 10^4$ K imply non-thermal emission is that free-free optical depth $\propto T_e^{-1.35}$ \citep{Seaquist&Bode08}, so it becomes harder for hotter gas to be optically thick at radio wavelengths. Therefore, while X-ray observations often imply the presence of shock-heated gas with $T_e  \gg 10^6$ K \citep[e.g.,][]{Mukai+08, Gordon+21}, the hot gas would need to have very large emission measures and masses in order to be optically thick at radio wavelengths and show high radio brightness temperatures. Such high emission measures would be inconsistent with observed X-ray luminosities (\citealt{Vlasov+16}; see also discussions in \citealt{Chomiuk+14tpyx}, \citealt{Weston+16a}, and \citealt{Linford+17}).

By this criterion, three of the novae plotted in Figure \ref{fig:bt_fast}---V838 Her, V5589 Sgr, and V392 Per---show strong evidence for synchrotron emission, with brightness temperatures $> 10^5$ K measured in early observations. V2672~Oph is also probably dominated by non-thermal emission, although constraints on its brightness temperature are not quite strong enough to conclusively discern. V1500~Cyg, V2491~Cyg, and U~Sco show brightness temperatures which could be explained with thermal emission, but cannot exclude synchrotron emission. 

These fastest radio novae have the highest expansion velocities in our sample, all with $v_2 \gtrsim 4000$ km s$^{-1}$ (Figure \ref{fig:tmax}). Such high velocities will naturally lead to more rapid radio evolution, regardless of whether the emission is thermal or non-thermal. We also expect higher velocity shocks to transfer more energy to relativistic electrons, as the energy in relativistic particles is usually parameterized as a fraction of the post-shock energy $\sim \rho v_{\rm sh}^2$, where $v_{\rm sh}$ is the velocity of the shock and $\rho$ is the density of gas being shocked. We might therefore expect synchrotron luminosities to be higher in novae with faster ejecta.

Evolved companions are over-represented amongst this class of fastest radio novae. Of the four novae in our sample known to have orbital periods $>$16 hr and companions evolving off the main sequence (see Table \ref{tab:props}), three of them have radio light curves that fall into this fastest category. These three novae (U~Sco, V5589~Sgr, and V392~Per) are also some of the optically fastest novae in our sample (i.e., shortest $t_2$), while the fourth nova V723~Cas is one of the slowest in our sample and is included in \S \ref{sec:slowpeak}.  It is not surprising that novae with sub-giant companions tend to have fast novae, as the evolved companion should power a high mass transfer rate which will trigger novae with relatively low ejecta masses \citep{Yaron+05, Kalomeni+16}. 
But notably, in at least the case of V5589~Sgr, the high brightness temperature is indicative of synchrotron emission with no evidence for a second thermal peak (contrary to what is observed for the novae in \S \ref{sec:doublepeak}). This lack of a thermal peak may be attributable to low ejecta mass ($M_{\rm ej} \lesssim 10^{-6}$ M$_{\odot}$).
As for the origin of the synchrotron emission, while we cannot exclude the possibility of shocks internal to the nova ejecta (as may be in V838~Her), such rapidly evolving non-thermal emission has primarily been seen in novae with red giant companions like V745~Sco, where the nova ejecta shock the red giant wind, accelerate particles to relativistic speeds, and create synchrotron emission \citep[e.g.,][]{Kantharia+16}. The synchrotron emission in V5589~Sgr and V392~Per may therefore be an indication of circumstellar material (CSM), presumably expelled by the  evolved companion. 

Naively, we do not expect significant wind CSM from a sub-giant donor, which should be transferring mass via Roche lob overflow \citep{Webbink+83, Kalomeni+16}.
Still, a sub-giant companion should have a higher mass transfer rate than a dwarf companion, and therefore may pose more opportunity for non-conservative mass transfer and the creation of detectable CSM around the binary (intermediate between the low-density environs of cataclysmic variables and the dense CSM around symbiotic stars). To date,  there is very little evidence for CSM around sub-giants. As the Galactic recurrent nova with the fastest recurrence time, U~Sco is one of the best studied novae and has a sub-giant companion, and yet it shows X-ray non-detections which place an upper limit on the density of the CSM (\citealt{Drake&Orlando10}; the data presented here also do not require synchrotron emission or shocks). It is notable that V5589~Sgr and V392 Per have longer orbital periods than U~Sco or V723~Cas, and therefore the companions are more evolved. The data presented here may be evidence of CSM density increasing as companion stars evolve across the sub-giant phase, and perhaps also evidence for diversity in CSM properties at a given orbital period. Clearly, radio observations of fast novae offer a unique and valuable window into the circumstellar environments of binary systems with mildly evolved donors.

Finally, we make a note of V838~Her, which shares rapid radio evolution and high brightness temperatures with V5589~Sgr and V392~Per (Figure \ref{fig:bt_fast}), but has a dwarf donor with an orbital period of 7.1 hr \citep{Ingram+92, Szkody&Ingram94}. While we cannot rule out the possibility that V838~Her has unusually dense CSM compared to its cataclysmic variable brethren, this seems unlikely and instead may point toward internal shocks shaping its radio light curve. As typical in nova studies, V838~Her serves as a reminder for any hypothesis we present to not get too tidy! We note that the expansion velocities in V838~Her are unusually high (Table \ref{tab:props}), which might be expected to produce more energetic shocks, regardless of whether they are  internal (within the ejecta) or external (with CSM).
The relative roles of internal and external shocks may be illuminated by folding in observations at other wavelengths like X-ray, which can constrain the absorbing column \citep[e.g.,][]{Gordon+21}.

\subsection{Double-peaked radio novae: combination synchrotron+thermal radio transients} \label{sec:doublepeak}

\begin{figure*}[t]
\begin{center}
\includegraphics[height=3.5in,angle=0]{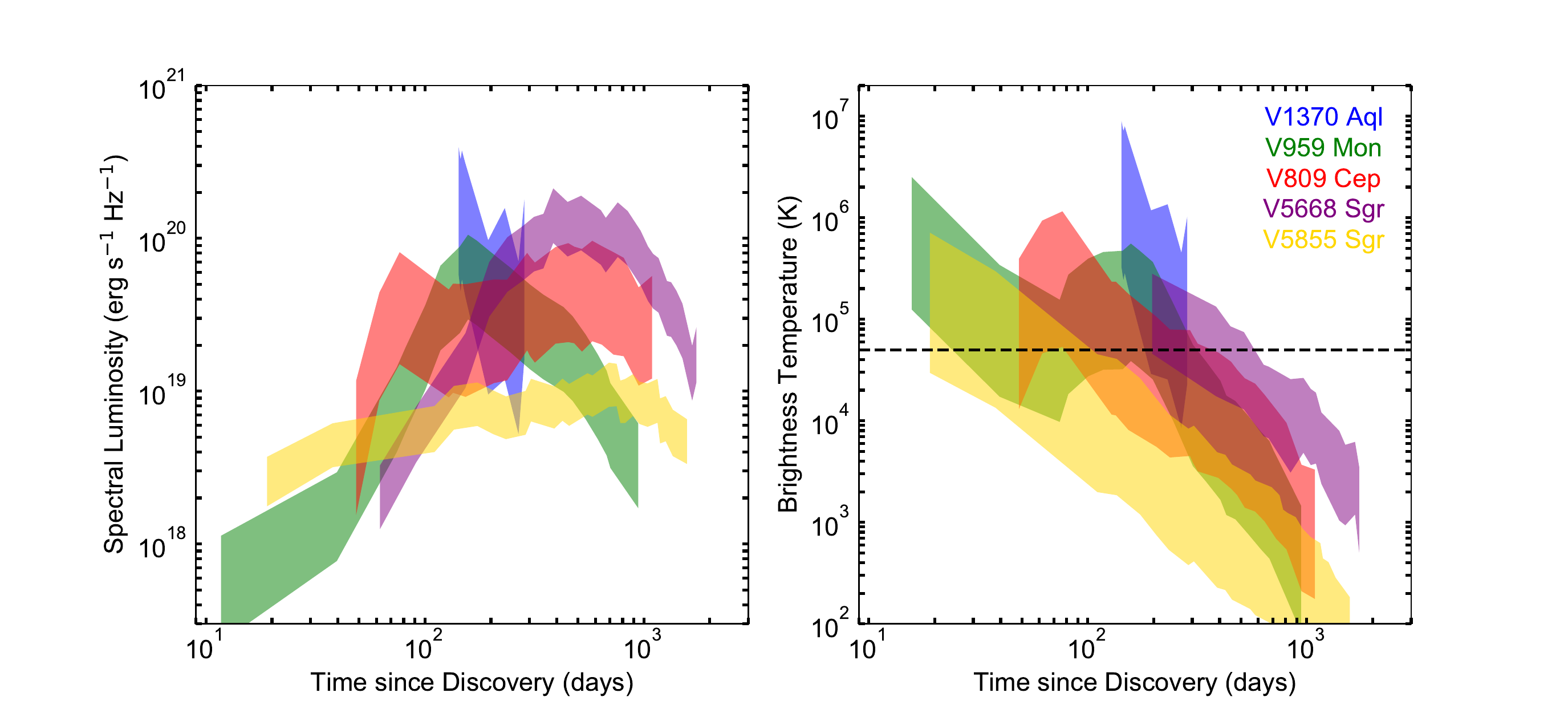}
\caption{As in Figure \ref{fig:bt_fast}, but here we plot five novae that show relatively slow radio light curves but also exhibit high brightness temperatures that are evidence for synchrotron emission. }
\label{fig:bt_slow1}
\end{center}
\end{figure*}

Three novae---V1723~Aql, V1324~Sco, and V357~Mus---show clearly double-peaked radio light curves, with the first peak occurring 50--80 days into eruption, and the second peak several hundred days after eruption (Figure \ref{fig:bt_double}). QU~Vul also shows clear indication of a similar early peak, although radio monitoring did not begin until day 206 and only the very end of the first peak was captured at 14.9/22.4 GHz; despite these limitations, \citet{Taylor+87} carried out an insightful analysis into the possible origins of the early radio maximum in QU~Vul. 
V705~Cas is a more ambiguous case borne of poorer observation quality, but it appears to show a 4.9 GHz maximum on day 108 before plateauing at a near constant flux value for the next 1000 days (Figure \ref{fig:lcnew2}. V809~Cep is another case which could be considered as a double-peaked radio light curve (Babul et al.\ 2021, in prep), but its early peak is only seen at lower frequencies (4.6/7.4 GHz; Figures \ref{fig:lcpub3} and \ref{fig:bt_slow1}), and the first peak is fainter than the later maximum.

V357~Mus is a never-before-published double-peaked radio nova, and by many criteria, the most dramatic. It reached a flux of 46 mJy at 9 GHz on day 64, which is an order of magnitude brighter than the later thermal maximum. This early peak reaches brightness temperatures $>3 \times 10^5$ K, which is an indication that it is dominated by non-thermal emission.  This first peak  strikingly coincides with the timing of the dust dip in the optical light curve (Figure \ref{fig:lcnew6}), which may be an indication of a common origin for synchrotron-emitting relativistic electrons and dust \citep{Derdzinski+17}.

The other novae in this class also show temporal coincidence between dust formation and the early non-thermal radio maximum.
V1324~Sco shares many traits with V357~Mus, reaching similar brightness temperatures  (Figure \ref{fig:bt_double}) and showing simultaneity between the first radio maximum and an optical dust dip (Figure \ref{fig:lcpub3}). V1723~Aql was unfortunately very poorly observed in the optical, but the data that do exist imply that the fading of the optical light curve is likely due to dust production \citep{Nagashima+13}, and therefore it too may show a temporal coincidence between an optical dust dip and an early radio maximum. The brightness temperature of V1723~Aql's early radio maximum is $> 9 \times 10^5$ K, and therefore one of the more clear-cut cases for non-thermal emission (see \citealt{Weston+16a} for more discussion). V705~Cas shows temporal coincidence between the 4.9 GHz maximum and the optical dust dip, but the case for a double-peaked radio light curve is more ambiguous. We note that not all novae with dust dips in their optical light curves show double radio peaks: V5668 Sgr (Figure \ref{fig:lcnew5}) can be described with a single radio maximum.  
Still, dramatic optical dust dips appear to be correlated with double-peaked radio light curves and in particular, an early synchrotron maximum. 
 We note that asymmetry and inclination effects could be important in shaping both the optical and radio synchrotron light curves, if dust preferentially forms in e.g., the orbital plane, or if synchrotron-emitting shocks peek out from the optically-thick thermal ejecta preferentially in some directions.
%One potential explanation for this correlation is non-spherical ejecta combined with inclination effects. If dust is concentrated in the orbital plane of the binary 

As discussed in \citet{Weston+16a} and \citet{Finzell+18}, the second radio maxima generally have brightness temperatures $T_B \lesssim 10^4$ K and are consistent with thermal emission from the warm expanding ejecta. In future work, we plan to fit thermal light curves to all novae presented here (where possible); in cases like V357~Mus and QU~Vul, this will be particularly valuable for decomposing the thermal and non-thermal contributions so that we may determine the energetics of the synchrotron emission and estimate the ejecta mass.

\subsection{Slower radio novae: mostly thermal but deceptively complex} \label{sec:slowpeak}

\begin{figure*}[t]
\begin{center}
\includegraphics[height=3.5in,angle=0]{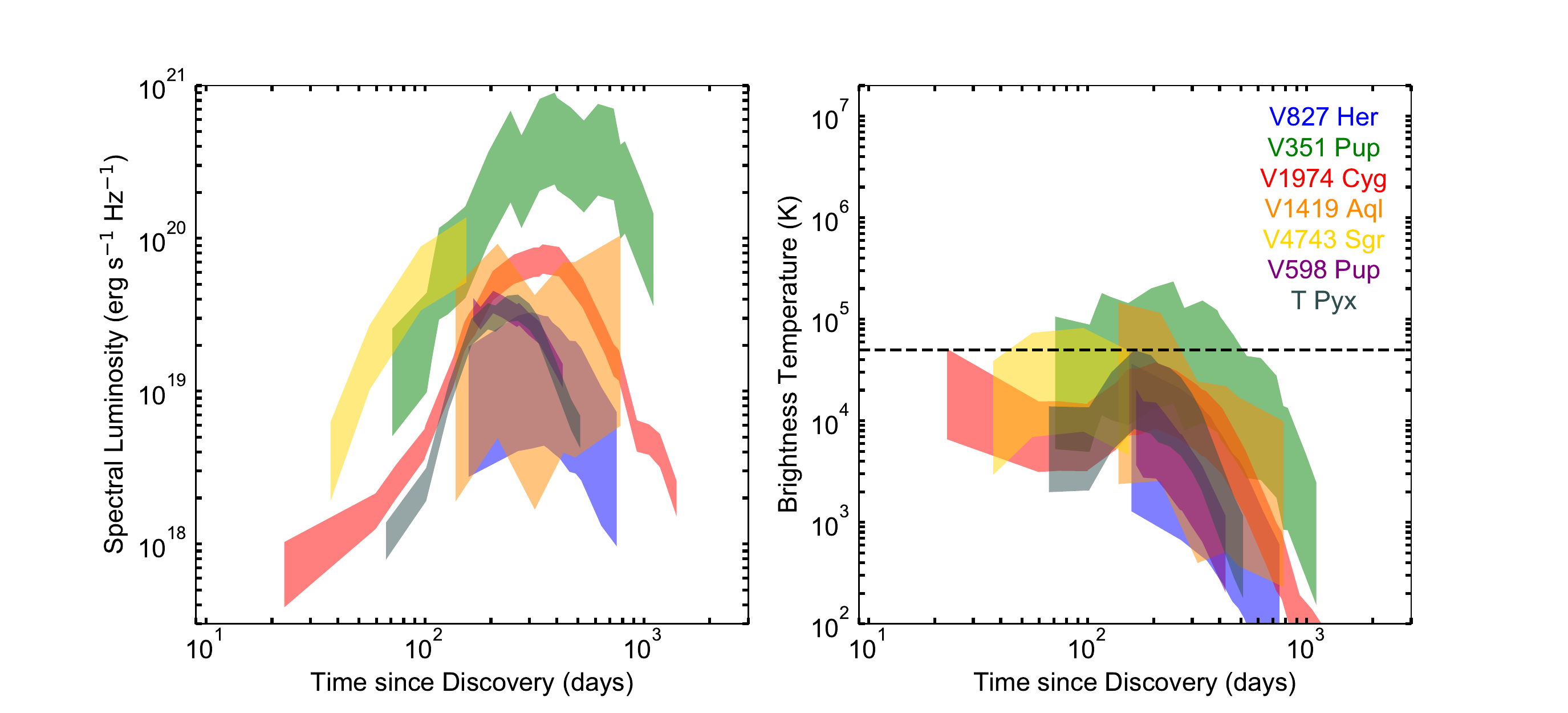}
\caption{As in Figure \ref{fig:bt_fast}, but here we plot seven novae that show slower evolution at radio wavelengths. }
\label{fig:bt_slow2}
\end{center}
\end{figure*}

\begin{figure*}[t]
\begin{center}
\includegraphics[height=3.5in,angle=0]{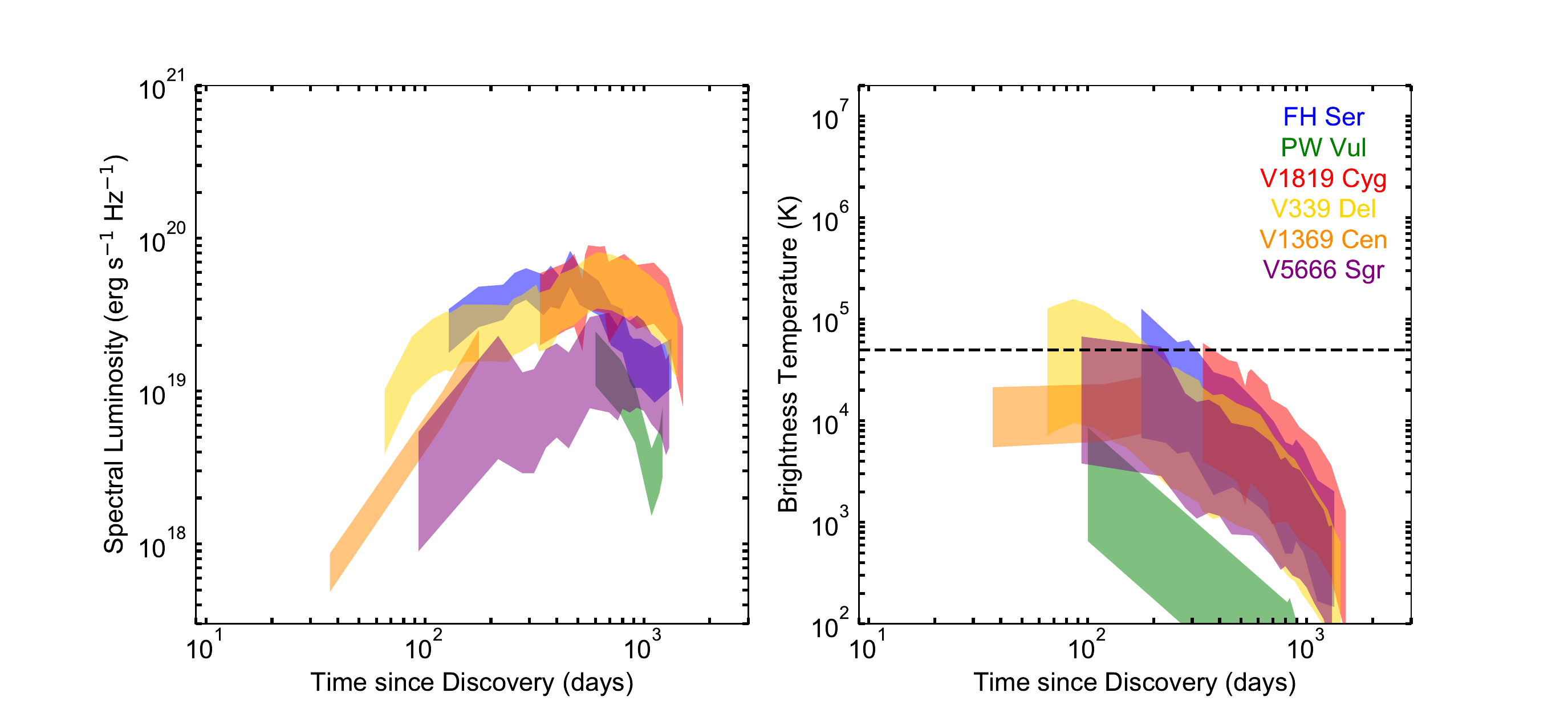}
\caption{ As in Figure \ref{fig:bt_fast}, but here we plot six more novae that show slower evolution at radio wavelengths.}
\label{fig:bt_slow3}
\end{center}
\end{figure*}

\begin{figure*}[t]
\begin{center}
\includegraphics[height=3.5in,angle=0]{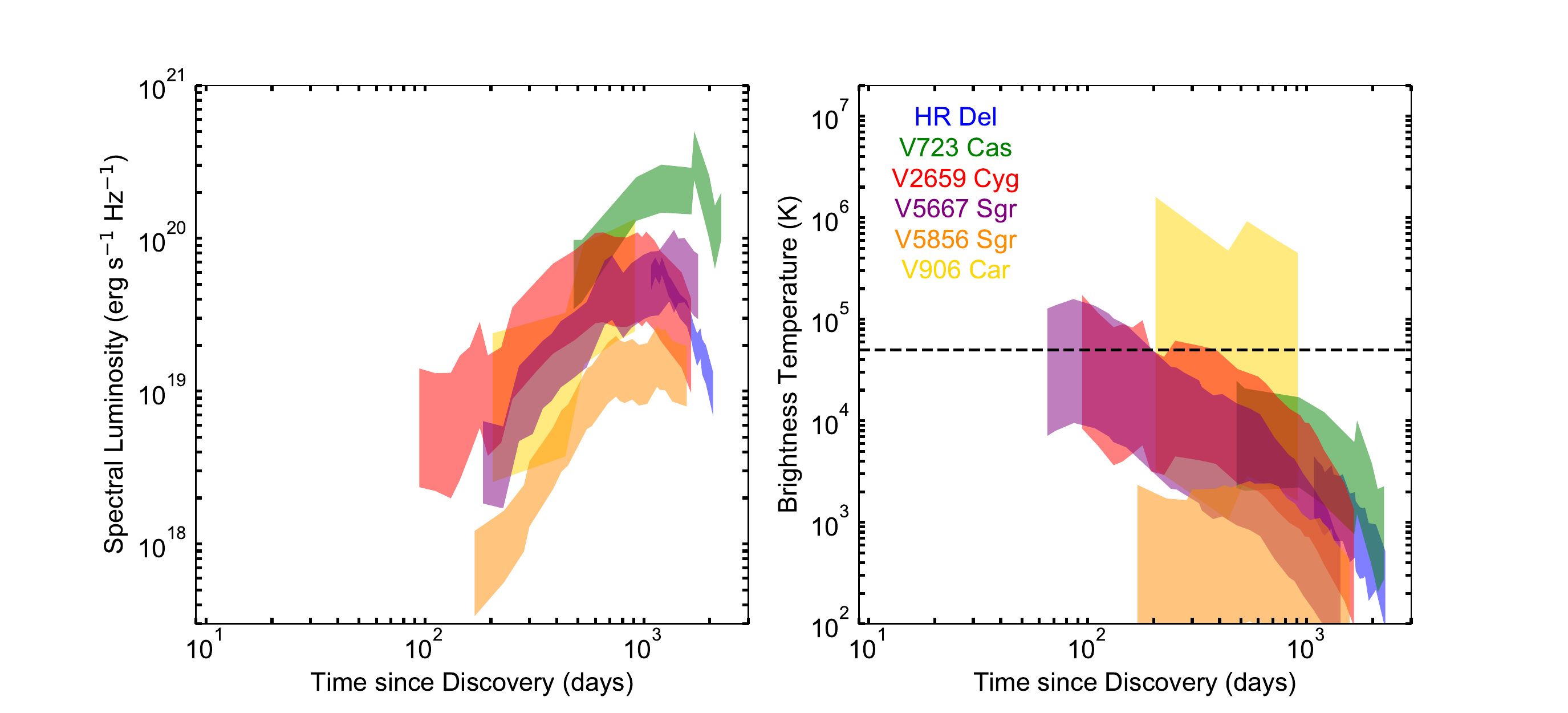}
\caption{As in Figure \ref{fig:bt_fast}, but here we plot six novae that show the slowest evolution at radio wavelengths. }
\label{fig:bt_slow4}
\end{center}
\end{figure*}

The largest group of radio light curves evolve more slowly ($t_{\rm max} \gtrsim 150$ days). A handful of these slower novae also show evidence for early synchrotron emission and are collected in Figure \ref{fig:bt_slow1}, while most are consistent with pure thermal emission and are plotted in Figure \ref{fig:bt_slow2}--\ref{fig:bt_slow4}.

The novae plotted in Figure \ref{fig:bt_slow1} further highlight the diversity of nova light curves, and in particular, how signatures of synchrotron emission can be diverse.
Of the novae with evidence for non-thermal emission plotted in  Figure \ref{fig:bt_slow1}, we have already discussed V809~Cep as a ``bridge" object showing some evidence for two distinct radio maxima (\S \ref{sec:doublepeak}). V1370~Aql has poor early-time and high-frequency coverage, but it becomes surprisingly bright at 1.5 GHz, which translates to $T_B > 10^5$ K. V959~Mon shows an early-time excess that implies high brightness temperature; this early-time emission also shows a spectral index indicative of synchrotron emission, and even higher brightness temperatures in high-resolution VLBI imaging presented by \citet{Chomiuk+14}. V5668~Sgr's radio light curve evolves relatively rapidly and shows brightness temperatures $>5 \times 10^4$ K; a likely possibility is that its primary radio maximum is a mixture of synchrotron and thermal emission (and the relative proportions change with time). It is notable that V5668~Sgr is the one remaining ``dust-dipper" nova in our sample that does not have a double-peaked radio light curve (\S \ref{sec:doublepeak}), and it too shows significant evidence for non-thermal emission. V5855~Sgr has a  peculiar radio light curve which rises quickly over $\sim$40 days, and then plateaus at a remarkably constant level (Figure \ref{fig:lcnew5}); the rapid rise to relatively bright levels translates to $T_B > 3 \times 10^4$ K, which is a marginal indication of non-thermal contributions. 

The remaining light curves, plotted in Figure \ref{fig:bt_slow2}--\ref{fig:bt_slow4} (grouped in order of increasing $t_{\rm max}$), are consistent with purely thermal emission from the ionized expanding ejecta, as their brightness temperatures are $\lesssim10^4$ K.  In some cases, the temporal and frequency coverage are of sufficient quality to exclude non-thermal signatures like those seen in Figures \ref{fig:bt_double}--\ref{fig:bt_slow1}, while in other cases observations are not sufficient to constrain the presence of e.g., an early time excess. We note that we cannot exclude non-thermal emission in these cases, and encourage further studies of the radio spectrum and/or high-resolution imaging to test for its presence---but given our limited analysis here, they seem to be explained with pure thermal emission.

However, even in these cases with a single radio peak and $T_B \lesssim10^4$ K,  many novae show subtle complexity in their radio light curves. Take for example, V339 Del (Figure \ref{fig:lcpub3}). The 4.6 GHz light curve rises relatively quickly ($\sim$100 days), and then flattens out for nearly two years; during this ``plateau" time, the slope/concavity of the light curve changes at least once. The overall impression from the light curve is that multiple emission components are contributing.  Even if all emission is thermal, there is still the likely possibility of multiple components with different densities and expansion velocities, perhaps from an equatorial torus and a bipolar wind \citep{Chomiuk+14, Aydi+20b}. Another example of potential thermal complexity is revealed by a comparison of V2659~Cyg and V5667~Sgr side by side (Figure \ref{fig:lcnew5}). Although these light curves have similar $t_{\rm max}$ and $S_{\rm max}$, the shape of their radio peaks is quite different, with V5667~Sgr remaining bright at 35 GHz out to $\sim$800 days, and V2659~Cyg  fading much faster. A potential explanation for these different light curve morphologies is nonspherical thermal ejecta \citep{Ribeiro+14}.

We again note that the 5/8.5 GHz light curves of most novae appear to peak with radio spectral luminosities of $10^{19}-10^{20}$ erg s$^{-1}$ Hz$^{-1}$.  Roughly, assuming $10^4$ K ejecta emitting free-free emission, $L_{\rm max} \propto M_{\rm ej}^{0.8}$ with weak dependence on the density distribution of the ejecta \citep{Seaquist&Bode08, Nelson+14}. The implication is that most of these slow novae have ejecta masses in the range, $\sim 10^{-5}$ M$_{\odot}$ to few $\times 10^{-4}$ M$_{\odot}$. 
One exception is V351~Pup, which shows an unusually high luminosity in Figure \ref{fig:bt_slow2}; a potential explanation is that the distance derived from expansion parallax by \citet{Wendeln+17} and used here may be erroneously large. Similarly, the radio luminosity of V5856~Sgr appears unusually low in Figure \ref{fig:bt_slow4}, which could be an indication of an erroneously small distance estimate (the one used here is estimated from three-dimensional reddening maps). As \emph{Gaia} geometric distances become available for more and more novae, such distance ambiguities will be largely clarified, enabling study of absolute nova energetics.

%======================

\section{Conclusions} \label{sec:conclusion}

We present multi-frequency radio light curves for 36 novae with dwarf or sub-giant companions, intentionally excluding novae with giant companions (but encouraging future study of their radio properties). The novae studied here span 50 years, a range of speed classes ($t_2 = 1-263$ days), optical brightnesses ($V_{\rm peak} \approx 16-2$ mag), and optical light curve morphologies (see Table \ref{tab:props} and Figures \ref{fig:lcpub1}--\ref{fig:lcnew6}).

The large sample and high-quality modern data presented here demonstrate that novae are far from simple thermal transients; synchrotron emission and multiple emitting components prove common. Using \emph{Gaia} distances and spectroscopically-measured velocities, we constrain the angular size of the nova and thereby estimate brightness temperatures from our flux density measurements. Of the 36 novae presented here, nine (25\%) show evidence for non-thermal emission in the form of brightness temperatures $> 5 \times 10^4$ K (\S \ref{sec:fastpeak}-\ref{sec:slowpeak}). We also expect that, on a deeper look which includes spectral analysis and/or high-resolution imaging, even more novae will exhibit definitive signatures of synchrotron emission (e.g., V959 Mon; \citealt{Chomiuk+14}). 

We compare radio and optical properties, and find that the peak optical brightness and the peak radio brightness are positively correlated. We also find that novae with faster $t_2$ values and expansion speeds show shorter time to radio maximum. The time elapsed between nova eruption and radio maximum varies greatly across our targets, from $\lesssim$16 days to 1697 days. 

In a few cases, the radio emission is consistent with pure synchrotron emission (V838~Her, V5589~Sgr, and V392~Per; Figure \ref{fig:bt_fast}); these are very rapidly evolving novae with high expansion velocities. V5589~Sgr and V392~Per also have mildly evolved companions, which may hint at a role for external shocks (between the nova ejecta and pre-existing CSM) in shaping shock signatures in these systems.

Several novae show two distinct peaks in their radio light curves, with the first occurring $\lesssim$100 days after eruption and showing $T_B> 5 \times 10^4$ K, and the second maximum occurring $\sim$few years after eruption and displaying $T_B \lesssim 10^4$ K (Figure \ref{fig:bt_double}). Building on \citet{Metzger+14}, \citet{Weston+16a}, and  \citet{Vlasov+16}, we identify the first maximum with synchrotron emission and the second with thermal emission from the bulk of the ejecta. Novae with strong dust formation episodes are concentrated amongst this class of double-peaked radio novae, and there is also striking temporal coincidence between the dust dip in the optical light curve and the early synchrotron maximum. These novae therefore provide evidence for a common site of dust production and relativistic particle acceleration in novae: shocks. 

Our basic analysis shows that radio emission from about half of our novae can be explained with pure thermal emission---that is, non-thermal components can be accommodated within the data, but are not required by a brightness temperature analysis (Figures \ref{fig:bt_slow2}--\ref{fig:bt_slow4}). But even in these pure-thermal cases, high-quality light curves display clear complexity in their shapes, sometimes showing several changes in slope/concavity (e.g., V339 Del; Figure \ref{fig:lcpub3}) and showing a range of durations around radio maximum (e.g., compare the light curves plotted in Figure \ref{fig:lcnew5}). It is possible that multiple, distinct thermal components are contributing to the radio light curves. 
Take for example, the popular scenario of a slowly-expanding equatorial disk or torus and a faster, more symmetric wind. Depending on their relative masses, velocities, and durations, the radio emission from the two thermal components can peak at distinct times/luminosities, and might explain the complex light curve shapes observed in e.g., V339 Del.

We hypothesize that the radio emission from all novae has both thermal free-free and non-thermal synchrotron components. However, the relative luminosity, timing, and duration of the emission can vary dramatically between novae, resulting in startling diversity of radio light curves. 
In some novae, the thermal and synchrotron components are well-spaced in time and comparable in flux (e.g., V1723 Aql, V1324~Sco; Figure \ref{fig:bt_double}), while in others the thermal and synchrotron contributions appear to blur together into a single radio maximum (e.g., V5668 Sgr, V5855 Sgr; Figure \ref{fig:bt_slow1}). Some novae appear dominated by synchrotron emission with little evidence for a thermal component (e.g., V838 Her, V392 Per; Figure \ref{fig:bt_fast}), while others are consistent with pure thermal emission (Figures \ref{fig:bt_slow2}--\ref{fig:bt_slow4}). In the future, we plan to study this diversity in the context of multi-wavelength data in order to understand what determines the shock energetics and non-thermal luminosity of nova explosions. In particular, it remains unclear what sets the GeV $\gamma$-ray luminosity of novae, which varies across at least two orders of magnitude \citep{Franckowiak+18, Chomiuk+21}, and the radio properties of novae in the \emph{Fermi} era may shed light on this diversity. 
%We note that radio synchrotron emission tracks relativistic electrons, while the \emph{Fermi $\gamma$-ray emission likely tracks relativistic ions \citep{Li+17, Aydi+21}, 

The thermal properties of the radio light curves presented here should also be systematically analyzed in light of constraints on velocity, filling factor, and geometry derived from multi-wavelength observations. We are embarking on such a study, which will produce the highest quality estimates of ejecta mass for a large and diverse sample of novae, which in turn will serve as an important test of nova theory (e.g., \citealt{Yaron+05}).

%======================

\acknowledgements
This paper is dedicated to the memory of Bob Hjellming, in gratitude for his ground-breaking work on novae and other radio stars.
We are also grateful for the help and support of NRAO staff, who were critical in obtaining these observations. We are grateful to Lars Bildsten, Bill Wolf, Ken Shen for insightful conversations. We thank Rami Alsaberi, Aliya-Nur Babul, Perica Manojlovi\'c, and Yong Zheng for their help in data acquisition, reduction, and analysis.

LC, EA, KVS, TF, and AK are grateful for support from NSF grants AST-1751874 and AST-1907790, NASA \emph{Fermi} grant 80NSSC20K1535, and a Cottrell fellowship of the Research Corporation. JDL acknowledges support from NASA \emph{Fermi} grant 80NSSC17K0511 and the NRAO ngVLA Community Studies Program. JLS and JHSW were supported by NSF grant AST-1816100 and Heising-Simons Foundation grant \#2017-246.
JS acknowledges support from the Packard Foundation. ECK acknowledges support from the G.R.E.A.T.\ research environment funded by {\em Vetenskapsr\aa det}, the Swedish Research Council, under project number 2016-06012, and support from The Wenner-Gren Foundations. 
V.A.R.M.R. acknowledges financial support from the  Funda\c{c}\~{a}o para a Ci\^encia e a Tecnologia (FCT) in the form of an exploratory project of reference IF/00498/2015/CP1302/CT0001, and from the Minist\'erio da Ci\^encia, Tecnologia e Ensino Superior (MCTES)  through national funds and when applicable co-funded EU funds under the project UIDB/EEA/50008/2020, and supported by Enabling Green E-science for the Square Kilometre Array Research Infrastructure (ENGAGE-SKA), POCI-01-0145-FEDER-022217, and PHOBOS, POCI-01-0145-FEDER-029932, funded by Programa Operacional Competitividade e Internacionaliza\c{c}\~ao (COMPETE 2020) and FCT, Portugal.

The National Radio Astronomy Observatory is a facility of the National Science Foundation operated under cooperative agreement by Associated Universities, Inc. The Australia Telescope Compact Array is part of the Australia Telescope National Facility which is funded by the Australian Government for operation as a National Facility managed by CSIRO. We acknowledge the Gomeroi people as the traditional owners of the Observatory site. This research has made use of the International Variable Star Index (VSX) database, operated at AAVSO, Cambridge, Massachusetts, USA.  We acknowledge with thanks the variable star observations from the AAVSO International Database contributed by observers worldwide and used in this research. This work has made use of data from the European Space Agency (ESA) mission {\it Gaia} (\url{https://www.cosmos.esa.int/gaia}), processed by the {\it Gaia} Data Processing and Analysis Consortium (DPAC,\url{https://www.cosmos.esa.int/web/gaia/dpac/consortium}). Funding for the DPAC has been provided by national institutions, in particular the institutions participating in the {\it Gaia} Multilateral Agreement.

{\it Facilities:}  \facility{VLA}, \facility{ATCA}, \facility{AAVSO}, \facility{Gaia}

\bibliography{radionovae.bib}

\clearpage
\begin{deluxetable}{lcccccccc}
\tablewidth{0 pt}
\tabletypesize{\footnotesize}
\setlength{\tabcolsep}{0.025in}
\tablecaption{ \label{tab:v1370aql}
Radio Observations of V1370 Aql}
\tablehead{UT Obs Date & MJD & $t-t_0$\tablenotemark{a} &  0.6 GHz S$_{\nu}$ & 1.5 GHz S$_{\nu}$ & 4.9 GHz S$_{\nu}$ & 14.9 GHz S$_{\nu}$ & Telescope \\ 
  & & (Days) & (mJy) & (mJy) & (mJy) & (mJy) & }
  1982 Mar 28 & 45055 & 59 & $8.4\pm2.0$ & & & & WSRT\tablenotemark{b}\\ 			
1982 Apr 10 & 45069.5 & 73 & & & $20.2\pm2.0$ & & WSRT\tablenotemark{b}\\
1982 Jun 18 & 45138.5 & 142 & &	$14.2\pm2.0$ & & & WSRT\tablenotemark{b}\\
1982 Jun 20 & 45140.5 & 144 & &	$12.2\pm2.0$ & & & WSRT\tablenotemark{b}\\
1982 Jun 21 & 45141.5 & 145 & &	$11.5\pm2.0$ & & & WSRT\tablenotemark{b}\\
1982 Jun 23 & 45143.5 & 147 & &	$13.4\pm2.0$ & & & WSRT\tablenotemark{b}\\
1982 Jun 28 & 45148.5 & 152 & &	$10.9\pm2.0$ & & & WSRT\tablenotemark{b}\\
1982 Aug 9 & 45190.5 & 194 & &	$3.0\pm2.0$ & & & WSRT\tablenotemark{b}\\
\enddata
\tablenotetext{a}{We take the time of discovery, 1982 Jan 27, as $t_0$.} % 	44996
\tablenotetext{b}{Data published by \citet{Snijders+87} at 1.41 GHz.}
\tablenotetext{c}{Data also published by \citet{Bode+87}.}
\end{deluxetable}

\begin{deluxetable}{lcccccccc}
\tablewidth{0 pt}
\tabletypesize{\footnotesize}
\setlength{\tabcolsep}{0.025in}
\tablecaption{ \label{tab:pwvul}
VLA Observations of PW~Vul}
\tablehead{UT Obs Date & MJD & $t-t_0$\tablenotemark{a} &  1.5 GHz S$_{\nu}$ & 4.9 GHz S$_{\nu}$ & 14.9 GHz S$_{\nu}$ & 22.4 GHz S$_{\nu}$ & Config \\ 
  & & (Days) & (mJy) & (mJy) & (mJy) & (mJy) & }
  1984 Aug 17.3 &  45929.3 &   20.3 & $<  1.14\pm 0.31$ & $<  0.33\pm 0.11$ & $<  0.82\pm 0.25$ & $<  5.06\pm 1.65$ &   D \\
1984 Aug 22.3 &  45934.3 &   25.3 &$< 19.42\pm 5.56$ & $<  0.40\pm 0.12$ & $<  0.79\pm 0.26$ & $<  4.91\pm 1.64$ &   D \\
1984 Aug 30.1 &  45942.1 &   33.1 &  	& $<  0.44\pm 0.15$ &  $<  1.32\pm 0.35$ &   	& D \\
1984 Nov  5.0 &  46009.0 &  100.0 &  	& $<  0.46\pm 0.14$ &  $  1.12\pm 0.24$ &  	& AD \\
1985 Jan 13.7 &  46078.7 &  169.7 &	&	& $<  4.31\pm 0.78$ &   & A \\
1986 Mar 20.5 &  46509.6 &  600.6 & 	 & $  2.85\pm 0.17$ & 	&	&  A \\
1986 Sep 21.2 &  46694.2 &  785.2 &  	& $  1.84\pm 0.12$ &  	&	& BC \\
1986 Oct 14.8 &  46717.8 &  808.8 & 	 & $  1.69\pm 0.12$ &  	&	& BC \\
1986 Oct 20.9 &  46723.9 &  814.9 & 	&	 & $  1.40\pm 0.31$ &  	& BC \\
1986 Nov 11.8 &  46745.8 &  836.8 &	&	 & $  1.71\pm 0.29$ &  	& C \\
1987 Jan 24.6 &  46819.7 &  910.7 & $<  1.29\pm 0.22$ & $  1.23\pm 0.09$ & $  1.06\pm 0.19$ &  	& CD \\
1987 Jul 17.0 &  46993.0 & 1084.0 & $  0.49\pm 0.13$ &  $  0.45\pm 0.07$ & $<  0.56\pm 0.19$ &   	& A \\
1987 Oct 18.0 &  47086.0 & 1177.0 &  	& $  0.61\pm 0.08$ &  $<  0.36\pm 0.12$ &  	& AB \\
1987 Nov 28.0 &  47127.0 & 1218.0 &  	& $  0.82\pm 0.14$ & $<  0.97\pm 0.27$ &   	& B \\
\enddata
\tablenotetext{a}{We take the time of discovery, 1984 Jul 28, as $t_0$.}
\end{deluxetable}

\begin{deluxetable}{ccccccccc}
\tablewidth{0 pt}
\tabletypesize{\footnotesize}
\setlength{\tabcolsep}{0.025in}
\tablecaption{ \label{tab:quvul}
VLA Observations of QU~Vul}
\tablehead{UT Obs Date & MJD & $t-t_0$\tablenotemark{a} &  1.5 GHz S$_{\nu}$ & 4.9 GHz S$_{\nu}$ & 8.4 GHz S$_{\nu}$ & 14.9 GHz S$_{\nu}$ & 22.4 GHz S$_{\nu}$ & Config \\ 
  & & (Days) & (mJy) & (mJy) & (mJy) & (mJy) & (mJy) & }
 1985 Jul 16.3 &  46262.3 &  206.3 & 			 	& $  4.10\pm 0.29$ & &$ 63.40\pm 6.59$ &   	& C\tablenotemark{b} \\
1985 Aug 21.0 &  46298.1 &  242.1 & $  1.43\pm 0.09$ & $  9.20\pm 0.50$ & &$ 36.40\pm 3.75$ & $ 58.60\pm 5.87$ &   C\tablenotemark{b} \\
1985 Sep 18.0 &  46326.1 &  270.1 & $  2.38\pm 0.14$ & $ 10.10\pm 0.59$ & &$ 32.80\pm 3.43$ & $ 49.50\pm 5.74$ &   C\tablenotemark{b} \\
1985 Oct  8.0 &  46346.0 &  290.0 &  	& $  9.00\pm 0.54$ &  &$ 35.80\pm 3.75$ &		&  AD\tablenotemark{b} \\
1985 Nov  2.9 &  46372.0 &  316.0 & $  2.24\pm 0.13$ & $ 10.20\pm 0.59$ & &$ 34.30\pm 3.55$ &  $ 52.10\pm 6.33$ &  CD\tablenotemark{b} \\
1985 Nov 27.9 &  46397.0 &  341.0 &$  2.21\pm 0.30$ & $ 10.68\pm 0.54$ & &$ 39.05\pm 3.91$ & $ 63.32\pm 6.39$ &   D \\
1985 Dec 18.8 &  46417.8 &  361.8 & $  3.00\pm 0.17$ & $ 11.10\pm 0.63$ & & $ 41.70\pm 4.34$ & $ 70.60\pm 8.02$ &   D\tablenotemark{b} \\
1986 Feb  4.7 &  46465.7 &  409.7 & $  2.60\pm 0.16$ & $ 12.00\pm 1.08$ & &$ 45.60\pm 5.15$ & $102.30\pm14.88$ &  AD\tablenotemark{b} \\
\enddata
\tablenotetext{a}{We take the time of discovery, 1984 Dec 22, as $t_0$.}
\tablenotetext{b}{Data published by \citet{Taylor+87}.}
\tablenotetext{b}{Data published by \citet{Taylor+88}.}
\end{deluxetable}

\begin{deluxetable}{ccccccccc}
\tablewidth{0 pt}
\tabletypesize{\footnotesize}
\setlength{\tabcolsep}{0.025in}
\tablecaption{ \label{tab:v1819cyg}
VLA Observations of V1819~Cyg}
\tablehead{UT Obs Date & MJD & $t-t_0$\tablenotemark{a} &  1.5 GHz S$_{\nu}$ & 4.9 GHz S$_{\nu}$ & 8.4 GHz S$_{\nu}$ & 14.9 GHz S$_{\nu}$ & Config \\ 
  & & (Days) & (mJy) & (mJy) & (mJy) & (mJy) & }
\startdata
1987 Jan 17.8 &  46812.8 &  166.8 & 			      & 			       & 			  & $<  2.81\pm 0.94$ &   C \\
1987 Jan 24.6 &  46819.7 &  173.7 & 			      & $<  0.27\pm 0.07$ &				   & $  0.81\pm 0.21$ &  CD \\
1987 Jul  5.3 &  46981.3 &  335.3 &  				      & $  0.55\pm 0.09$ &   			   & 				&  A \\
1987 Jul 18.4 &  46994.4 &  348.4 & 				      &				     &				   & $  1.24\pm 0.20$ &   A \\
1987 Oct 18.0 &  47086.0 &  440.0 & 			      & $  0.66\pm 0.08$ &  				   & $  3.19\pm 0.50$ &  AB \\
1987 Nov 28.0 &  47127.1 &  481.1 & 			       & $  0.73\pm 0.12$ &   			   & 				&  B \\
1988 Jan  9.6 &  47169.6 &  523.6 & 			       & $  0.51\pm 0.08$ &				   &				 &  B\\
1988 Jan 24.6 &  47184.6 &  538.6 &  $<  0.96\pm 0.32$ & $  0.73\pm 0.11$ &  			   &  $  3.48\pm 0.44$ &   B \\
\enddata
\tablenotetext{a}{We take the time of discovery, 1986 Aug 4, as $t_0$.}
\end{deluxetable}

\begin{deluxetable}{ccccccccc}
\tablewidth{0 pt}
\tabletypesize{\footnotesize}
\setlength{\tabcolsep}{0.025in}
\tablecaption{ \label{tab:v827her}
VLA Observations of V827~Her}
\tablehead{UT Obs Date & MJD & $t-t_0$\tablenotemark{a} &  1.5 GHz S$_{\nu}$ & 4.9 GHz S$_{\nu}$ & 8.4 GHz S$_{\nu}$ & 14.9 GHz S$_{\nu}$ & Config \\ 
  & & (Days) & (mJy) & (mJy) & (mJy) & (mJy) & }
\startdata
1987 Jul  2.3 &  46978.3 &  158.3 &  				& $  1.25\pm 0.08$ &   			& $  3.41\pm 0.38$ &   A \\
1987 Jul  7.3 &  46983.3 &  163.3 & $<  0.58\pm 0.11$ & 					&		&				& A \\
1987 Oct 18.0 &  47086.0 &  266.0 &  				 & $  1.93\pm 0.20$ & 		&$  6.73\pm 1.03$ &  AB \\
1987 Nov 28.0 &  47127.1 &  307.1 & 				& $  2.02\pm 0.22$ &		& $  6.14\pm 0.96$ &   B \\
1988 Jan  9.6 &  47169.6 &  349.6 & 			       & $  2.00\pm 0.12$ &   		 & $  4.49\pm 0.50$ &   B \\
1988 Jan 24.6 &  47184.6 &  364.6 & $<  0.75\pm 0.18$ & $  1.95\pm 0.16$ &			& $  4.03\pm 0.49$ &   B \\
1988 Feb 11.5 &  47202.5 &  382.5 & $<  1.01\pm 0.19$ & $  1.80\pm 0.11$ & 			&$  2.85\pm 0.32$ &  BD \\
1988 Mar 12.5 &  47232.5 &  412.5 & $<  1.01\pm 0.21$  & $  1.65\pm 0.10$ & $  2.57\pm 0.15$ & $  2.48\pm 0.31$ &  CD \\
1988 Apr 26.4 &  47277.5 &  457.5 & $<  0.74\pm 0.25$ & $  1.36\pm 0.10$ & $  1.81\pm 0.13$ & $  1.77\pm 0.27$ &  CD \\
1988 May 26.3 &  47307.3 &  487.3 & $<  1.69\pm 0.31$ & $  1.34\pm 0.11$ & $  1.63\pm 0.11$ & $  1.55\pm 0.29$ &  CD \\
1988 Jun 24.4 &  47336.4 &  516.4 & $<  1.01\pm 0.29$ & $  1.21\pm 0.10$ & $  1.42\pm 0.09$ &  		& CD \\
1988 Oct 24.8 &  47458.9 &  638.9 & $ 0.37\pm 0.12$ & $  0.64\pm 0.08$ & $  0.64\pm 0.05$ & $<  0.65\pm 0.22$ &   A \\
1989 Feb 14.6 &  47571.6 &  751.6 & 			 & $  0.45\pm 0.04$ &  			&				& AB \\
\enddata
\tablenotetext{a}{We take the time of discovery, 1987 Jan 25, as $t_0$.}
\end{deluxetable}

\begin{deluxetable}{ccccccccc}
\tablewidth{0 pt}
\tabletypesize{\footnotesize}
\setlength{\tabcolsep}{0.025in}
\tablecaption{ \label{tab:v838her}
VLA Observations of V838~Her}
\tablehead{UT Obs Date & MJD & $t-t_0$\tablenotemark{a} &  1.5 GHz S$_{\nu}$ & 4.9 GHz S$_{\nu}$ & 8.4 GHz S$_{\nu}$ & 14.9 GHz S$_{\nu}$ & Config \\ 
  & & (Days) & (mJy) & (mJy) & (mJy) & (mJy) & }
\startdata
1991 Mar 28.6 &  48343.5 &    4.5 &		 & $<  0.18\pm 0.06$ & $<  0.16\pm 0.05$ & $<  0.82\pm 0.18$ &   D \\
1991 Apr  5.4 &  48351.5 &   12.5 &  		 & $  1.17\pm 0.10$ & $  1.48\pm 0.11$ & $  2.08\pm 0.29$  &   D \\
1991 Apr 10.5 &  48356.5 &   17.5 &  	& $  3.69\pm 0.23$ & $  4.54\pm 0.24$ & $  5.79\pm 0.65$ &   D \\
1991 Apr 18.3 &  48364.5 &   25.5 & $  3.34\pm 0.42$ & $  4.79\pm 0.25$ & $  5.62\pm 0.29$ & $  7.48\pm 0.78$ &   D \\
1991 Apr 27.6 &  48373.5 &   34.5 & $  1.18\pm 0.19$ & $  1.72\pm 0.18$ & $  1.75\pm 0.18$ & $  2.89\pm 0.46$ &   D \\
1991 May 31.2 &  48407.5 &   68.5 & $  0.72\pm 0.05$ & $  0.86\pm 0.08$ & $  1.43\pm 0.09$ & $  2.71\pm 0.39$&  AD \\
1991 Jun  4.2 &  48411.5 &   72.5 & & & & $  2.40\pm 0.27$ &  AD \\
1991 Jul  7.2 &  48444.5 &  105.5 & $<  0.37\pm 0.07$ & $  0.65\pm 0.06$ & $  0.54\pm 0.06$ & $  0.39\pm 0.10$ &  AD \\
1991 Aug 14.3 &  48482.5 &  143.5 &  & $  0.41\pm 0.06$ & $  0.25\pm 0.05$ & $<  0.55\pm 0.12$ &   A \\
1991 Sep  8.9 &  48507.5 &  168.5 &  & $  0.22\pm 0.04$ & $  0.17\pm 0.02$ & &   A \\
1991 Sep 26.0 &  48524.5 &  185.5 &  & & $  0.12\pm 0.01$ & & AB \\
1991 Oct 31.8 &  48560.5 &  221.5 & & & $<  0.07\pm 0.02$ & & AB \\
\enddata
\tablenotetext{a}{We take the time of discovery, 1991 Mar 24, as $t_0$.}
\end{deluxetable}

\begin{deluxetable}{ccccccccc}
\tablewidth{0 pt}
\tabletypesize{\footnotesize}
\setlength{\tabcolsep}{0.025in}
\tablecaption{ \label{tab:v1974cyg}
VLA Observations of V1974~Cyg}
\tablehead{UT Obs Date & MJD & $t-t_0$\tablenotemark{a} &  1.5 GHz S$_{\nu}$ & 4.9 GHz S$_{\nu}$ & 8.4 GHz S$_{\nu}$ & 14.9 GHz S$_{\nu}$ & 22.4 GHz S$_{\nu}$ & Config \\ 
  & & (Days) & (mJy) & (mJy) & (mJy) & (mJy) & (mJy) & }
\startdata
 1992 Feb 25.8 & 48677.8 &    6.8 &   & $<  0.30\pm 0.07$ & $<  0.18\pm 0.06$ & & & C \\
 
 1992 Mar 6.4 & 48687.4 &   16.4 &   & $<  0.44\pm 0.15$ & $<  0.39\pm 0.08$ & $<  2.10\pm 0.37$  & $<  2.95\pm 0.63$ & C \\
 
 1992 Mar 12.8 & 48693.8 &   22.8 &  & $  0.22\pm 0.07$  & $  0.49\pm 0.07$ & $<  1.06\pm 0.33$  & $<  2.67\pm 0.45$ & C \\
 
 1992 Mar 30.7 & 48711.7 &   40.7 &  & & $  0.38\pm 0.06$  & $  1.42\pm 0.33$  & $  2.02\pm 0.47$ & C \\
 
 1992 Apr 18.5 & 48730.5 &   59.5 &  & $  0.55\pm 0.05$  & $  1.43\pm 0.08$ & & $  8.04\pm 0.87$ & C \\
 
 1992 May 2.5 & 48744.5 &   73.5 &  & $  0.85\pm 0.07$ & $  2.30\pm 0.13$ & $  6.74\pm 0.72$ & $ 14.57\pm 1.49$ & C\\
 
 1992 May 28.5 & 48770.5 &   99.5 &  & $  1.51\pm 0.09$ & $  4.29\pm 0.22$ & $ 12.79\pm 1.30$  & $ 26.30\pm 2.66$ & CD \\

 1992 May 29.5 & 48771.5 &  100.5 &   & $  1.59\pm 0.10$ & $  4.75\pm 0.25$ & $ 14.45\pm 1.48$  & $ 30.19\pm 3.05$ & CD \\
\enddata
\tablenotetext{a}{We take the time of discovery, 1992 Feb 19, as $t_0$.}
\end{deluxetable}

\begin{deluxetable}{ccccccccc}
\tablewidth{0 pt}
\tabletypesize{\footnotesize}
\setlength{\tabcolsep}{0.025in}
\tablecaption{ \label{tab:v705cas}
VLA Observations of V705~Cas}
\tablehead{UT Obs Date & MJD & $t-t_0$\tablenotemark{a} &  1.4 GHz S$_{\nu}$ & 4.9 GHz S$_{\nu}$ & 8.4 GHz S$_{\nu}$ & 14.9 GHz S$_{\nu}$ & 22.4 GHz S$_{\nu}$ & Telescope \\ 
  & & (Days) & (mJy) & (mJy) & (mJy) & (mJy) & (mJy) & }
\startdata
1993 Dec 24.3 &  49345.5 &   17.5 & & $<  0.50\pm 0.13$ & $<  0.29\pm 0.10$ & $<  1.00\pm 0.33$ & & VLA/D\\
1994 Feb 16.2 &  49399.5 &   71.5 &  & $<  0.44\pm 0.15$ & $  0.46\pm 0.12$ &  $<  1.48\pm 0.49$ &  $<  3.73\pm 1.24$ & VLA/AD \\
1994 Mar 25.0 &  49436.5 &  108.5 & $  0.75\pm 0.15$ &  $  3.37\pm 0.35$ &  $  2.55\pm 0.27$ &  $<  1.12\pm 0.31$ &  $<  2.13\pm 0.71$ & VLA/A\\
1994 Jul 23.7 &  49556.5 &  228.5 & $  0.66\pm 0.09$ &  $  1.92\pm 0.12$ &  $  3.87\pm 0.21$ &  $  4.74\pm 0.61$ &  $< 12.40\pm 4.13$ & VLA/B\\
1994 Sep 11.6 &  49606.5 &  278.5 & $<  0.80\pm 0.27$ &  $  2.19\pm 0.14$ &  $  4.40\pm 0.24$ &$  4.70\pm 0.56$ &  $<  3.00\pm 1.00$ & VLA/B$^b$\\

1995 Jul 22.5 &  49920.5 &  592.5 &  & $  1.5\pm 0.3$ & & & & MERLIN$^b$\\
1996 Jan 24.0 &  50106.5 &  778.5 & $  0.6\pm 0.1$ &  & & & & MERLIN$^b$\\
1996 Dec  4.5 &  50421.5 & 1093.5 &  & $  2.0\pm 0.3$ & & & & MERLIN$^b$\\
1997 Jan 27.0 &  50475.5 & 1147.5 &  & $  1.9\pm 0.5$ & & & & MERLIN$^b$\\
1998 Mar  4.5 &  50876.5 & 1548.5 &  & $  0.7\pm 0.2$ & & & & MERLIN$^b$\\
1998 Apr 10.5 &  50913.5 & 1585.5 & $<  0.12$ &  & & & & MERLIN$^b$\\
1998 Dec  8.0 &  51155.5 & 1827.5 &  & $<  0.30$ & & & & MERLIN$^b$ \\

\enddata
\tablenotetext{a}{We take the time of discovery, 1993 Dec 7, as $t_0$.}
\tablenotetext{b}{Data published by \citet{Eyres+00}.}
\end{deluxetable}

\begin{deluxetable}{ccccccccc}
\tablewidth{0 pt}
\tabletypesize{\footnotesize}
\setlength{\tabcolsep}{0.025in}
\tablecaption{ \label{tab:v1419aql}
VLA Observations of V1419~Aql}
\tablehead{UT Obs Date & MJD & $t-t_0$\tablenotemark{a} & 4.9 GHz S$_{\nu}$ & 8.4 GHz S$_{\nu}$ & 14.9 GHz S$_{\nu}$ & 22.4 GHz S$_{\nu}$ & Config \\ 
  & & (Days) & (mJy) & (mJy) & (mJy) & (mJy) & }
\startdata
 1993 Jun 19.3 & 49157.3 &   36.3 &  $<  0.32\pm 0.10$ & $<  0.28\pm 0.08$  & $<  1.23\pm 0.31$ & &C \\
 1993 Sep 29.1 & 49259.1 &  138.1 &  $  0.39\pm 0.11$ & $  0.80\pm 0.09$ & $  1.71\pm 0.35$  & $  2.91\pm 0.65$ & CD \\
 1993 Dec 14.8 & 49335.8 &  214.8 &  $  0.83\pm 0.12$  & $  0.86\pm 0.10$ & $  2.13\pm 0.43$ & $  5.40\pm 0.54$ & D \\
 1993 Feb 14.7 & 49397.7 &  276.7 & & $<  1.08\pm 0.29$ & & & D  \\
 1994 Mar 27.5 & 49438.5 &  317.5 &  $  0.34\pm 0.10$ & $  0.41\pm 0.09$ & & & A \\
 1994 Jul 17.3 & 49550.3 &  429.3  & $  0.65\pm 0.07$ & $  0.95\pm 0.07$ & $  2.46\pm 0.35$ & $  4.48\pm 1.05$ & B\\
1994 Sep 11.0 &  49606.0 &  485.0  & $  0.63\pm 0.09$  & $  1.61\pm 0.11$ & $  2.90\pm 0.45$ & & B \\
1995 Jul 5.2 &  49903.2 &  782.2 &  $  0.97\pm 0.11$ & $  1.29\pm 0.10$  & $  1.33\pm 0.31$ & & A\\

\enddata
\tablenotetext{a}{We take the time of discovery, 1993 May 14, as $t_0$.}
\end{deluxetable}

\begin{deluxetable}{ccccccccc}
\tablewidth{0 pt}
\tabletypesize{\footnotesize}
\setlength{\tabcolsep}{0.025in}
\tablecaption{ \label{tab:usco}
VLA Observations of U~Sco}
\tablehead{UT Obs Date & MJD & $t-t_0$\tablenotemark{a} & 0.6 GHz S$_{\nu}$ & 1.5 GHz S$_{\nu}$ & 4.9 GHz S$_{\nu}$ & 8.4 GHz S$_{\nu}$  & Config \\ 
  & & (Days) & (mJy) & (mJy) & (mJy) & (mJy) & }
\startdata
1987 outburst: \\
1987 Jun  2.2 &  46948.2 & 18.3 & & 	 & $<  0.48\pm 0.16$ & 	&	 D \\
1987 Jul  5.2 &  46981.2 & 51.2 & & 	 & $<  0.31\pm 0.10$ &   	 &	 A \\
& \\
1999 outburst: & \\
1999 Mar  4.6 &  51241.6 &    8.6 &  &		& $<  0.19\pm 0.06$ &   &		 D \\
1999 Mar  5.4 &  51242.4 &    9.4 & &$<  0.35\pm 0.10$ & $<  0.10\pm 0.03$ & $<  0.08\pm 0.03$ &  D \\
1999 Mar  6.4 &  51243.4 &   10.4 & & $<  0.19\pm 0.05$ & $<  0.10\pm 0.02$ & $<  0.07\pm 0.02$ &   D \\
\enddata
\tablenotetext{a}{We take the times of discovery---1987 May 15, 1999 Feb 24, and 2010 Jan 28---as $t_0$ for the 1987, 1999, and 2010 outbursts respectively.}
\tablenotetext{b}{Central frequency of this observation is 1.28 GHz.}
\tablenotetext{c}{Data published by \citet{Anupama+13}.}
\end{deluxetable}

\begin{deluxetable}{ccccccccc}
\tablewidth{0 pt}
\tabletypesize{\footnotesize}
\setlength{\tabcolsep}{0.025in}
\tablecaption{ \label{tab:v4743sgr}
VLA Observations of V4743~Sgr}
\tablehead{UT Obs Date & MJD & $t-t_0$\tablenotemark{a} & 4.9 GHz S$_{\nu}$ & 8.5 GHz S$_{\nu}$ & 22.5 GHz S$_{\nu}$ & Config \\ 
  & & (Days) & (mJy) & (mJy) & (mJy) & }
\startdata
 2002 Oct 11.9 & 52558.9 &   21.9 &   $<  0.24\pm 0.05$   & $<  0.21\pm 0.04$  & $<  0.91\pm 0.18$ & C\\
 2002 Oct 27.0 & 52574.0 &   37.0 &  & $  0.21\pm 0.04$ & & C\\
 2002 Nov 14.9 & 52592.9 &   55.9 &   & $  1.01\pm 0.06$& & C \\
 2002 Dec 24.6 & 52632.6 &   95.6 &   $  1.94\pm 0.12$ & $  3.31\pm 0.18$ & $ 11.68\pm1.19$ & C\\
  2003 Feb 21.6 & 52691.6 &  154.6 &   $  3.05\pm 0.19$ & $  5.10\pm 0.32$  & $ 12.85\pm 1.30$ & D\\
  \enddata
\tablenotetext{a}{We take the time of discovery, 2002 Sep 20, as $t_0$.}
\end{deluxetable}

\begin{deluxetable}{ccccccccc}
\tablewidth{0 pt}
\tabletypesize{\footnotesize}
\setlength{\tabcolsep}{0.025in}
\tablecaption{ \label{tab:v598pup}
VLA Observations of V598~Pup}
\tablehead{UT Obs Date & MJD & $t-t_0$\tablenotemark{a} & 8.5 GHz S$_{\nu}$ & 22.5 GHz S$_{\nu}$ & Config \\ 
  & & (Days) & (mJy) & (mJy)  & }
\startdata
 2007 Nov 18.3 & 54422.3 &  166.3 &   $  9.96\pm 0.50$ & & B \\
 2007 Dec 1.4 & 54435.4 &  179.4 &  $  8.72\pm 0.46$  & $ 15.39\pm 1.56$ & B\\
 2007 Dec 27.0 & 54461.0 &  205.0 &   $ 11.13\pm 0.56$  & $ 18.27\pm 1.86$ & B\\
 2008 Jan 17.2 & 54482.2 &  226.2 &  $ 10.38\pm 0.52$  & $ 19.15\pm 1.93$ & B\\
2008 Feb 23.1&  54519.1 &  263.1 &  $  9.06\pm 0.48$ & & BC\\
 2008 Feb 29.1 & 54525.1 &  269.1 &  $  9.28\pm 0.47$ & $ 12.26\pm 1.25$ & BC \\
 2008 Mar 29.0 & 54554.0 &  298.0 &   $  8.01\pm 0.41$ & & C\\
 2008 May 2.9 & 54588.9 &  332.9 &   $  7.01\pm 0.36$  & $  7.65\pm 0.79$ & C \\
 2008 May 31.7& 54617.7 &  361.7 & $  5.62\pm 0.29$  & $  6.57\pm 0.70$ & C\\
  2008 Aug 3.7 & 54681.7 &  425.7 &  $  3.66\pm 0.22$ & $<  3.52\pm 1.17$ & D\\
\enddata
\tablenotetext{a}{We take the time of discovery, 2007 Jun 5, as $t_0$.}
\end{deluxetable}

\begin{deluxetable}{ccccccccc}
\tablewidth{0 pt}
\tabletypesize{\footnotesize}
\setlength{\tabcolsep}{0.025in}
\tablecaption{ \label{tab:v2491cyg}
VLA Observations of V2491~Cyg}
\tablehead{UT Obs Date & MJD & $t-t_0$\tablenotemark{a} & 4.9 GHz S$_{\nu}$ & 8.5 GHz S$_{\nu}$ & 43.3 GHz S$_{\nu}$ & Config \\ 
  & & (Days) & (mJy) & (mJy) & (mJy)  & }
\startdata
 2008 Apr 28.6 & 54584.6 &   18.6 & & $<  0.22\pm 0.05$ & $  0.92\pm 0.26$ & C\\
 2008 May 21.4 & 54607.4 &   41.4 &   & $  1.46\pm 0.09$ & & C \\
2008 May 29.3 &  54615.3 &   49.3 &   & $  1.19\pm 0.08$ & $  5.08\pm 0.60$ & C \\
 2008 Jun 6.6 & 54623.6 &   57.6 &  $  0.74\pm 0.11$ & & & CD \\
 2008 Jul 9.4 & 54656.4 &   90.4 &   & $  0.79\pm 0.06$ & $  1.52\pm 0.34$ & D\\
 2008 Aug 23.0 & 54701.0 &  135.0 &  & $  0.42\pm 0.05$  & $<  0.83\pm 0.22$ & D\\
 2008 Aug 24.0 & 54702.0 &  136.0 &  $  0.35\pm 0.06$ & & & D\\
 2008 Sep 19.2 & 54728.2 &  162.2 &  $<  0.64\pm 0.11$  & $  0.33\pm 0.06$ & $<  1.25\pm 0.42$ & D \\
\enddata
\tablenotetext{a}{We take the time of discovery, 2008 Apr 10, as $t_0$.}
\end{deluxetable}

\begin{deluxetable}{ccccccccc}
\tablewidth{0 pt}
\tabletypesize{\footnotesize}
\setlength{\tabcolsep}{0.025in}
\tablecaption{ \label{tab:v2672oph}
VLA Observations of V2672~Oph}
\tablehead{UT Obs Date & MJD & $t-t_0$\tablenotemark{a} & 8.5 GHz S$_{\nu}$ & 22.4 GHz S$_{\nu}$ & Config \\ 
  & & (Days) & (mJy) & (mJy)  & }
\startdata
 2009 Sep 1.1 & 55075.1 &   16.1 &  $  0.55\pm 0.04$ & & C\\
 2009 Sep 3.2 & 55077.2 &   18.2 & & $<  3.20\pm 0.81$ & C \\
 2009 Sep 13.2 & 55087.2 &   28.2 &   $  0.32\pm 0.06$  & $<  4.33\pm 1.36$ & C\\
 2009 Sep 29.1 & 55103.1 &   44.1 &  $  0.12\pm 0.04$ & & CD\\
2009 Nov 7.9 &  55142.9 &   83.9 &  & $<  0.38\pm 0.13$ & D\\
\enddata
\tablenotetext{a}{We take the time of discovery, 2009 Aug 16, as $t_0$.}
\end{deluxetable}

\begin{deluxetable}{ccccccccc}
\tablewidth{0 pt}
\tabletypesize{\footnotesize}
\setlength{\tabcolsep}{0.025in}
\tablecaption{ \label{tab:v1369cen}
ATCA Observations of V1369~Cen}
\tablehead{UT Obs Date & MJD & $t-t_0$\tablenotemark{a} & 5.5 GHz S$_{\nu}$ & 9 GHz S$_{\nu}$ & 17 GHz S$_{\nu}$ & 19 GHz S$_{\nu}$ & 34 GHz S$_{\nu}$ & Config \\ 
  & & (Days) & (mJy) & (mJy) & (mJy) &  (mJy) & (mJy) & }
\startdata
2013 Dec 05.8 & 56631.8 & 3.8 &   &   & & & $0.10\pm0.03$ & 750B \\
2013 Dec 16.8 & 56642.8 & 14.8 &  &  & & & $0.38\pm0.04$ & 750B\tablenotemark{b} \\
2014 Jan 7.8 & 56664.8 & 36.8 & $0.55\pm0.04$ & $0.73\pm0.05$ & & & $3.5\pm0.2$ & 750B \\
2014 Apr 1.4 & 56748.4 & 120.4 & $6.5\pm0.3$ & $16.7\pm0.8$ & & & $152\pm15$ & H168\tablenotemark{c}\\
%2014-04-10 - 5/9/17/19/33/35 GHz - 6A
%2014-04-18 -5/9/33/35 - 6A
2014 May 27.3 & 56804.3 & 176.3 &  $16.5\pm0.8$ & $29.9\pm1.5$ & $76.7\pm3.8$ & $95.0\pm4.8$ & & 1.5D

\enddata
\tablenotetext{a}{We take the time of discovery, 2013 Dec 2, as $t_0$.} %mjd 56628
\tablenotetext{b}{Published in \citet{Bannister+13}. $^c$Published in \citet{Bannister+14}.}
\end{deluxetable}

\begin{deluxetable}{cccccccccccc}
\tablewidth{0 pt}
\tabletypesize{\footnotesize}
\setlength{\tabcolsep}{0.025in}
\tablecaption{ \label{tab:v5666sgr}
VLA Observations of V5666~Sgr}
\tablehead{UT Obs Date & MJD & $t-t_0$\tablenotemark{a} & Config & $\nu$ & S$_{\nu}$ & $\nu$ & S$_{\nu}$ & $\nu$ & S$_{\nu}$ \\ 
  & & (Days) & & (GHz) & (mJy) & (GHz) & (mJy) & (GHz) & (mJy)}
\startdata

2014 Feb 19.5 & 56707.5 &   24.5 & AB &  4.55 & $<  0.03\pm 0.01$ &  7.38 & $<  0.04\pm 0.01$  & 28.20 & $<  0.09\pm 0.03$ \\
 & & & & 36.50 & $<  0.13\pm 0.04$ \\

2014 Mar 6.4 &  56722.4 &   39.4 & A & 4.55 & $<  0.06\pm 0.01$ &  7.38 & $<  0.04\pm 0.01$  & 28.20 & $<  0.11\pm 0.04$ \\
 & & & & 36.50 & $<  0.21\pm 0.05$ \\
 
2014 Mar 23.6 & 56739.6 &   56.6 & A &  4.55 & $<  0.06\pm 0.01$ &  7.38 & $<  0.04\pm 0.01$ & 29.50 & $  0.17\pm 0.03$ \\
 & & & & 35.00 & $  0.34\pm 0.05$ \\
 
 2014 Apr 12.5 & 56759.5 &   76.5 & A & 4.55 & $<  0.05\pm 0.01$ &  7.38 & $<  0.06\pm 0.01$ & 29.50 & $  0.30\pm 0.04$ \\
 & & & & 35.00 & $  0.32\pm 0.04$ \\
\enddata
\tablenotetext{a}{We take the time of discovery, 2014 Jan 26, as $t_0$.}
\end{deluxetable}

\begin{deluxetable}{cccccccccccc}
\tablewidth{0 pt}
\tabletypesize{\footnotesize}
\setlength{\tabcolsep}{0.025in}
\tablecaption{ \label{tab:v2659cyg}
VLA Observations of V2659~Cyg}
\tablehead{UT Obs Date & MJD & $t-t_0$\tablenotemark{a} & Config & $\nu$ & S$_{\nu}$ & $\nu$ & S$_{\nu}$ & $\nu$ & S$_{\nu}$ & $\nu$ & S$_{\nu}$ \\ 
  & & (Days) & & (GHz) & (mJy) & (GHz) & (mJy) & (GHz) & (mJy) & (GHz) & (mJy)}
\startdata

 2014 Apr 5.4 & 56752.4 &    6.4 & A & 4.55 & $<  0.03\pm 0.01$             
 &  7.38 & $<  0.04\pm 0.01$ & 29.50 & $<  0.12\pm 0.03$ & 35.00 & $<  0.12\pm 0.04$ \\             
                                   
 2015 Apr 22.4 & 56769.4 &   23.4 & A  & 4.55 & $<  0.04\pm 0.01$                                               
  &  7.38 & $<  0.04\pm 0.01$ 
 & 29.50 & $<  0.06\pm 0.02$                                              
& 35.00 & $<  0.07\pm 0.02$ \\        
                                        
 2015 May 10.4 & 56787.4 &  41.4 & A & 4.55 & $<  0.06\pm 0.01$                                                
&  7.38 & $<  0.03\pm 0.01$                                                
& 29.50 & $<  0.07\pm 0.02$                                                
& 35.00 & $<  0.08\pm 0.02$ \\      
                                       
2014 May 18.3 & 56795.3 &   49.3 & A & 4.55 & $<  0.05\pm 0.01$                                         
&  7.38 & $<  0.06\pm 0.01$                                               
& 29.50 & $  0.07\pm 0.02$                                                
 & 35.00 & $  0.08\pm 0.02$ \\           
                                      
 2015 May 24.4 & 56801.4 &   55.4 & A & 4.55 & $<  0.03\pm 0.01$                                              
 &  7.38 & $  0.05\pm 0.01$                                                 
 & 29.50 & $  0.13\pm 0.02$                                               
 & 35.00 & $  0.15\pm 0.02$ \\     
 
2014 Jun 15.2 & 56823.2 &   77.2 & AD &  4.55 & $<  0.12\pm 0.02$                                                
&  7.38 & $  0.08\pm 0.01$                                                
 & 13.50 & $  0.13\pm 0.01$                                                 
& 17.40 & $  0.13\pm 0.02$ \\
  & & & &    29.50 & $  0.25\pm 0.03$                                                
& 35.00 & $  0.34\pm 0.04$ \\
\enddata
\tablenotetext{a}{We take the time of discovery, 2014 Mar 30, as $t_0$.}
\end{deluxetable}

\begin{deluxetable}{cccccccccccc}
\tablewidth{0 pt}
\tabletypesize{\footnotesize}
\setlength{\tabcolsep}{0.025in}
\tablecaption{ \label{tab:v5667sgr}
VLA Observations of V5667~Sgr}
\tablehead{UT Obs Date & MJD & $t-t_0$\tablenotemark{a} & Config & $\nu$ & S$_{\nu}$ & $\nu$ & S$_{\nu}$ & $\nu$ & S$_{\nu}$ & $\nu$ & S$_{\nu}$ \\ 
  & & (Days) & & (GHz) & (mJy) & (GHz) & (mJy) & (GHz) & (mJy) & (GHz) & (mJy)}
\startdata

2015 Feb 19.6 & 57072.6 &    7.6 & B & 5.00 & $<  0.03\pm 0.01$                                                                   
 &  7.00 & $<  0.03\pm 0.01$                                                                    
& 29.50 & $<  0.06\pm 0.02$                                                                    
& 35.00 & $<  0.07\pm 0.02$ \\  
                                                                  
 2015 Mar 4.6 & 57085.6 &   20.6 & B & 5.00 & $<  0.03\pm 0.01$                                                                    
&  7.00 & $<  0.03\pm 0.01$                                                                    
& 29.50 & $<  0.09\pm 0.02$                                                                    
& 35.00 & $<  0.08\pm 0.03$ \\         
                                                           
 2015 Mar 26.5 & 57107.5 &   42.5 & B &  1.26 & $<  0.43\pm 0.09$                                                                    
&  1.80 & $<  0.15\pm 0.04$                                                                    
&  5.00 & $<  0.03\pm 0.01$                                                                   
 &  7.00 & $<  0.03\pm 0.01$ \\                                                                   
   & & & & 29.50 & $<  0.07\pm 0.02$                                                                    
 & 35.00 & $<  0.07\pm 0.02$ \\  
                                                                  
 2015 Apr 21.4 & 57133.4 &   68.4 & B &  1.26 & $<  0.34\pm 0.11$                                                                   
&  1.80 & $<  0.19\pm 0.05$                                                                   
 &  5.00 & $<  0.03\pm 0.01$                                                                   
 &  7.00 & $<  0.05\pm 0.01$ \\
\enddata
\tablenotetext{a}{We take the time of discovery, 2015 Feb 12, as $t_0$.}
\end{deluxetable}

\begin{deluxetable}{cccccccccccc}
\tablewidth{0 pt}
\tabletypesize{\footnotesize}
\setlength{\tabcolsep}{0.025in}
\tablecaption{ \label{tab:v5668sgr}
VLA Observations of V5668~Sgr}
\tablehead{UT Obs Date & MJD & $t-t_0$\tablenotemark{a} & Config & $\nu$ & S$_{\nu}$ & $\nu$ & S$_{\nu}$ & $\nu$ & S$_{\nu}$ & $\nu$ & S$_{\nu}$ \\ 
  & & (Days) & & (GHz) & (mJy) & (GHz) & (mJy) & (GHz) & (mJy) & (GHz) & (mJy)}
\startdata
2015 Mar 17.5 &  57098.5 &    2.5 &B & 1.26 & $<  1.05\pm 0.23$                                                                                                   
&  1.74 & $<  0.33\pm 0.11$                                                                                                    
&  5.00 & $<  0.04\pm 0.01$                                                                                                    
 &  7.00 & $<  0.03\pm 0.01$ \\                                                                                                   
    & & & &    29.50 & $<  0.25\pm 0.08$                                                                                               
    & 35.00 & $<  0.13\pm 0.04$ \\    
                                                                                                
2015 Mar 26.5 & 57107.5 &   11.5 & B & 1.26 & $<  0.52\pm 0.16$                                                                                                    
&  1.80 & $<  0.33\pm 0.07$                                                                                                    
&  5.00 & $<  0.04\pm 0.01$                                                                                                    
 &  7.00 & $<  0.04\pm 0.01$ \\                                                                                                   
    & & & &  29.50 & $<  0.08\pm 0.03$                                                                                                   
& 35.00 & $<  0.09\pm 0.03$ \\
\enddata
\tablenotetext{a}{We take the time of discovery, 2015 Mar 15, as $t_0$.}
\end{deluxetable}

\begin{deluxetable}{cccccccccccc}
\tablewidth{0 pt}
\tabletypesize{\footnotesize}
\setlength{\tabcolsep}{0.025in}
\tablecaption{ \label{tab:v5855sgr}
VLA Observations of V5855~Sgr}
\tablehead{UT Obs Date & MJD & $t-t_0$\tablenotemark{a} & Config & $\nu$ & S$_{\nu}$ & $\nu$ & S$_{\nu}$ & $\nu$ & S$_{\nu}$ & $\nu$ & S$_{\nu}$ \\ 
  & & (Days) & & (GHz) & (mJy) & (GHz) & (mJy) & (GHz) & (mJy) & (GHz) & (mJy)}
\startdata

2016 Nov 4.8 & 57696.8 &   15.8 & A &  1.26 & $<  0.20\pm 0.07$                                                                  
&  1.74 & $<  0.18\pm 0.06$                                                                       
&  5.00 & $<  0.07\pm 0.02$                                                                       
&  7.00 & $<  0.08\pm 0.02$ \\  
                                                                      
 2016 Nov 8.0 & 57700.0 &   19.0 & A & 1.26 & $<  0.21\pm 0.05$                                                                      
 &  1.74 & $<  0.15\pm 0.05$                                                                       
 &  5.00 & $  0.14\pm 0.02$                                                                        
&  7.00 & $  0.11\pm 0.02$ \\    
                                                                     
 2016 Nov 26.9 & 57718.9 &   37.9 & A &  1.26 & $<  0.24\pm 0.06$                                                                        
 &  1.74 & $<  0.32\pm 0.07$                                                                      
&  5.00 & $  0.25\pm 0.02$                                                                        
 &  7.00 & $  0.32\pm 0.02$ \\
                                                                         
 2017 Feb 7.6 & 57791.6 &  110.6 & AD &   1.26 & $<  3.13\pm 0.70$                                                                        
 &  1.74 & $<  0.96\pm 0.21$                                                                        
 &  5.00 & $  0.32\pm 0.03$                                                                         
 &  7.00 & $  0.53\pm 0.04$ \\
\enddata
\tablenotetext{a}{We take the time of discovery, 2015 Oct 20, as $t_0$.}
\end{deluxetable}

\begin{deluxetable}{cccccccccccc}
\tablewidth{0 pt}
\tabletypesize{\footnotesize}
\setlength{\tabcolsep}{0.025in}
\tablecaption{ \label{tab:v5856sgr}
VLA Observations of V5856~Sgr}
\tablehead{UT Obs Date & MJD & $t-t_0$\tablenotemark{a} & Config & $\nu$ & S$_{\nu}$ & $\nu$ & S$_{\nu}$ & $\nu$ & S$_{\nu}$ & $\nu$ & S$_{\nu}$ \\ 
  & & (Days) & & (GHz) & (mJy) & (GHz) & (mJy) & (GHz) & (mJy) & (GHz) & (mJy)}
\startdata
2016 Nov 11.9 &  57703.9 &   17.9 & A &  1.26 & $<  0.56\pm 0.13$                                                                                                           
&  1.74 & $<  0.25\pm 0.08$                                                                                                           
&  5.00 & $<  0.04\pm 0.01$                                                                                                           
 &  7.00 & $<  0.04\pm 0.01$ \\   
                                                                                                       
 2016 Nov 25.8 & 57717.8 &   31.8 & A &  13.50 & $<  0.04\pm 0.01$                                                                                                           
  & 16.50 & $<  0.05\pm 0.01$                                                                                                           
  & 29.50 & $<  0.08\pm 0.03$                                                                                                           
  & 35.00 & $<  0.16\pm 0.03$ \\  
                                                                                                        
 2016 Nov 25.9 & 57718.9 &   32.9 & A & 1.26 & $<  0.40\pm 0.13$                                                                                                           
  &  1.74 & $<  0.25\pm 0.08$                                                                                                           
  &  5.00 & $<  0.06\pm 0.02$                                                                                                           
  &  7.00 & $<  0.05\pm 0.02$ \\
\enddata
\tablenotetext{a}{We take the time of discovery, 2015 Oct 25, as $t_0$.}
\end{deluxetable}

\begin{deluxetable}{cccccccccc}
\tablewidth{0 pt}
\tabletypesize{\footnotesize}
\setlength{\tabcolsep}{0.025in}
\tablecaption{ \label{tab:v357mus}
ATCA Observations of V357~Mus}
\tablehead{UT Obs Date & MJD & $t-t_0$\tablenotemark{a} & Config & $\nu$ & S$_{\nu}$ & $\nu$ & S$_{\nu}$ & $\nu$ & S$_{\nu}$ \\ 
  & & (Days) & & (GHz) & (mJy) & (GHz) & (mJy) & (GHz) & (mJy) }
\startdata
 2018 Jan 18.8 & 58136.8 &    4.8 & 750A &  5.5 & $<  0.06$  
  &  9.0 & $<  0.04$ \\
 
 2018 Feb 4.9 & 58153.9 &   22.8 & 750A & 5.5 & $  0.10\pm 0.02$  
 &  9.0 & $  0.22\pm 0.02$ \\
 
 2018 Feb 22.8 & 58171.8 &   38.0 & 750B & 5.5 & $  0.44\pm 0.03$  
 &  9.0 & $  1.16\pm 0.06$ \\
 
 2018 Mar 19.5 & 58196.5 &   64.0 & EW352 &  1.3 & $<0.7$ 
&  1.8 & $  4.0\pm 0.4$ 
 &  2.4 & $  7.8\pm 0.8$ \\
& & & &  2.9 & $ 11.6\pm 1.2$ 
& 5.5 & $ 31.3\pm 1.6$  
 &  9.0 & $ 45.9\pm 2.3$ \\
 
 2018 Apr 3.3 & 58211.3 &   79.0 & EW352 & 5.5 & $ 11.9\pm 1.4$  
 &  9.0 & $ 13.6\pm 1.7$ \\
 
 2018 May 13.2 & 58251.2 &  119.0 & 6D &  5.5 & $  2.1\pm 0.4$  
 &  9.0 & $  4.1\pm 0.8$ \\
 
 2018 Jun 16.2 & 58285.2 &  153.2 & 1.5A &  5.5 & $  1.5\pm 0.2$  
&  9.0 & $  2.9\pm 0.7$ \\
\enddata
\tablenotetext{a}{We take the time of discovery, 2018 Jan 14, as $t_0$.}
\end{deluxetable}

\begin{deluxetable}{cccccccccc}
\tablewidth{0 pt}
\tabletypesize{\footnotesize}
\setlength{\tabcolsep}{0.025in}
\tablecaption{ \label{tab:v906car}
ATCA Observations of V906~Car}
\tablehead{UT Obs Date & MJD & $t-t_0$\tablenotemark{a} & Config & $\nu$ & S$_{\nu}$ & $\nu$ & S$_{\nu}$ & $\nu$ & S$_{\nu}$ \\ 
  & & (Days) & & (GHz) & (mJy) & (GHz) & (mJy) & (GHz) & (mJy) }
\startdata
2018 Apr 3.3 &  58211.3 & 18.3 & EW352 & 5.5 & $<0.3$ & 9.0 & $<0.1$ & & \\
% & & & & & & & & \\

2018 May 13.2 &  58251.2 &  58.2 & 6D &  5.5 & $0.4\pm0.1$ & 9.0 & $0.5\pm0.1$  & & \\             
% & & & & & & & & \\

2018 June 16.2 &  58285.2 & 92.2 & 1.5A & 5.5 & $0.3\pm0.1$& 9.0 & $0.6\pm0.1$  & & \\
% & & & & & & & & \\

2018 July 13.6 &  58312.6 & 119.6 & H75 & 5.5 & $0.3\pm0.1$ & 9.0 & $0.4\pm0.1$  & & \\
% & & & & & & & & \\
	
2018 Aug 25.0 &  58355.0 &162.0 & H214 & 5.5 & $1.3\pm0.1$ & 9.0 & $1.2\pm0.1$  & & \\
% & & & & & & & & \\

2018 Oct 5.9 &  58396.9 &203.9 & 6A & 1.3 & $<0.8$ & 1.8 & $0.5\pm0.2$ & 2.4 & $0.5\pm0.1$  \\
 & & & & 2.9 & $0.6\pm0.2$ & &  & & \\
% & & & & & & & & \\

2018 Oct 10.9 &  58401.9 & 208.9 & 6A  & 5.5 & $1.7\pm0.2$ & 9.0 & $1.7\pm0.3$& & \\
\enddata
\tablenotetext{a}{We take the time of discovery, 2018 Mar 16, as $t_0$.}
\end{deluxetable}

\begin{deluxetable}{cccccccccccc}
\tablewidth{0 pt}
\tabletypesize{\footnotesize}
\setlength{\tabcolsep}{0.025in}
\tablecaption{ \label{tab:v392per}
VLA Observations of V392~Per}
\tablehead{UT Obs Date & MJD & $t-t_0$\tablenotemark{a} & Config & $\nu$ & S$_{\nu}$ & $\nu$ & S$_{\nu}$ & $\nu$ & S$_{\nu}$ & $\nu$ & S$_{\nu}$ \\ 
  & & (Days) & & (GHz) & (mJy) & (GHz) & (mJy) & (GHz) & (mJy) & (GHz) & (mJy)}
\startdata
 2018 Apr 30.9 & 58238.9 &    1.9 & A & 1.26 & $<  0.16\pm 0.05$                                                                                             
  &  1.74 & $<  0.15\pm 0.04$                                                                                             
  &  5.00 & $<  0.05\pm 0.01$                                                                                             
 &  7.00 & $<  0.04\pm 0.01$ \\       
                                                                  
 2018 May 8.7 & 58246.7 &    9.7 & A & 5.00 & $  0.23\pm 0.02$  
 &  7.00 & $  0.30\pm 0.02$  
 & 13.50 & $  0.43\pm 0.05$  
 & 16.50 & $  0.34\pm 0.04$ \\
    & & & & 29.50 & $  0.97\pm 0.10$  
 & 35.00 & $  0.91\pm 0.09$ \\
 
 2018 May 12.6 & 58250.6 &   13.6 & A &  1.26 & $<  0.26\pm 0.05$  
  &  1.74 & $  0.15\pm 0.04$  
  &  5.00 & $  0.50\pm 0.03$  
  &  7.00 & $  0.75\pm 0.04$ \\
 
 2018 May 17.7 & 58255.7 &   18.7 & A & 13.50 & $  7.38\pm 0.74$                              
  & 16.50 & $  8.32\pm 0.83$ 
  & 29.50 & $ 11.85\pm 1.19$  
  & 35.00 & $ 12.99\pm 1.30$ \\
 
 2018 May 18.6 & 58256.6 &   19.6 &  A & 1.26 & $  0.63\pm 0.06$  
  &  1.74 & $  0.95\pm 0.06$  
  &  5.00 & $  3.23\pm 0.16$  
  &  7.00 & $  4.20\pm 0.21$ \\
 
 2018 May 23.0 & 58261.0 &   24.0 & A &  1.26 & $  1.58\pm 0.10$  
  &  1.74 & $  2.02\pm 0.11$  
  &  5.00 & $  4.52\pm 0.23$  
  &  7.00 & $  5.89\pm 0.29$ \\
 
 2018 May 31.6 &58269.6 &   32.6 &A &  1.26 & $  3.47\pm 0.18$  
  &  1.74 & $  3.89\pm 0.20$  
  &  5.00 & $  5.23\pm 0.26$  
  &  7.00 & $  5.99\pm 0.30$ \\
\enddata
\tablenotetext{a}{We take the time of discovery, 2018 Apr 29, as $t_0$.}
\end{deluxetable}

%=================================================

\clearpage
\appendix

Some novae are observed for a few epochs at radio wavelengths, but then observations are dropped---sometimes due to non-detections, sometimes due to scheduling difficulties. The resulting coverage is not sufficient to illuminate the light curve evolution, but could be useful for future  studies, so we include these data here. While we strive to include data for all novae that were observed more than twice with the VLA (and a few that were observed even more sparsely), we cannot guarantee that all such targets are fully captured here.

\section{OS~And}
OS~And was observed with the VLA in six epochs following its 1986 outburst under programs AH185 and AH254 (PI R.\ Hjellming). No confident detections were obtained; the results are collated in Table \ref{tab:osand}.

\section{QV~Vul}
QV~Vul was observed with the VLA in five epochs following its 1987 outburst under programs AH254 and AH301 (PI R.\ Hjellming). No confident detections were obtained, as detailed  in Table \ref{tab:qvvul}.

\section{V5587~Sgr}
V5587~Sgr was observed with the Jansky VLA in three epochs following its 2011 outburst under programs VLA/10B-200 and 11A-254. As tabulated in Table \ref{tab:v5587sgr}, all observations yielded non-detections.

\section{V5588~Sgr}
V5588~Sgr was observed with the Jansky VLA in six epochs following its 2011 outburst under programs VLA/11A-254 and 11A-271. As tabulated in Table \ref{tab:v5588sgr}, it shows the start of a rise at high frequency (33 GHz), while the 5.9 GHz observations yield mostly non-detections (the data are consistent with a marginal detection on 2011 May 15). All detections were consistent with an unresolved point source.

\section{V1312~Sco}
V1312~Sco was observed with the Jansky VLA in four epochs following its 2011 outburst under program VLA/11A-280. As tabulated in Table \ref{tab:v1312sco}, all observations yielded non-detections.

\section{V2676~Oph}
V2676~Oph was observed with the Jansky VLA in four epochs following its 2012 outburst under programs VLA/11B-170, 12A-479, and 12A-483. As tabulated in Table \ref{tab:v2676oph}, it shows the start of a rise at high frequency (33 GHz), while the 5.9 GHz observation yields a marginal detection. All detections were consistent with an unresolved point source.

\section{V2677~Oph}
V2677~Oph was observed with two Jansky VLA observations following its 2012 outburst under program VLA/12A-483. It was observed on 2012 Jun 26.5 at C band, when it  was detected with a flux density of $0.20\pm0.01$ mJy at 5.0 GHz and $0.27\pm0.01$ mJy at 6.8 GHz. Ka band observations were subsequently obtained on 2012 Jun 28.1 at Ka band, yielding a detection of  $0.45\pm0.11$ mJy at 33.0 GHz.
Both observations were obtained in the B configuration, and V2677~Oph appeared as a point source.

\section{V1533 Sco}
V1533~Sco was observed with the Jansky VLA in two epochs following its 2013 outburst under program VLA/13A-455. The first took place on 2013 Jun 12.3, and the second on 2013 Jul 12.2; both observed at C band  and Ka band. All observations yielded non-detections, placing upper limits $<$0.01 mJy at 5.9 GHz and $<$0.25 at 34 GHz.

\section{V5853~Sgr}
V5853~Sgr  was observed with the Jansky VLA in three epochs following its 2016 outburst under program VLA/16A-318. 
%really four but the fourth wasn't reduced alas
As tabulated in Table \ref{tab:v5853sgr}, it shows the start of a rise at high frequency (33 GHz), while the last 7 GHz observation yields a marginal detection. All detections were consistent with an unresolved point source.

%\section{V612 Sct}
%wiki page didn't copy so i don't include it here, :(

\begin{deluxetable}{ccccccccc}
\tablewidth{0 pt}
\tabletypesize{\footnotesize}
\setlength{\tabcolsep}{0.025in}
\tablecaption{ \label{tab:osand}
VLA Observations of OS~And}
\tablehead{UT Obs Date & MJD & $t-t_0$\tablenotemark{a} & 4.9 GHz S$_{\nu}$ & 14.9 GHz S$_{\nu}$ & Config \\ 
  & & (Days) & (mJy) & (mJy) & & }
\startdata
1986 Dec 10.2 &  46774.2 &    5.2 &  $<  0.25\pm 0.06$ & $  <0.18\pm 0.06$ &   C \\
%is this really a detection at u band?
1986 Dec 21.0 &  46785.1 &   16.1 &		& $<  0.54\pm 0.17$ &   C \\
1986 Dec 24.1 &  46788.1 &   19.1 & $<  0.21\pm 0.07$ &   	& C \\
1987 Jan 17.8 &  46812.8 &   43.8 & $<  0.41\pm 0.14$ & $<  0.53\pm 0.14$ &   C \\
1987 Jan 24.6 &  46819.7 &   50.7 & $<  0.19\pm 0.06$ &  $<  0.55\pm 0.18$ &  CD \\
1987 Jul  5.3 &  46981.3 &  212.3 & $<  0.23\pm 0.08$ &  	& A \\
\enddata
\tablenotetext{a}{We take the time of discovery, 1986 Dec 5, as $t_0$.}
\end{deluxetable}

\begin{deluxetable}{ccccccccc}
\tablewidth{0 pt}
\tabletypesize{\footnotesize}
\setlength{\tabcolsep}{0.025in}
\tablecaption{ \label{tab:qvvul}
VLA Observations of QV~Vul}
\tablehead{UT Obs Date & MJD & $t-t_0$\tablenotemark{a} & 1.5 GHz S$_{\nu}$ & 4.9 GHz S$_{\nu}$ & 8.4 GHz S$_{\nu}$ & 14.9 GHz S$_{\nu}$& Config \\ 
  & & (Days) & (mJy) & (mJy) & (mJy) & (mJy) & }
\startdata
1987 Nov 28 &  47127 &   13.5 &  & $<  0.30\pm 0.10$ & & $<  0.78\pm 0.24$ & B  \\
1988 Jan  9 &  47169 &   55.5 &   & $<  0.41\pm 0.08$ & & $<  1.00\pm 0.21$ &  B\\
1988 Jan 24  &  47184 &   70.5 &   & $  0.76\pm 0.23$ &  & $<  0.93\pm 0.31$ &  B \\
1988 Oct 24  &  47458 &  344.5 & $<  0.31\pm 0.10$ & $<  0.28\pm 0.09$ & $<  0.19\pm 0.05$ & $<  0.64\pm 0.21$ &  A\\
1989 Feb 14 &  47571 &  457.5 &  & $<  0.13\pm 0.04$ & & & AB\\
\enddata
\tablenotetext{a}{We take the time of discovery, 1987 Nov 15, as $t_0$.}
\end{deluxetable}

\begin{deluxetable}{ccccccccc}
\tablewidth{0 pt}
\tabletypesize{\footnotesize}
\setlength{\tabcolsep}{0.025in}
\tablecaption{ \label{tab:v5587sgr}
VLA Observations of V5587~Sgr}
\tablehead{UT Obs Date & MJD & $t-t_0$\tablenotemark{a} & 5.9 GHz S$_{\nu}$ & 33.1 GHz S$_{\nu}$ & Config \\ 
  & & (Days) & (mJy) & (mJy) & & }
\startdata
2011 Feb 1.6 & 55593.6 & 7.6  & $<0.03\pm0.01$ & & BC \\
2011 Feb 8.5 & 55600.5 & 14.5 & & $< 0.14\pm0.05$ & BC \\
2011 Mar 5.5 & 55625.5 & 39.5 & $< 0.03\pm0.01$ & $<0.27\pm0.09$ & B \\

\enddata
\tablenotetext{a}{We take the time of discovery, 2011 Jan 25, as $t_0$.}
\end{deluxetable}

\begin{deluxetable}{ccccccccc}
\tablewidth{0 pt}
\tabletypesize{\footnotesize}
\setlength{\tabcolsep}{0.025in}
\tablecaption{ \label{tab:v5588sgr}
VLA Observations of V5588~Sgr}
\tablehead{UT Obs Date & MJD & $t-t_0$\tablenotemark{a} & 5.9 GHz S$_{\nu}$ & 33.1 GHz S$_{\nu}$ & Config \\ 
  & & (Days) & (mJy) & (mJy) & & }
\startdata
2011 Apr 21.5 & 55672.5 & 25.5 & $< 0.03\pm0.01$ & & B \\
2011 Apr 30.3 & 55681.3 & 34.3 &  &  $<0.21\pm0.05$ & B\\
2011 May 1.3 & 55682.3 & 35.3 &  $<0.03\pm0.01$  & & B\\
2011 May 14.5 & 55695.5  & 48.5 & & $0.39\pm0.06$ & AB\\ 
2011 May 15.4 & 55696.4  & 49.4 &   $0.05\pm0.01$ &  &AB \\
2011 Jun 2.3 & 55714.3  &  67.3 & $<0.04\pm0.01$  & $<0.13\pm0.04$ & AB \\
2011 Jun 15.4 & 55727.4	& 80.4 &  &  $0.84\pm0.04$ & A\\
2011 Jul 27.1 & 55769.1 & 122.1 &  $<0.03\pm0.01$ & &A \\
\enddata
\tablenotetext{a}{We take the time of discovery, 2011 Mar 27, as $t_0$.}
\end{deluxetable}

\begin{deluxetable}{ccccccccc}
\tablewidth{0 pt}
\tabletypesize{\footnotesize}
\setlength{\tabcolsep}{0.025in}
\tablecaption{ \label{tab:v1312sco}
VLA Observations of V1312~Sco}
\tablehead{UT Obs Date & MJD & $t-t_0$\tablenotemark{a} & 6 GHz S$_{\nu}$ & 33 GHz S$_{\nu}$ & Config \\ 
  & & (Days) & (mJy) & (mJy) & & }
\startdata
2011 Jul 12.1 & 55754.1 & 41.1 & & $<0.16$ & A\\
2011 Jul 26.1 &  55768.1 & 55.1 & $<0.03$ & &A\\
2011 Aug 9.1 & 55782.1 & 69.1 & &  $<0.11$ & A\\
2011 Aug 19.1 & 55792.1 & 79.1 & $<$0.02 & & A\\

\enddata
\tablenotetext{a}{We take the time of discovery, 2011 Jun 1, as $t_0$.}
\end{deluxetable}

\begin{deluxetable}{ccccccccc}
\tablewidth{0 pt}
\tabletypesize{\footnotesize}
\setlength{\tabcolsep}{0.025in}
\tablecaption{ \label{tab:v2676oph}
VLA Observations of V2676~Oph}
\tablehead{UT Obs Date & MJD & $t-t_0$\tablenotemark{a} & 5.9 GHz S$_{\nu}$ & 32.0 GHz S$_{\nu}$ & Config \\ 
  & & (Days) & (mJy) & (mJy) & & }
\startdata
2012 Apr 23.5 & 56040.5 & 29.5 & &	$<0.10\pm0.03$ &	C	\\		
2012 Jun 26.3 & 56104.3 & 93.3 & $0.04\pm0.01$ & &	B\\
2012 Jun 28.1 & 56106.1 & 95.1 &  & $<0.28\pm0.09$ &	B	\\
2012 Jul 22.3 & 56130.3 & 119.3 &  & $0.46\pm0.08$ &	B \\
\enddata
\tablenotetext{a}{We take the time of discovery, 2012 Mar 25, as $t_0$.}
\end{deluxetable}

\begin{deluxetable}{ccccccccc}
\tablewidth{0 pt}
\tabletypesize{\footnotesize}
\setlength{\tabcolsep}{0.025in}
\tablecaption{ \label{tab:v5853sgr}
VLA Observations of V5853~Sgr}
\tablehead{UT Obs Date & MJD & $t-t_0$\tablenotemark{a} & 5 GHz S$_{\nu}$ & 7 GHz S$_{\nu}$ & 29.5 GHz S$_{\nu}$ & 35.0 GHz S$_{\nu}$ & Config \\ 
  & & (Days) & (mJy) & (mJy) & (mJy) & (mJy) & }
\startdata
2016 Aug 30.6 & 57630.6 & 24.6 &$<0.05\pm0.02$ & $<0.04\pm0.01$ & $<0.11\pm0.04$ & $<0.15\pm0.04$	& B\\	
2016 Sep 18.0 & 57649.0	& 43.0 &  $0.04\pm0.01$ & $<0.03\pm0.01$ & $0.11\pm0.03$ & $0.23\pm0.03$ & AB\\	
2016 Oct 5.1 & 57666.1 & 60.1 &  $<0.05\pm0.02$ & $	<0.05\pm0.02$ & $0.44\pm0.03$ & $0.65\pm0.04$ & A\\
\enddata
\tablenotetext{a}{We take the time of discovery, 2016 Aug 6, as $t_0$.}
\end{deluxetable}

\end{document}